\documentclass[a4paper,onecolumn,11pt,longbibliography,accepted=2026-03-23]{quantumarticle}
\pdfoutput=1

\usepackage[utf8]{inputenc}
\usepackage[english]{babel}
\usepackage[T1]{fontenc}
\usepackage{hyperref}
\usepackage{amsmath,amssymb,amsthm,mathrsfs,amsfonts,dsfont}
\usepackage[dvips]{graphicx}
\usepackage{braket}
\usepackage{bm}
\usepackage{enumerate}
\usepackage{color}
\usepackage[normalem]{ulem}
\usepackage{booktabs}
\usepackage{siunitx}
\usepackage[square,sort,compress,numbers]{natbib}
\usepackage{tabularx}
\usepackage{widetable}
\usepackage{silence}
\usepackage[table]{xcolor}
\usepackage[labelformat=simple]{subcaption}

\newcolumntype{C}{>{\centering\arraybackslash}X}
\newcolumntype{P}{>{\centering\arraybackslash}p{3cm}}

\newtheorem{theorem}{Theorem}
\newtheorem*{theorem*}{Theorem}

\usepackage{algorithm}
\usepackage{algpseudocode}

\begin{document}
\title{Efficient and high-performance routing of lattice-surgery paths on three-dimensional lattice}

\author{Kou Hamada}
\email{zkouaaa@g.ecc.u-tokyo.ac.jp}
\affiliation{Department of Mathematical Informatics, Graduate School of Information Science and Technology, The University of Tokyo, 7-3-1 Hongo, Bunkyo-ku, Tokyo 113-0033, Japan}

\author{Yasunari Suzuki}
\email{yasunari.suzuki@riken.jp}
\affiliation{NTT Computer and Data Science Laboratories, NTT Corporation, Musashino 180-8585, Japan}
\affiliation{JST, PRESTO, 4-1-8 Honcho, Kawaguchi, Saitama, 332-0012, Japan}
\affiliation{Center for Quantum Computing, RIKEN, 2-1 Hirosawa, Wako, Saitama 351-0198, Japan}

\author{Yuuki Tokunaga}
\email{tokunaga@ist.hokudai.ac.jp}
\affiliation{NTT Computer and Data Science Laboratories, NTT Corporation, Musashino 180-8585, Japan}
\affiliation{Faculty of Information Science and Technology, Hokkaido University, Sapporo 060-0814, Japan}

\begin{abstract}
  Encoding logical qubits with surface codes and performing multi-qubit logical operations with lattice surgery is one of the most promising approaches to demonstrate fault-tolerant quantum computing.
  Thus, a method to efficiently schedule a sequence of lattice-surgery operations is vital for high-performance fault-tolerant quantum computing.
  A possible strategy to improve the throughput of lattice-surgery operations is splitting a large instruction into several small instructions, such as Bell state preparation and measurements, and executing a part of them in advance. However, scheduling methods to fully utilize this idea have yet to be explored.
  In this paper, we propose a fast and high-performance scheduling algorithm for lattice-surgery instructions leveraging this strategy. We achieved this by converting the scheduling problem of lattice-surgery instructions to a graph problem of embedding 3D paths into a 3D lattice, which enables us to explore efficient scheduling by solving path search problems in the 3D lattice. Based on this reduction, we propose a method to solve the path-finding problems, the look-ahead Dijkstra projection. We numerically show that this method reduced the execution time of benchmark programs generated from quantum phase estimation algorithms by 3.8 times compared with a naive method based on greedy algorithms. Our study establishes the relation between the lattice-surgery scheduling and graph search problems, which leads to further theoretical analysis on compiler optimization of fault-tolerant quantum computing.
\end{abstract}
\maketitle

\section{Introduction}
Quantum computing can efficiently solve several vital scientific problems~\cite{shor1999polynomial,kitaev1995quantum,lloyd1996universal}, but its high error rates prevent us from a practical demonstration of quantum advantage. We can overcome this problem with a quantum error-correction technology, i.e., by creating a logical qubit with a negligible error rate by using several noisy qubits. Surface codes~\cite{kitaev1997quantum,bravyi1998quantum,fowler2012surface} are known as promising candidates for quantum error-correcting codes since they show high error-correction performance and can be implemented with qubits having nearest-neighbor interactions on a two-dimensional~(2D) grid lattice. Lattice surgery~\cite{horsman2012surface,fowler2018low} plays a key role in the fault-tolerant quantum computation with surface-code logical qubits: they allow multi-qubit Pauli measurements in a fault-tolerant manner by connecting target logical qubits through ancillary logical-qubit space, and they can be performed in parallel as long as their paths do not overlap. The current state-of-the-art resource estimation stands on architectures with surface codes and lattice surgery~\cite{beverland2022assessing,babbush2018encoding,yoshioka2022hunting}.

To run quantum algorithms with a short execution time, we should maximize the throughput of quantum instructions~\cite{javadiabhari2014scaffcc,Qiskit,sivarajah2020t,watkins2023high} by optimizing their scheduling. Unfortunately, finding the optimal scheduling of lattice-surgery instructions is known to be NP-hard~\cite{herr2017optimization,molavi2023compilation}. Thus, fast and near-optimal approximation strategies for scheduling lattice-surgery instructions are demanded. For frequently used modules such as magic-state distillation circuits, there are dedicated designs and logical-qubit mappings~\cite{fowler2018low,litinski2019game,litinski2019magic}. On the other hand, for general purposes, we need an algorithm that can accept an arbitrary lattice-surgery sequence.
One possible strategy is to map problems into well-known NP-hard instances and find high-quality solutions with fast solvers. Lao~\textit{et al.}~\cite{lao2018mapping} map problems into the quadratic assignment problem, and Molavi~\textit{et al.}~\cite{molavi2023compilation} into SAT problems. While this method would show near-optimal solutions, the target size of the compilation is limited by the capability of solvers. Thus, efficient and high-performance scheduling methods based on other concepts are still in demand.

One of the promising approaches for general and high-performance optimization is to split instructions that are difficult to parallelize into several small fragments and execute parts of them in advance. Beverland~\textit{et al.}~\cite{beverland2022surface} proposed a method based on this idea. They propose methods to find a way to split and schedule instructions by finding the edge-disjoint path on a 2D lattice and converting the solution to a lattice-surgery sequence for multiple execution cycles. While this strategy is shown to be effective, the proposed method sticks to specific logical-qubit allocation patterns while possible logical-qubit allocations can be arbitrary~\cite{babbush2018encoding,beverland2022assessing,yoshioka2022hunting}, which loses the versatility in compilation targets. Also, the range of optimization in this algorithm is limited to instructions within a narrow time window. Therefore, we need a strategy that can fully leverage the idea of instruction split and preceding execution of them.

Here, we propose an efficient and high-performance lattice-surgery scheduling algorithm that utilizes the strategy. The key step of our proposal is establishing the correspondence between a path in a 3D lattice, an executable sequence of lattice-surgery instructions, and achievable actions on target logical qubits. Thanks to this correspondence, we can rephrase the optimization of a lattice-surgery sequence as a problem to embed several 3D paths in a 3D grid lattice, in which we can utilize graph optimization methods. On the basis of this result, we show several polynomial-time approximation algorithms to solve this problem. We evaluated these algorithms with numerical simulations and found that one among them, named look-ahead Dijkstra projection, significantly improves throughput with a reasonable compilation time. We numerically showed that the look-ahead Dijkstra projection improves throughput by 3.8 times compared with a naive solution in a benchmark with realistic applications, \texttt{SELECT} modules in qubitization-based quantum phase estimation~\cite{low2019hamiltonian,babbush2018encoding}.

Our results not only optimize the runtime of quantum algorithms but also show that the complicated lattice-surgery compilation can be reduced to simple graph problems, where we can naturally utilize the knowledge of path-search algorithms. Since our algorithm works for general logical-qubit mapping and connectivity, our methods can be extended to other fault-tolerant quantum computing~(FTQC) designs based on lattice surgery or distributed quantum computing.
While our benchmark focuses on the case of two-body lattice-surgery instructions, we also showed that our method can be straightforwardly extended to multi-body lattice surgery in Appendix~\ref{app:manybody_lattice_surgery}.
Thus, our results offer a base for versatile and high-performance compilation for fault-tolerant quantum computing.

Our contribution can be summarized as follows:
\begin{itemize}
  \item We extend the lattice-surgery scheduling problem from a 2D routing problem to a 3D routing problem. Since our formulation strictly contains the original problem as a special case, it always improves the achievable solutions equal to or superior to the conventional 2D approaches.
  \item We propose an efficient and high-performance scheduling method for our 3D formulation, called look-ahead Dijkstra projection. This algorithm leverages the properties of the extended formalism with only a modest increase in time complexity compared to simple baseline algorithms.
  \item We numerically evaluate the performance of the proposed algorithm and demonstrate a 3.8-times speedup on benchmark programs derived from practical applications compared with a baseline method. Although the compilation time increases by a factor of 7.0 in the worst scenario, it still completes within a few tens of seconds even for practical-scale applications. We also show that other sources of overhead, such as increased logical error rates, remain modest or negligible.
\end{itemize}

\section{Preliminary}

\subsection{Surface codes and logical operations}
Since qubits suffer from large error rates, we need quantum error correction to suppress their error rate to an arbitrarily small value. Surface codes are known as the most promising quantum error-correcting codes, which encode the information of a qubit into a square cell of qubits. The width of the cell corresponds to a code distance, and by increasing the code distance, we can exponentially suppress the error rates if physical error rates are smaller than the value known as thresholds~\cite{kitaev1997quantum,fowler2012towards,acharya2022suppressing}. A cell of surface codes has two types of boundaries: $X$-type and $Z$-type, where Pauli-$X$ operations along with $X$-type boundary constitute logical $X$ operations, and the same for Pauli-$Z$.

\begin{figure}[tb]
  \centering
  \includegraphics[width=0.6\columnwidth]{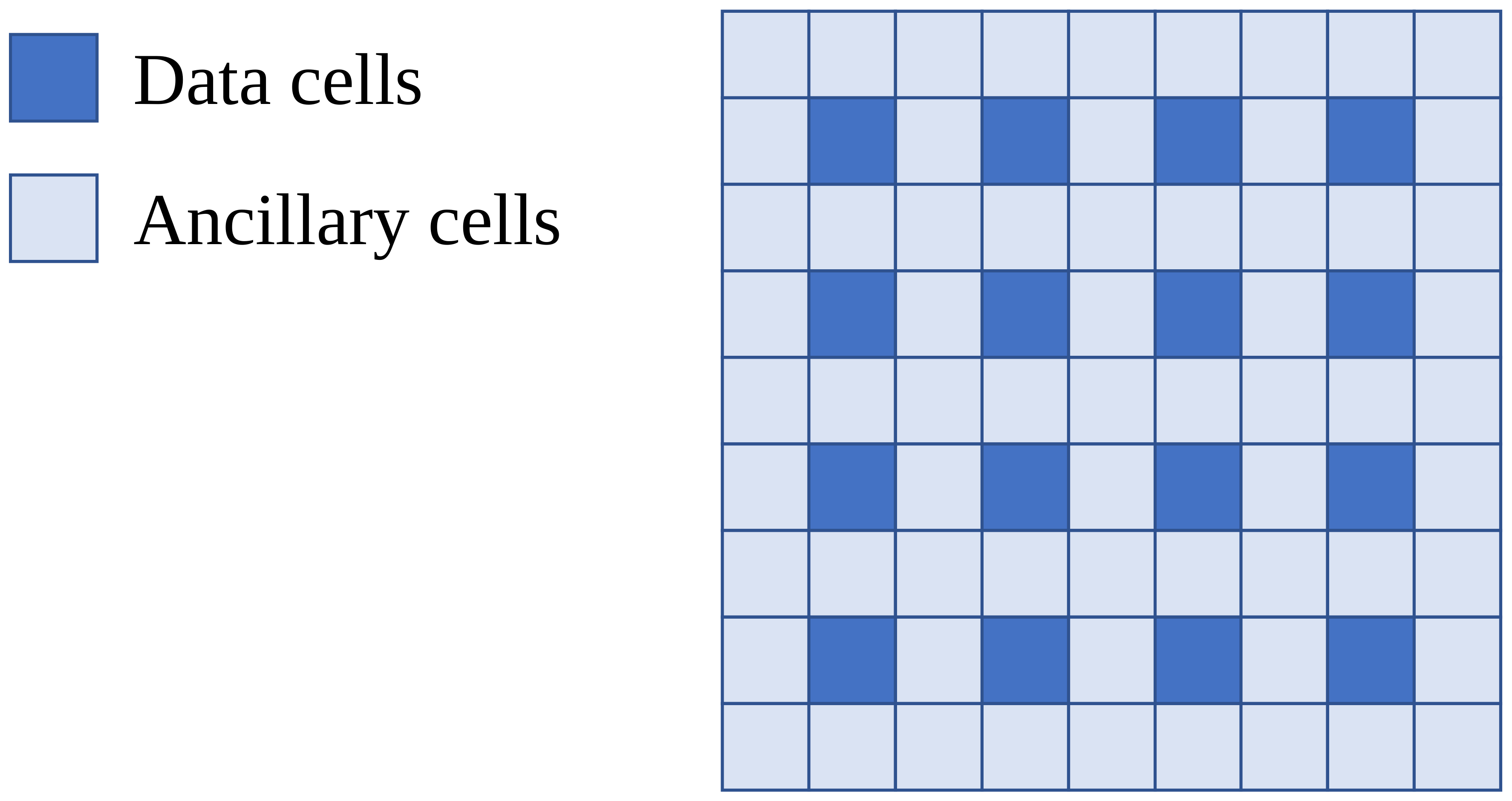}
  \caption{Data allocation in a qubit plane.}
  \label{fig:qubit_plane}
\end{figure}

We suppose that qubits are integrated on a large two-dimensional plane, which we call a qubit plane. We assume that the qubit plane is divided into several surface-code logical-qubit cells. During the computation, several cells are used for keeping single-qubit information, which we call {\it data cells}. The other cells are available as working space for logical operations on the data cells, which we call {\it ancillary cells}. Fig.\,\ref{fig:qubit_plane} shows an example of a qubit plane, where dark cells are data cells and light cells are ancillary cells. As surface codes have the two types of boundaries, each cell also has two types of directions: top and bottom are $X$-boundaries, and the others are $Z$-boundaries; or top and bottom are $Z$-boundaries, and the others are $X$-boundaries.

To execute quantum algorithms, we need to perform universal quantum operations on encoded logical qubits fault-tolerantly. The most standard set of logical operations is as follows. We can initialize a cell in a $Z$-basis ($\ket{0}, \ket{1}$) or $X$-basis ($\ket{+}, \ket{-}$) and can measure them in $X$- and $Z$-bases destructively. These operations can be achieved with a single code cycle, i.e., with effectively negligible time compared with the other operations. Note that the direction of $X/Z$-type boundaries can be freely chosen at the initialization timing. We can also fault-tolerantly perform single-qubit logical operations, i.e., Hadamard gates, phase gates with a twist, T gates via magic-state injection, distillation, and teleportation~\cite{fowler2012towards,brown2017poking,fowler2018low}. Finally, we can achieve multi-qubit logical operations via lattice surgery~\cite{horsman2012surface}, which acts on code space as multi-qubit Pauli measurements.

Hereafter, the discussion will focus exclusively on logical qubits. Thus, the term ``qubit'' refers to a logical qubit unless specified otherwise.

\subsection{Scheduling of lattice surgery}
Among the universal operations, the scheduling of lattice-surgery instructions tends to be a complicated task since multiple cells are relevant to them. Thus, in this paper, we focus on the scheduling of lattice-surgery instructions. A lattice-surgery instruction is achieved by connecting target logical-qubit cells. If the target cells are not neighboring, we can use ancillary cells to connect them. Once connected, we need to repeat syndrome measurements for $d$ times, where $d$ is a code distance, to guarantee reliable parity-check measurements. Thus, it is convenient to define this duration of $d$ code cycles for lattice surgery as a time unit, which is called a {\it code beat}. The types of connected boundaries determine the basis of Pauli measurements. We can perform Pauli-$Z$ measurements if a $Z$-boundary of a cell is connected and Pauli-$X$ for $X$-boundary.

Throughout this paper, we assume the following for simplicity.
\begin{enumerate}
  \item We ignore the time for executing instructions other than the lattice surgery. In other words, we assume all the instructions are lattice surgery.
  \item While lattice surgery can act on an arbitrary number of cells, we only consider lattice-surgery instructions acting on two logical qubits with the same types of boundaries.
\end{enumerate}
The extension to many-body lattice surgery is detailed in Appendix~\ref{app:manybody_lattice_surgery}. The extensions to general cases will be left as future work and discussed in Sec.\,\ref{sec:conclusion}.

\begin{figure}[tb]
  \centering
  \includegraphics[width=0.7\columnwidth]{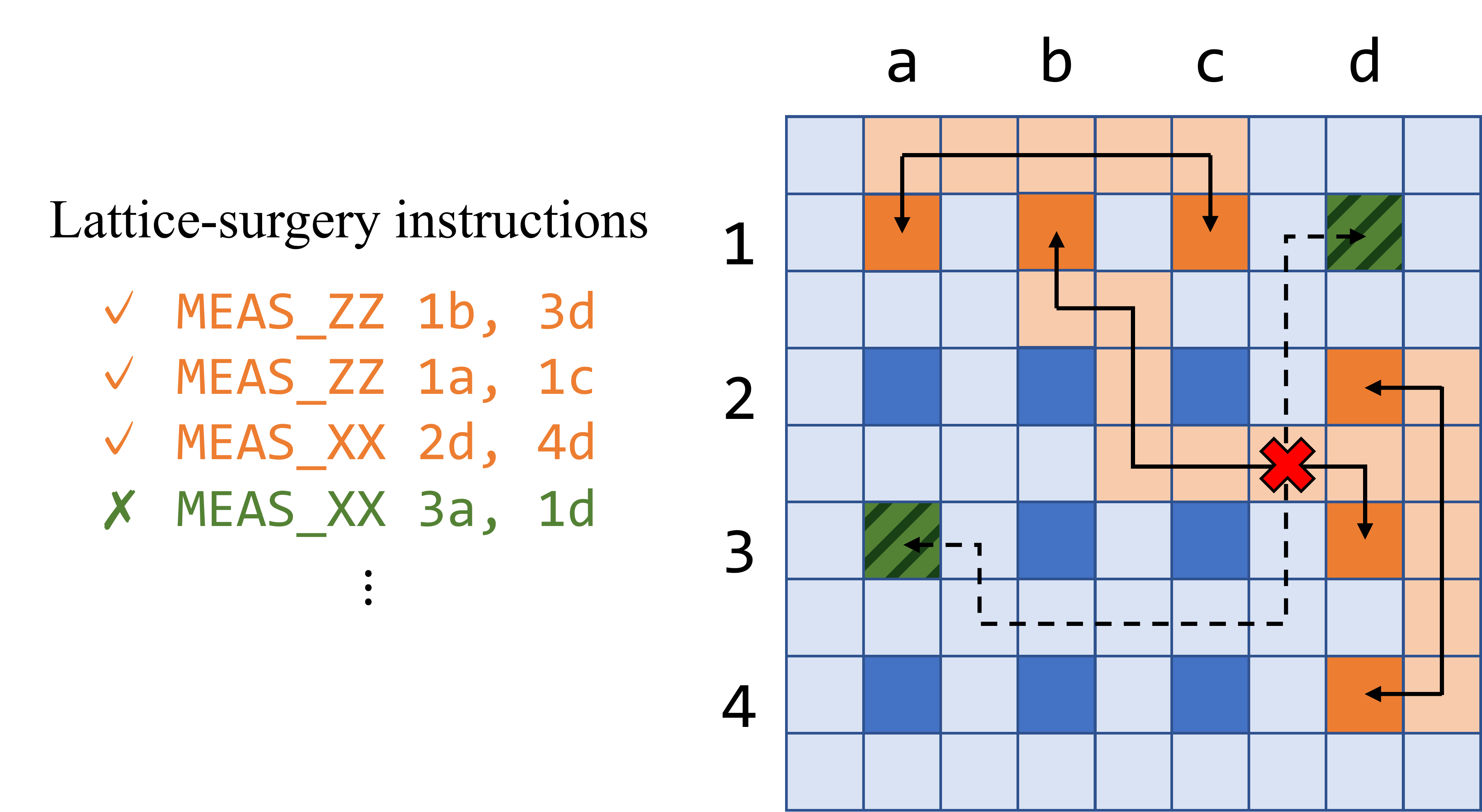}
  \caption{Example of not-parallelizable lattice-surgery instructions due to path conflict.}
  \label{fig:bfs}
\end{figure}

To maximize the computational speed, we should parallelize the instructions as much as possible. Suppose that lattice-surgery instructions are provided as a sequence, each of which describes two target logical qubits and the type of boundaries. Fig.\,\ref{fig:bfs} shows the example of parallel execution of lattice surgeries. In this figure, three pairs of orange logical qubits are connected simultaneously by using ancillary cells, which means three lattice-surgery instructions are performed at the same time. On the other hand, we cannot perform lattice surgery on two hatched green cells in this code beat since there are no paths between them. After a code beat, the cells occupied by orange lattice-surgery paths become available, and we can connect hatched green cells.

This example illustrates that optimizing the ordering and routing of lattice surgery is an important factor in maximizing the performance of quantum computing. Unfortunately, the optimal scheduling of the sequence of lattice surgery is known to be NP-hard, i.e., the optimal scheduling is computationally difficult to find within a reasonable time~\cite{herr2017optimization}. Thus, an efficient algorithm that can find the near-optimal paths of lattice-surgery instructions is strongly demanded.

\subsection{Baseline scheduling method: BFS}
As a baseline method, we introduce a simple procedure to schedule the sequence of lattice-surgery instructions. The most naive strategy for scheduling instructions is to use breadth-first search~(BFS) as a subroutine to find one of the shortest paths. We call the procedure {\it BFS}, which is defined as follows.
\begin{enumerate}
  \item Mark all the ancillary cells as available.
  \item Pop an instruction from the instruction queue.
  \item Find one of the shortest paths connecting two logical qubits with specified boundaries. If found, mark the cells on the path as unavailable, and then go to 2. If not, increase the code beat and go to 1.
\end{enumerate}
The pseudo-code of this procedure is shown in Alg.\,\ref{alg:bfs}. Here, we keep whether the ancillary cells are available or not in a map object $A$. $\textsc{FindShortestPath}$ will return one of the shortest paths from $v_1$ to $v_2$ on graph $G$ for a lattice surgery instruction $o = (v_1, v_2, d)$ so that the path does not use surface-code cells marked $1$ in $A$ and $v_1, v_2$ are connected to the path from $d\in \{X,Z\}$ boundaries. If there is no path between them, the function returns an empty list.
We can find the shortest paths in a graph $G$ with $V$ vertices and $E$ edges in $O(V + E)$ time using the BFS algorithm. In this procedure, while choosing the shortest path will not necessarily lead to optimal results, we expect this choice to be reasonable since it will consume the fewest cells.

\begin{figure}[tb]
  \centering
  \includegraphics[width=0.45\columnwidth]{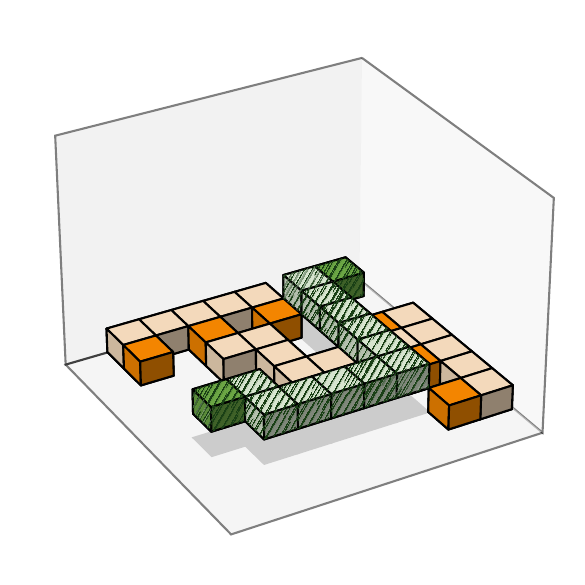}
  \caption{3D lattice representation of lattice-surgery instructions. This example corresponds to the sequence in Fig.\,\ref{fig:bfs}.}
  \label{fig:lattice_surgery_in_3d}
\end{figure}

The routed lattice-surgery sequence can be visualized in the 3D lattice, where the XY-plane corresponds to the qubit plane and the Z-axis to the time flow. Fig.\,\ref{fig:lattice_surgery_in_3d} shows an example converted from that of Fig.\,\ref{fig:bfs}. A 3D cell, or a \textit{voxel}, at the $(x,y,t)$ position in the 3D lattice corresponds to the surface-code cell at the coordinate $(x,y)$ at the $t$-th code beat. When the cell is used for encoding a logical qubit or temporally for logical operations, it is drawn as a colored voxel in the 3D lattice. When we assume logical qubits are persistently allocated at the same position, the cell of encoded logical qubits is visualized as a pillar from the initialization to the destructive measurements. The lattice surgery can be visualized as a path connecting two target pillars at a certain time slice.

\begin{algorithm}[tb]
  \caption{BFS scheduling}\label{alg:bfs}
  \begin{algorithmic}[1]
    \Require Logical-qubit connectivity graph $G = (V, E)$.
    \Require List of $m$ instructions of two-body lattice surgery. Each instruction $o \in O$ consists of $o = (v_1, v_2, d)$ where $v_1, v_2 \in V$ and $d \in \{X,Z\}$.
    \Ensure Scheduling paths and timings $S$ and total runtime $t$.
    \State $S \gets []$
    \State Define $A$ as a map from $V$ to $\{0,1\}$ and initialize all the elements with $0$
    \State $t \gets 0$
    \State $i \gets 0$
    \While{$i < m$}
    \While{$i < m$}
    \State $o \gets O[i]$
    \State $p \gets $ \Call{FindShortestPath}{$o, G, A$}
    \If{$p$ is empty}
    \State \textbf{break}
    \EndIf
    \State Append $(p,t)$ to $S$
    \For{$v$ in $p$}
    \State $A[v] \gets 1$
    \EndFor
    \State $i \gets i+1$
    \EndWhile
    \For{$v$ in $V$}
    \State $A[v] \gets 0$
    \EndFor
    \State $t \gets t+1$
    \EndWhile
    \State return $(S, t)$
  \end{algorithmic}
\end{algorithm}

\subsection{Baseline scheduling method: Look-ahead BFS}
\label{sec:look_ahead_bfs}

When the instructions are provided as a 1D sequence, we can change the order of the instructions so that the actions of sequences do not change and achieve a more efficient scheduling. A method to find unexecuted instructions is to iterate over all the unexecuted instructions and check if each unexecuted instruction can be executed beforehand, that is, maintain a flag to indicate whether each cell has any previous unexecuted instructions to be executed during the iteration and check if either of the two logical qubits has any previous instructions. The modified algorithms can be found as follows.
\begin{enumerate}
  \item Mark all the instructions unexecuted.
  \item Mark all the ancillary cells as available.
  \item Mark all the data cells as ready.
  \item Perform Steps 5 and 6 for all the unexecuted instructions.
  \item If the two target logical qubits are both marked as ready, find one of the shortest paths connecting two logical qubits with specified boundaries. If found, mark the cells on the path as unavailable, and mark the instruction executed. If not, do nothing.
  \item Mark the two target logical qubits as unready.
  \item Increase the code beat and go to 2.
\end{enumerate}
The pseudo-code of this procedure is shown in Alg.\,\ref{alg:look_ahead_bfs}.

In this algorithm, we use an object $D$ with the following data structure for time-efficient scheduling. This object manages an instruction dependency graph as a directed acyclic graph, where each node has pointers to child nodes and has a counter of the unexecuted parent nodes. The function \textsc{GetExecutableIndexList} returns all the indices of instructions that have no unexecuted parent nodes in ascending order, and \textsc{UpdateDependencyGraph} marks the $i$-th instruction as executed, i.e., decrements the count of unexecuted parent nodes for each child node. Assuming each instruction acts on a constant number of logical qubits, we can create this data structure with $O(m)$, and these functions have $O(\log m)$ running time per instruction by using a priority queue: \textsc{GetExecutableIndexList} runs in $O(\left|L\right| \log m)$ time and \textsc{UpdateDependencyGraph} runs in $O(\log m)$ time, where $\left|L\right|$ denotes the number of executable instructions. Thus, the runtime to maintain the instruction dependency is negligible compared with the runtime of searching for a path.

\begin{algorithm}[tb]
  \caption{Look-ahead BFS scheduling}\label{alg:look_ahead_bfs}
  \begin{algorithmic}[1]
    \Require Logical-qubit connectivity graph $G = (V, E)$.
    \Require List of $m$ instructions of two-body lattice surgery. Each instruction $o \in O$ consists of $o = (v_1, v_2, d)$ where $v_1, v_2 \in V$ and $d \in \{X,Z\}$.
    \Require Instruction dependency graph $D$. See main text for details.
    \Ensure Scheduling paths and timings $S$ and total runtime $t$.
    \State Define $S$ as a length-$m$ list to store the paths and timings.
    \State Define $A$ as a map from $V$ to $\{0,1\}$ and initialize all the elements with $0$.
    \State $t \gets 0$
    \While{not \Call{IsEmpty}{$D$}}
    \State $L \gets$ \Call{GetExecutableIndexList}{$D$}
    \For{$i$ in $L$}
    \State $o \gets O[i]$
    \State $p \gets $ \Call{FindShortestPath}{$o, G, A$}
    \If{$p$ is not empty}
    \State $S[i] \gets (p,t)$
    \State \Call{UpdateDependencyGraph}{$D, i$}
    \For{$v$ in $p$}
    \State $A[v] \gets 1$
    \EndFor
    \EndIf
    \EndFor
    \For{$v$ in $V$}
    \State $A[v] \gets 0$
    \EndFor
    \State $t \gets t+1$
    \EndWhile
    \State return $(S, t)$
  \end{algorithmic}
\end{algorithm}

\section{Scheduling of lattice surgery using entanglement}
\subsection{Overview}
In this paper, we propose a polynomial-time algorithm for efficiently scheduling lattice surgery. The outline of our idea is as follows. First, we split a lattice-surgery instruction on two data cells into a sequence of lattice-surgery instructions on the data and working cells. Then, we can execute a part of the separated instructions in advance to improve the throughput. A difficult point of this approach is finding an efficient splitting and scheduling method for instructions. Here, we found that the optimization of lattice-surgery scheduling using this mechanism can be converted into the path search problem in the 3D lattice satisfying several conditions. Thus, we can propose an efficient lattice-surgery scheduling algorithm by proposing path-packing algorithms on the 3D lattice. In this section, we explain this conversion and propose four path-finding algorithms.

This section is constructed as follows. In Sec.\,\ref{subsec:examples}, we show a simple example to motivate readers and clarify our idea using a specific situation. In Sec.\,~\ref{subsec:time_direction}, we theoretically show that a 3D path satisfying several conditions can be systematically translated to a sequence of lattice-surgery instructions that result in a desired logical operation, such as long-range lattice-surgery or CNOT gates.
Sec.\,\ref{subsec:solution_improvement} evaluates the effectiveness of 3D routing by establishing the upper and lower bounds for performance improvement.
Based on the obtained results, in Sec.\,\ref{subsec:method}, we will propose a method to search for 3D paths and pack them. The time complexities of the baseline methods and our proposed methods are analyzed in Sec.\,\ref{sec:time_complexity}. Their performances are numerically benchmarked in Sec.\,\ref{sec:evaluation}.

\subsection{Motivating examples}
\label{subsec:examples}

\begin{figure*}[tb]
  \centering
  \includegraphics[width=\textwidth]{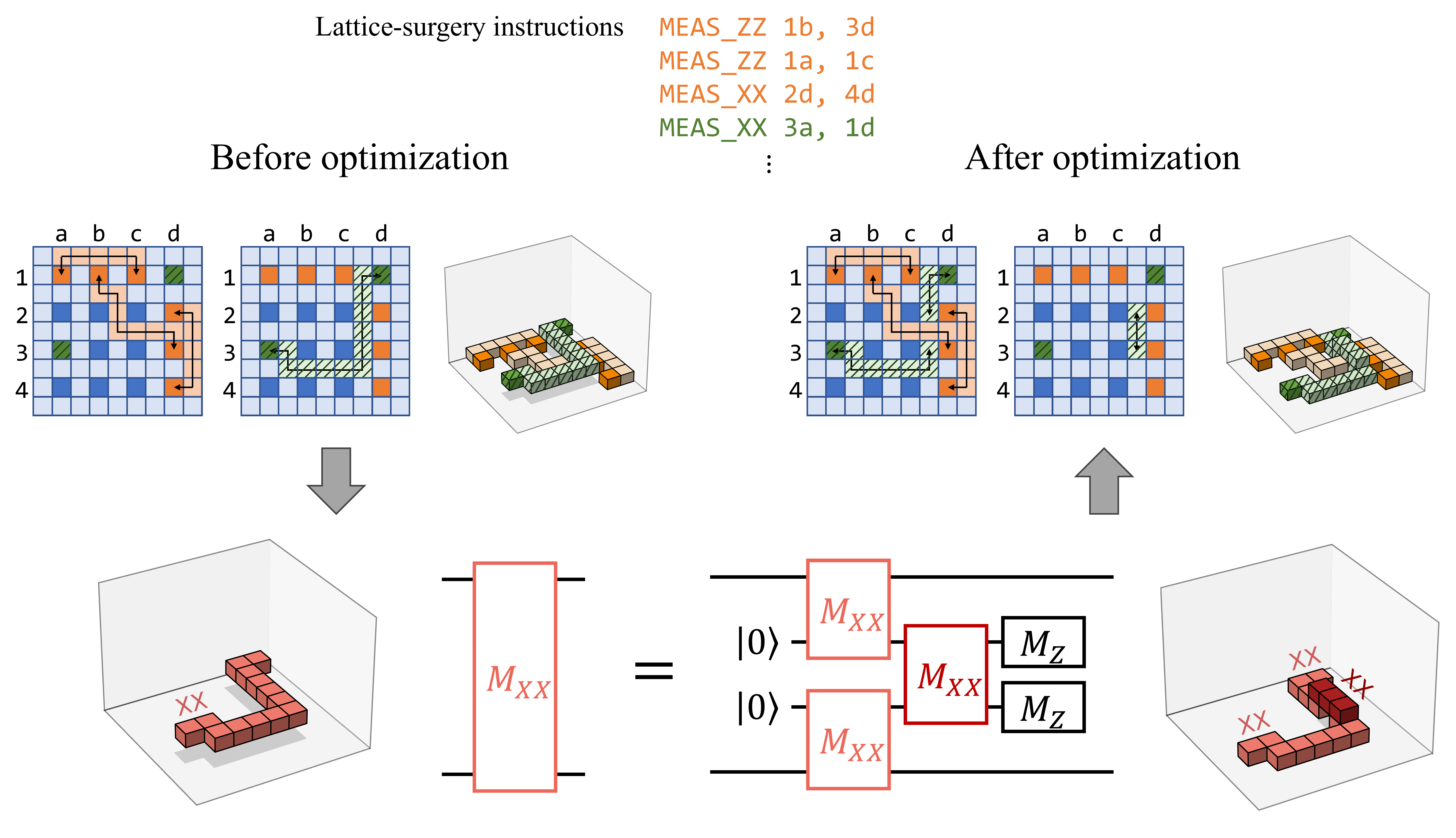}
  \caption{Overview of the optimization by routing lattice-surgery paths more flexibly in a 3D lattice. The equality of quantum circuits holds up to appropriate feedback of Pauli operations.}
  \label{fig:overview}
\end{figure*}

Here, with a simple example, we show that splitting of lattice-surgery instructions enables efficient use of resources, i.e., voxels in the 3D lattice. An overview of our example is shown in Fig.\,\ref{fig:overview}. We consider the same situation as Fig.\,\ref{fig:bfs}. In this situation, the BFS or look-ahead BFS algorithm cannot schedule all the routing at the same code beat, and a hatched path between green cells must wait for the next code beat. The 3D-lattice representation of this scheduling is shown in the left half of the figure. In this case, voxels below green paths are left unused and lose the parallelism of scheduling.

To mitigate this degradation, we can divide a not-executable instruction into multiple instructions and execute parts of them. As shown in the bottom of Fig.\,\ref{fig:overview}, the action of Pauli-$XX$ measurement instruction is equal to that of a chain of three Pauli-$XX$ measurements. This instruction can be achieved with three horizontal routings in the 3D lattice shown in the right-bottom figure. After this modification, voxels unused in the original scheduling are now utilized for instructions, and we can create free voxels for future instructions. Thus, with this technique, we can route a lattice-surgery path more flexibly in the 3D lattices. In other words, if we find a lattice-surgery instruction that cannot be efficiently routed, we can convert the instruction into a chain of them so that a 3D path achieves the chain of instructions and compactly fits into the currently available voxels. This modification enables more efficient use of voxels and allows us to leave more free voxels for future instructions.
To utilize this mechanism, we show a general relationship between 3D paths and separated lattice-surgery instructions in the next subsection.

\subsection{3D routing for lattice surgery}
\label{subsec:time_direction}

\begin{figure}[tb]
  \centering
  \begin{minipage}[b]{0.3\linewidth}
    \centering
    \includegraphics[width=\linewidth]{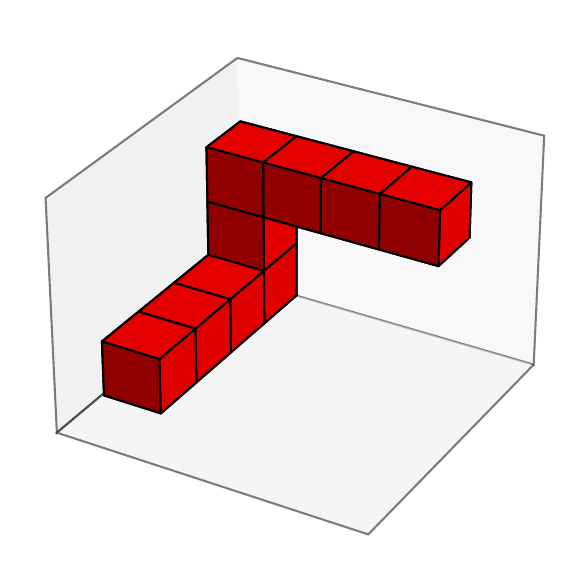}
    \subcaption{}
    \label{fig:kink}
  \end{minipage}
  \hfill
  \begin{minipage}[b]{0.3\linewidth}
    \centering
    \includegraphics[width=\linewidth]{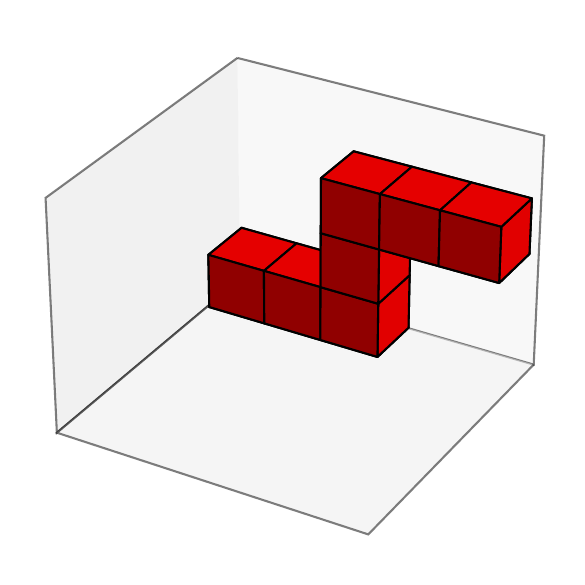}
    \subcaption{}
    \label{fig:not_kink_1}
  \end{minipage}
  \hfill
  \begin{minipage}[b]{0.3\linewidth}
    \centering
    \includegraphics[width=\linewidth]{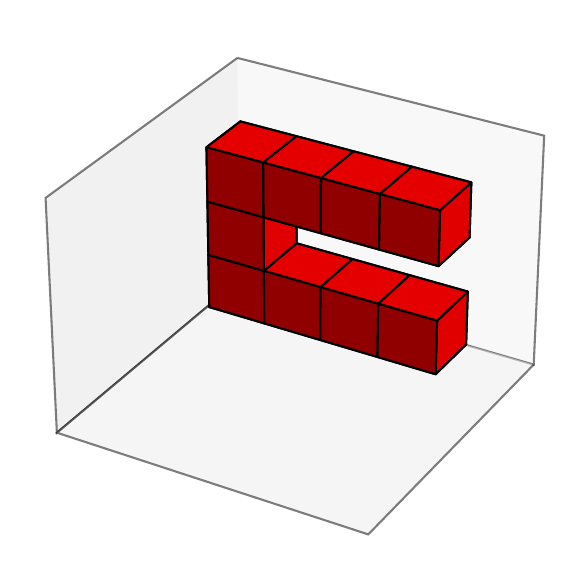}
    \subcaption{}
    \label{fig:not_kink_2}
  \end{minipage}
  \caption{Kink and not-kink examples. In these examples, (a) is an example of a kink, while (b) and (c) are not kinks.}
\end{figure}

In this section, we show that a 3D path satisfying several conditions can be converted to a sequence of lattice-surgery instructions having the desired action on the target data cells. To describe the conditions, we introduce the concept of \emph{kink}, which refers to a bending point in the path whose direction is rotated by 90 degrees before and after it moves in the time direction (i.e., Z-direction). For example, the bending point shown in Fig.\,\ref{fig:kink} is a kink while bending points shown in Figs.\,\ref{fig:not_kink_1} and \ref{fig:not_kink_2} are not kinks. Using the concept of kink, we finally show the following theorem.

\begin{theorem}
  \label{thm:path}
  If there is a 3D path that connects $X$-($Z$-)boundaries of cells and has an even number of kinks, then there is a sequence of lattice-surgery instructions that consumes the resource corresponding to voxels in the path and results in logical Pauli-$XX$(-$ZZ$) measurements. Also, if the path connects different types of boundaries and has an odd number of kinks, there is a sequence that results in a logical CNOT.
\end{theorem}

The above theorem implies that we can convert the scheduling optimization into the problem of packing several 3D paths connecting specified cells into a 3D lattice while minimizing its height.

\subsubsection{Conversion of 3D paths to quantum circuits}
First, we show a procedure for converting a 3D path connecting two logical qubits into a quantum circuit composed of lattice-surgery instructions.
Suppose that a 3D path is given connecting two logical qubits with the same types of boundaries. Any 3D path can be split into two types of segments: horizontal and vertical, which are regions moving in the XY-plane and along the Z-axis, respectively.

Each vertical segment is converted to the allocation and deallocation of an ancillary surface-code cell at the position of the segment at the lowest and highest timing of the segment. We note that this corresponds to allocating an intermediate logical qubit to the existing ancillary cell and does not necessitate any additional physical qubits. The ancillary cell is initialized with the state $\ket{0}$ or $\ket{+}$ and eventually measured in the $Z$ or $X$ basis. The basis for the initialization and measurement is determined from the two horizontal segments connected to the vertical segment, which is discussed in the next paragraph. These operations can be executed on a shorter time scale than lattice-surgery operations. Thus, they do not occupy voxels in the 3D lattice.

\begin{figure}[tb]
  \begin{minipage}[b]{0.48\columnwidth}
    \centering
    \includegraphics[width=0.95\columnwidth]{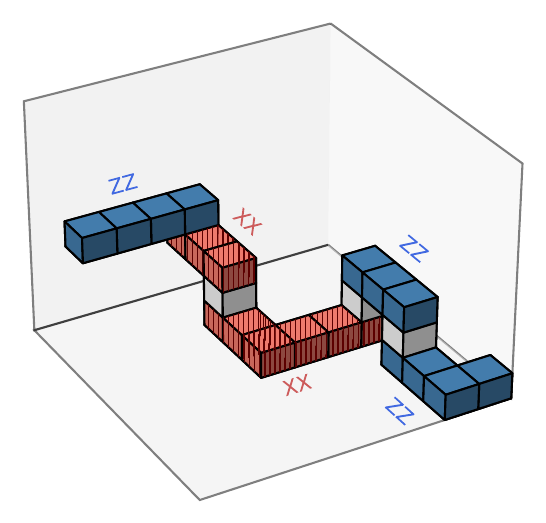}
    \vspace{-4mm}
    \subcaption{Example of assigning measurements to a 3D path.}
    \label{fig:path}
  \end{minipage}
  \hfill
  \begin{minipage}[b]{0.48\columnwidth}
    \centering
    \includegraphics[width=\columnwidth]{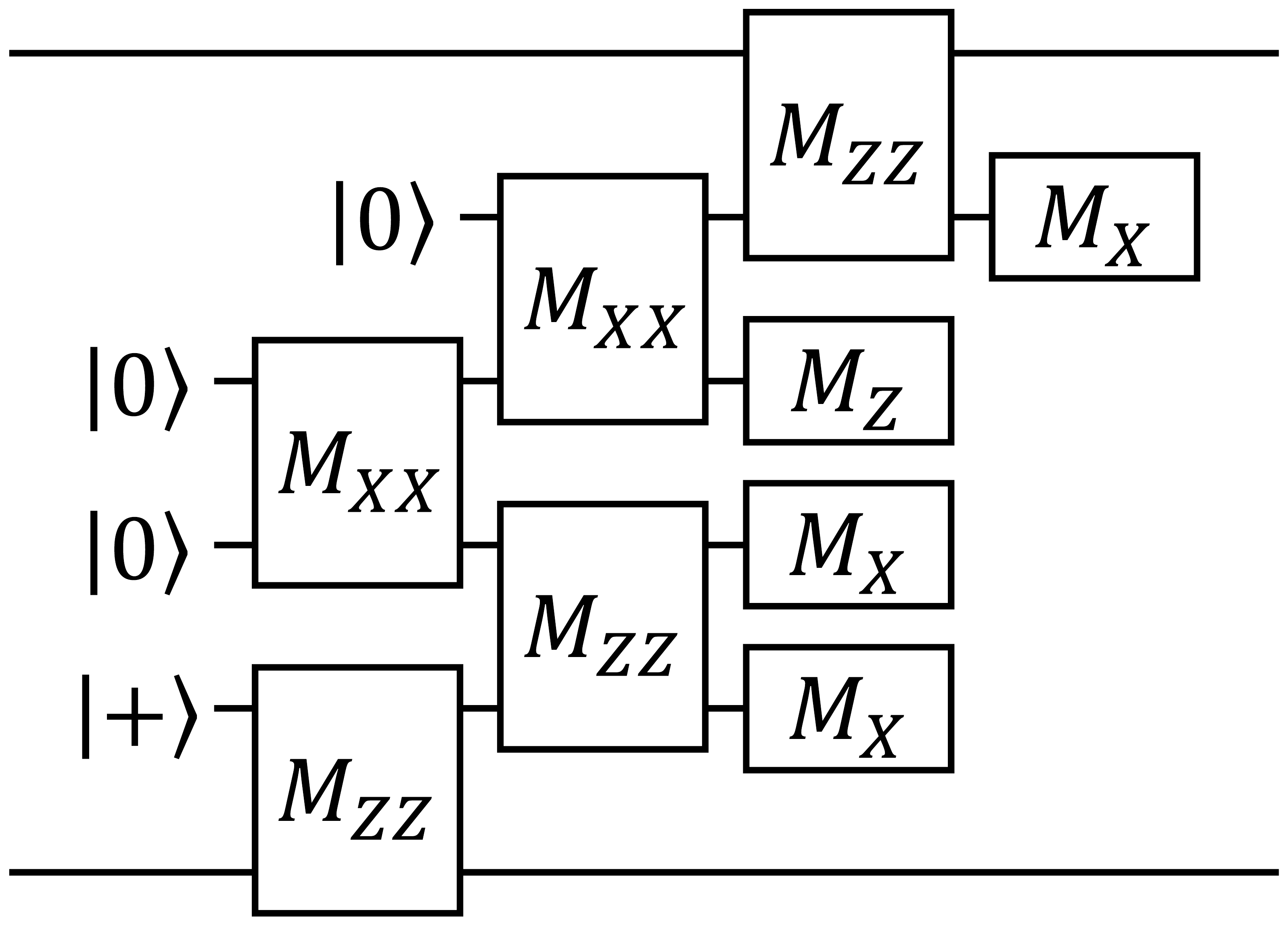}
    \vspace{-3mm}
    \subcaption{Quantum circuit converted from the 3D path.}
    \label{fig:quantum_circuit}
  \end{minipage}
  \caption{Conversion of a 3D path to a quantum circuit.}
\end{figure}

For each horizontal segment, we assign the $XX$ or $ZZ$ measurement. The basis is determined to be consistent with the direction of cells at the endpoint of the 3D path as follows. Suppose that the 3D path is divided into a sequence of horizontal and vertical segments. Note that the first and last segments are horizontal. We denote the boundary type of the starting point of the 3D path as $A \in \{X, Z\}$. Then, we assign the horizontal path connected to the starting point as the Pauli-$AA$ measurement. This assignment determines the rotating directions of the ancillary cell corresponding to the first vertical segment. Then, whether this vertical segment is a kink or not determines the type of connected boundary of the second horizontal segment. If a vertical segment is a kink, the type of measurements assigned to horizontal segments before and after the vertical segment will be different. Otherwise, i.e., if the direction changes by 0 or 180 degrees before and after the vertical segment, the type of measurements assigned to horizontal segments before and after the vertical segment will be the same. For the next horizontal segment, we assume that it connects the same type of boundary. The measurement types of subsequent horizontal segments and the rotations of vertical-segment cells are assigned by repetitively applying the above rule.
This rule will be consistent with the rotation of the other endpoint of the 3D path if the 3D path connects the same type of boundaries and has an even number of kinks. It also becomes consistent if the path connects different types of boundaries and has an odd number of kinks.
A concrete example is shown in Fig.\,\ref{fig:path}. In this example, there are two kinks.

Next, we assign the initialization and measurement basis for each vertical segment. Each ancillary cell is initialized in a basis different from the basis of the first two-body measurement: an ancillary cell whose first two-body measurement is a Pauli-$XX$ (-$ZZ$) measurement should be initialized to $\ket{0}$ ($\ket{+}$), respectively.
Similarly, each ancillary cell is measured in a basis different from the basis of the last two-body measurement: an ancillary cell whose last two-body measurement is a Pauli-$XX$ (-$ZZ$) measurement should be measured in the $Z$ ($X$) basis, respectively.
Consequently, the pre-processing and post-processing of the ancillary cells in a given 3D path can be uniquely determined. The quantum circuit corresponding to the 3D path in Fig.\,\ref{fig:path} is shown in Fig.\,\ref{fig:quantum_circuit}.

\subsubsection{Simplifying Quantum Circuits}

\begin{figure*}
  \centering
  \begin{minipage}[b]{0.32\linewidth}
    \centering
    \includegraphics[width=\linewidth]{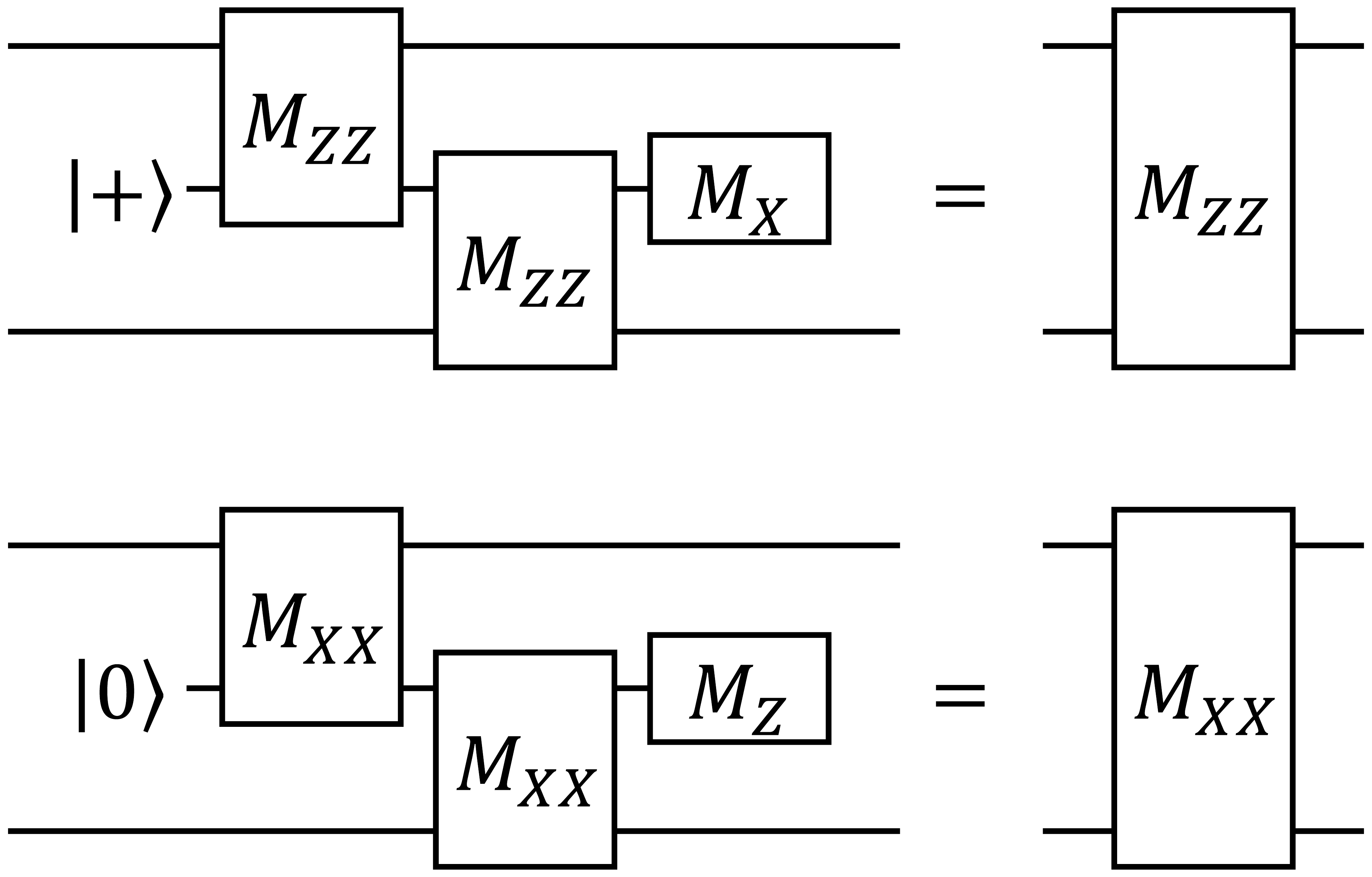}
    \subcaption{}
    \label{fig:qc_same}
  \end{minipage}
  \hfill
  \begin{minipage}[b]{0.32\linewidth}
    \centering
    \includegraphics[width=\linewidth]{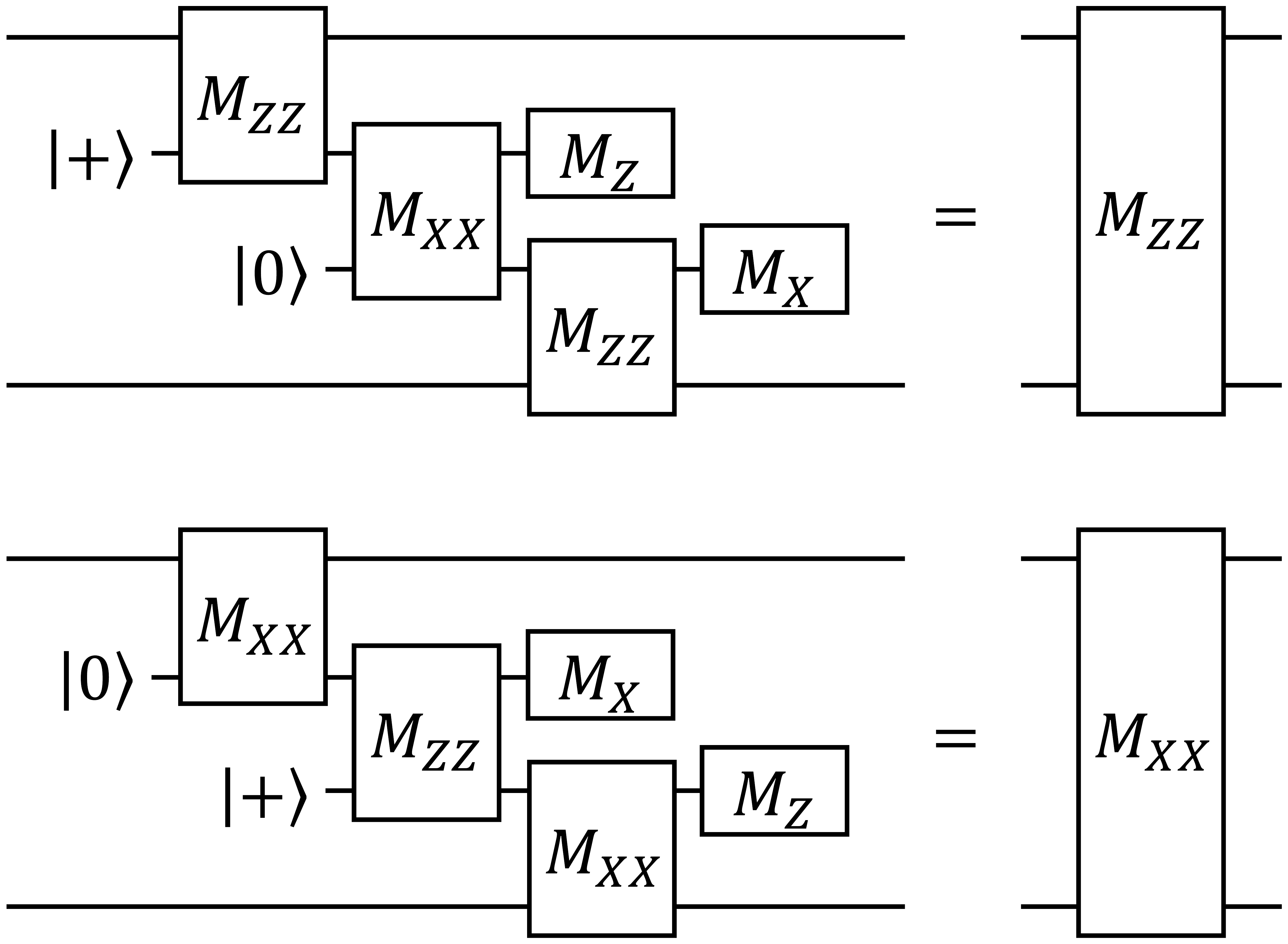}
    \subcaption{}
    \label{fig:qc_between}
  \end{minipage}
  \hfill
  \begin{minipage}[b]{0.32\linewidth}
    \centering
    \includegraphics[width=\linewidth]{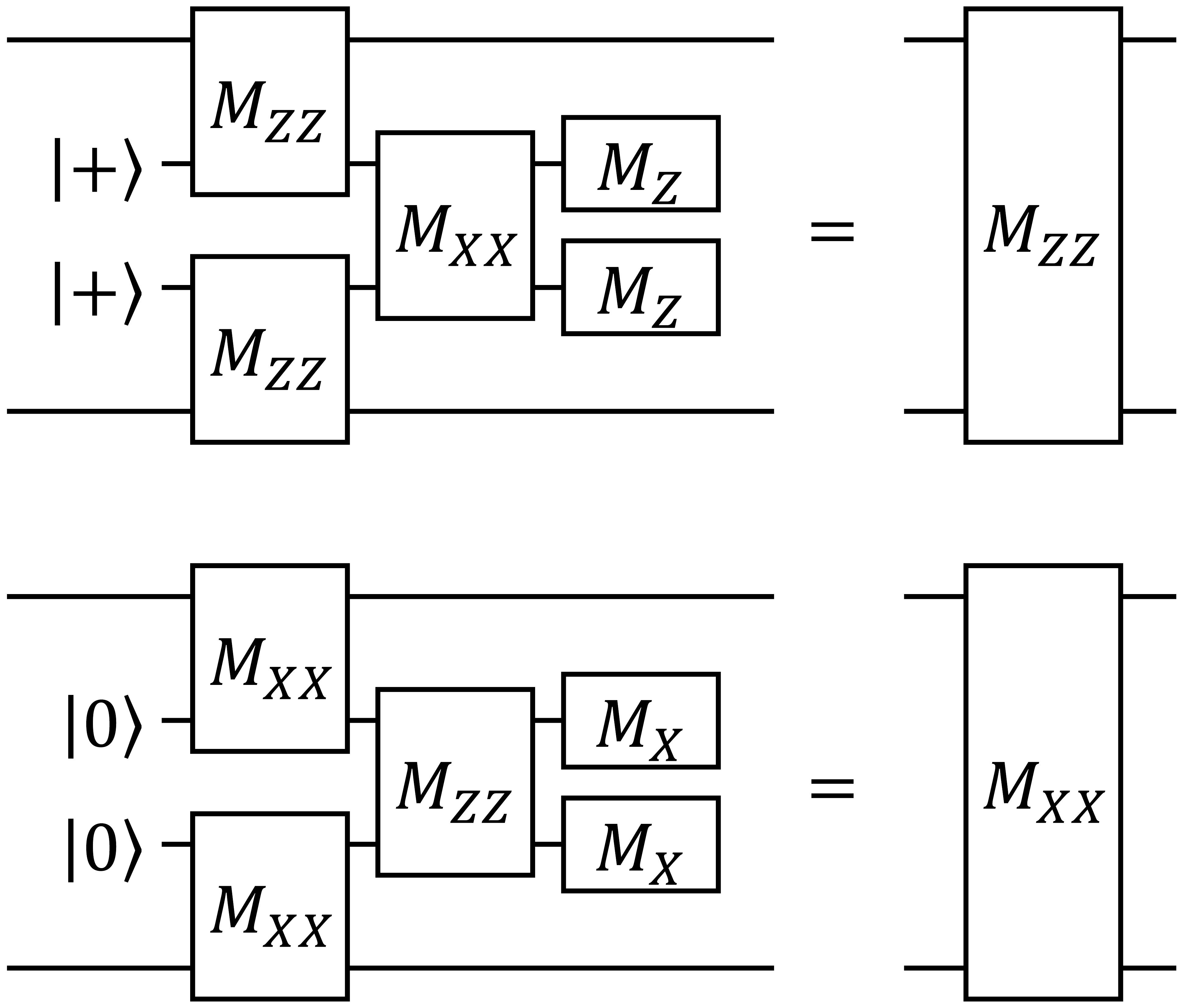}
    \subcaption{}
    \label{fig:qc_convex}
  \end{minipage}
  \vspace{1em}
  \\
  \begin{minipage}[b]{0.32\linewidth}
    \centering
    \includegraphics[width=\linewidth]{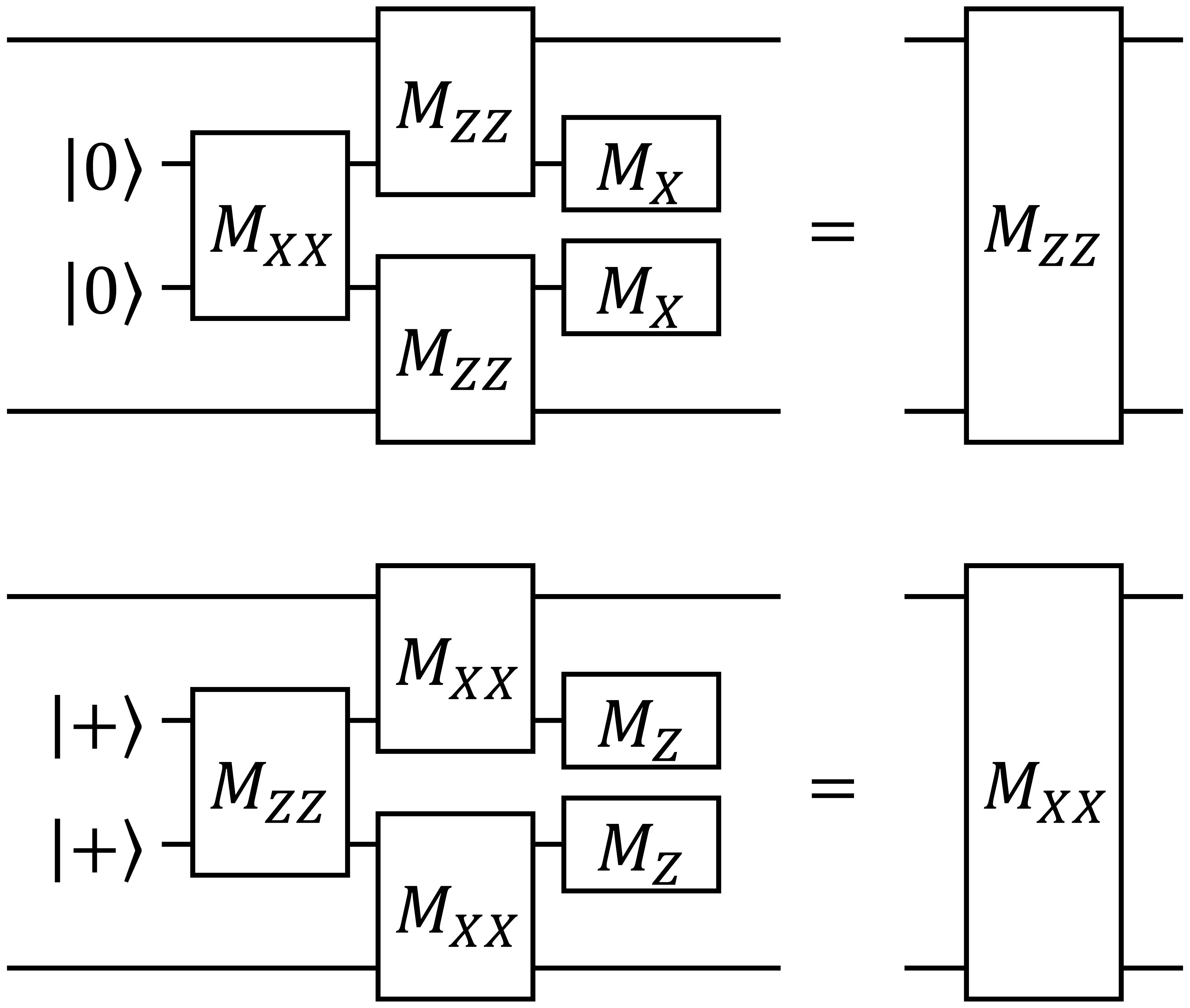}
    \subcaption{}\label{fig:qc_concave}
  \end{minipage}
  \hfill
  \begin{minipage}[b]{0.32\linewidth}
    \centering
    \includegraphics[width=\linewidth]{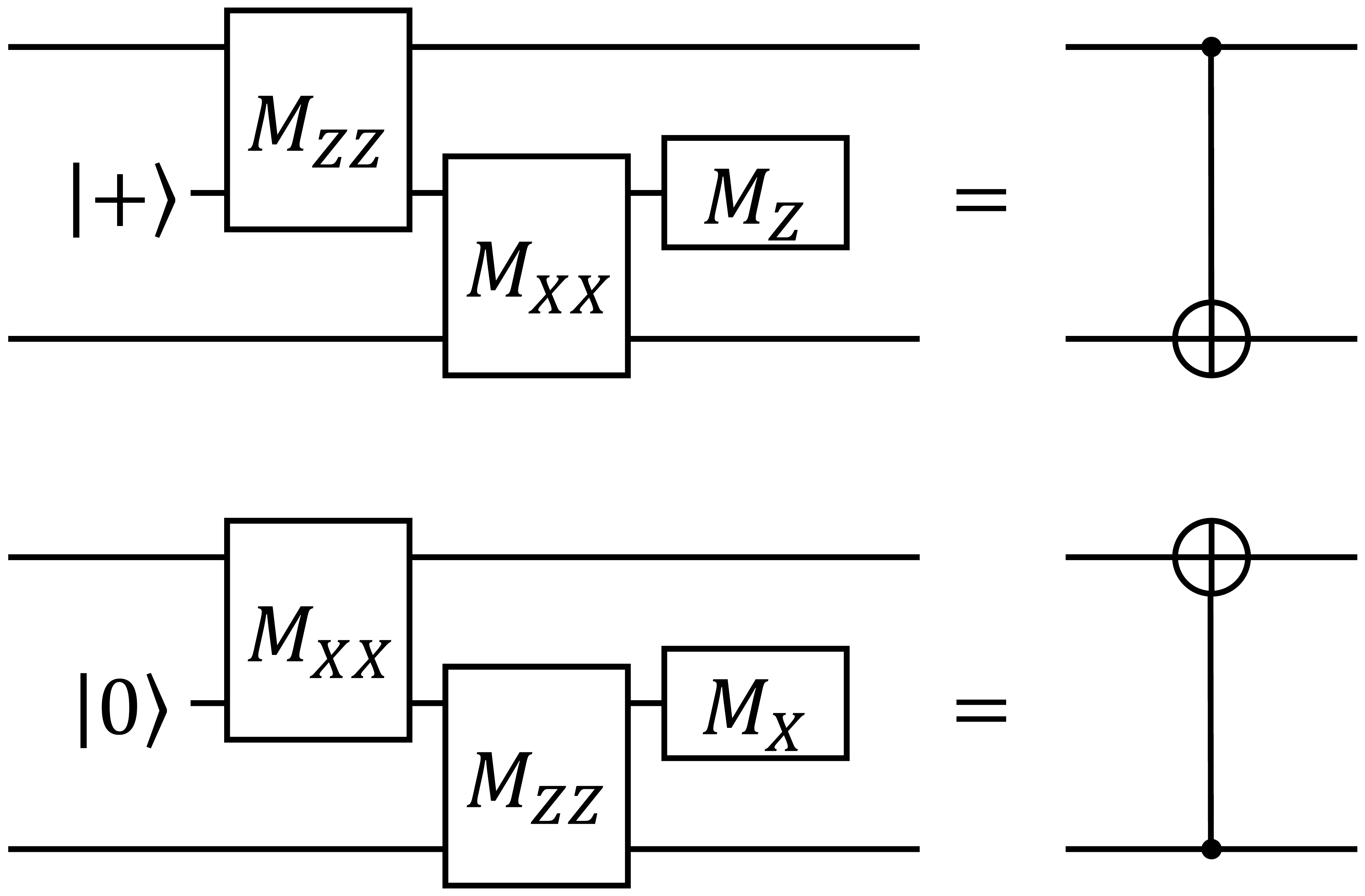}
    \subcaption{}\label{fig:qc_cnot}
  \end{minipage}
  \hfill
  \hspace{0.32\linewidth}
  \caption{Equivalent processes of two-qubit measurements. (a) Two two-body measurements with the same basis are equal to a single two-body measurement. (b,c,d) Three alternating two-body measurements are equal to a single two-body measurement. (e) Two two-body measurements with different bases are equal to a CNOT gate.}
  \label{fig:qc_equivalent}
\end{figure*}
We show that the quantum circuit converted from a 3D path with the presented rule can be simplified to a two-body measurement or a CNOT operation on the data cells at the path endpoints in accordance with whether the number of kinks is even or odd. The quantum circuit converted from a 3D path is a chain of Pauli-$XX$ measurements~($M_{XX}$) and Pauli-$ZZ$ measurements~($M_{ZZ}$), as shown in Fig.\,\ref{fig:quantum_circuit}. We use the following rules to simplify such a circuit with the feedback of appropriate logical Pauli operations.
\begin{enumerate}
  \item
        \label{item:rule_same}
        As shown in Fig.\,\ref{fig:qc_same}, a chain of measurements of the same type can be simplified to a single measurement of the same type.

  \item
        \label{item:rule_between}
        As shown in Fig.\,\ref{fig:qc_between}, a chain of two measurements of the same type with another in-between measurement of the other type can be simplified to a single measurement of the same type as the two measurements.

  \item
        \label{item:rule_convex}
        As shown in Fig.\,\ref{fig:qc_convex}, a chain of two measurements of the same type followed by another measurement of the other type can be simplified to a single measurement of the same type as the two measurements.

  \item
        \label{item:rule_concave}
        As shown in Fig.\,\ref{fig:qc_concave}, a chain of two measurements of the same type following another measurement of the other type can be simplified to a single measurement of the same type as the two measurements.

  \item
        \label{item:rule_cnot}
        As shown in Fig.\,\ref{fig:qc_cnot}, a chain of two measurements of the different types can be simplified to a CNOT operation.
\end{enumerate}
These rules can be verified by simple calculations. Note that while the final state must be corrected by the Pauli operation in accordance with the measurement results, the time required for this correction is negligible since this is achieved via the Pauli frame~\cite{fowler2012towards,fowler2018low}.

Putting these rules together, we can show the proposed theorem by simplifying the circuit with a chain of $M_{XX}$ and $M_{ZZ}$ as follows.
\begin{enumerate}
  \item By repeatedly applying rule \ref{item:rule_same}, all the sub-sequences consisting of the same type of measurements can be merged into one, so that the different types of measurements alternate in the chain. Note that kinks correspond to changes in measurements. Hence, after the simplification, if the number of kinks is even, the number of measurements is odd; if the number of kinks is odd, the number of measurements is even.
  \item Using the rules of \ref{item:rule_between}, \ref{item:rule_convex}, and \ref{item:rule_concave}, we can reduce the number of measurements by two if the length of a chain is longer than two. Therefore, a circuit composed of an even number of measurements can be reduced to two two-body measurements with alternating bases, and a circuit of an odd number of measurements to a single two-body measurement.
  \item If a circuit of two measurements is obtained, it is equivalent to a CNOT operation by rule \ref{item:rule_cnot}, and the direction of the CNOT is determined from the boundary types of the endpoints. If a circuit of a single two-body measurement is obtained, it is equivalent to an $XX$ or $ZZ$ measurement, and the type of measurement is the same as the boundary types of the endpoints.
\end{enumerate}
As a consequence, the circuit can be simplified to a two-body measurement if the path has an even number of kinks, and to a CNOT operation if the path has an odd number of kinks, which leads to the theorem.

\subsection{Solution improvement by 3D routing}
\label{subsec:solution_improvement}

\begin{figure}[tb]
  \begin{minipage}[b]{0.48\columnwidth}
    \centering
    \includegraphics[width=0.95\columnwidth]{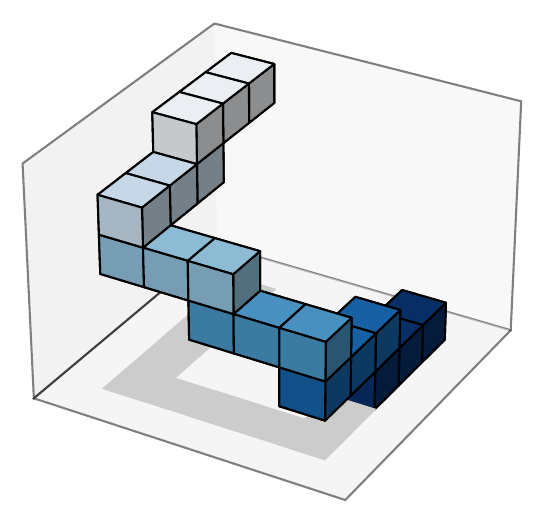}
    \vspace{-4mm}
    \subcaption{2D routing with $m$ code beats.}
  \end{minipage}
  \hfill
  \begin{minipage}[b]{0.48\columnwidth}
    \centering
    \includegraphics[width=\columnwidth]{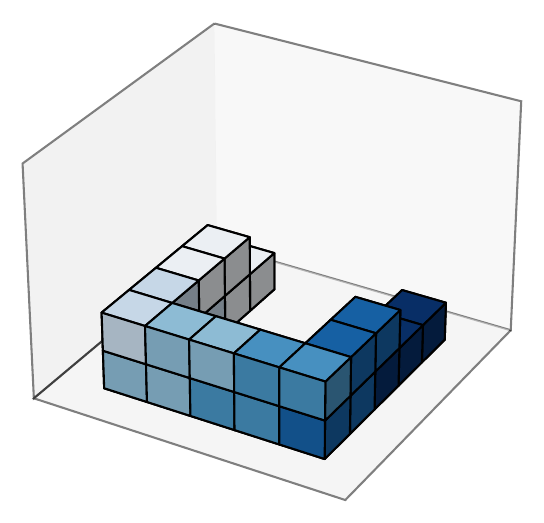}
    \vspace{-3mm}
    \subcaption{3D routing with $2$ code beats.}
  \end{minipage}
  \caption{Instruction sequence with a stair-like structure, to which 3D routing is most effective.}
  \label{fig:3d_maximum_improvement}
\end{figure}

This section demonstrates the capabilities and limitations of 3D routing through two contrasting examples. The maximum benefits of 3D routing are observed when the input instruction sequence has a stair-like structure (Fig.\,\ref{fig:3d_maximum_improvement}). For instance, consider an instruction sequence of length $m$ where the $i$-th instruction operates on logical qubits $i$ and $i + 1$. Then, 3D routing achieves a $\Theta(m)$-fold improvement, as it completes the process in $2$ code beats while 2D routing requires $m$ code beats.  We emphasize that the subsequent proposed methods aim to replicate this improvement.

Conversely, the benefits vanish when the instruction sequence exhibits a hub-and-spoke structure. Specifically, when the $i$-th instruction connects logical qubits $1$ and $i + 1$, the solution obtained by 2D routing is already optimal, meaning 3D routing offers no additional acceleration.

These examples establish the upper and lower bounds for the performance improvement provided by 3D routing. They also clearly demonstrate that the achievable gain is highly dependent on the input structure. Our numerical experiments in Sec.\,\ref{sec:evaluation} will quantify the performance improvement by the proposed 3D routing methods on two realistic inputs with distinct structural characteristics.

\subsection{Scheduling algorithms}
\label{subsec:method}
In this section, we propose methods to schedule a sequence of lattice-surgery instructions using the proved theorem. First, we ignore the kink condition for simplicity and discuss four scheduling methods that achieve the 3D routing of lattice-surgery operations. The first two methods, {\it 3D BFS} and {\it 3D Dijkstra}, are simple extensions of the baseline methods. While they provide scheduling with higher throughput than the baseline method, they need 3D path-finding algorithms and consume longer compilation time. To mitigate this drawback without losing the throughput of instructions, we propose a method named {\it Dijkstra projection}. This algorithm provides high-throughput scheduling comparable to the 3D variants and achieves reduced time complexity. We also introduce the \textit{look-ahead Dijkstra projection}, which integrates the instruction look-ahead technique to further optimize its scheduling performance. Then, we consider the kink condition and show modifications to the Dijkstra projection algorithms by which the paths satisfy the kink condition. The baseline and proposed scheduling methods are listed and compared in Tabs.\,\ref{tab:comparison-path} and \ref{tab:comparison-method}.
The time complexities in these tables are analyzed in Sec.\,\ref{sec:time_complexity}.

\begin{table*}[t]
  \centering
  \caption{Comparison of path search subroutines between the scheduling methods. These properties remain the same when instruction look-ahead is integrated. As for the time complexity, $n$ denotes the number of cells in the qubit plane, and $h$ denotes the height of the 3D lattice.}
  \label{tab:comparison-path}
  \vspace{-3mm}
  \small
  \newcommand{\gathermargin}{0.4cm}
  \begin{widetable}{\textwidth}{ccc@{\hskip-\gathermargin}cc}
    \toprule
    Method & Path search space & \multicolumn{2}{c}{Path search time complexity} & Utilize 3D path \\
    \midrule
    BFS & 2D lattice & {\hskip \gathermargin} $O(n)$ & (Fast) & No \\
    3D BFS & 3D lattice & {\hskip \gathermargin} $O(n h)$ & (Slow) & Yes \\
    3D Dijkstra & 3D lattice & {\hskip \gathermargin} $O(n h \log nh)$ & (Slow) & Yes \\
    Dijkstra projection & 2D lattice & {\hskip \gathermargin} $O(n \log n)$ & (Fast) & Yes \\
    \bottomrule
  \end{widetable}
\end{table*}

\begin{table*}[t]
  \centering
  \caption{Comparison of the baseline and proposed scheduling methods. As for the time complexity, $n$ denotes the number of cells in the qubit plane, and $m$ denotes the number of instructions.}
  \label{tab:comparison-method}
  \vspace{-3mm}
  \small
  \begin{widetable}{\textwidth}{ccccc}
    \toprule
    Method & Time complexity & Execution time & Solution quality & Kink condition \\
    \midrule
    BFS & $O(n m)$ & Shortest & Low & Not required \\
    LA BFS & $O(n^2 m + m \log m)$ & Very Short & Moderate & Not required \\
    3D BFS & $O(n m^2)$ & Long & High & Not satisfied \\
    3D Dijkstra & $O(n m^2 \log n m)$ & Long & Very High & Not satisfied \\
    Dijkstra proj. & $O(n m \log n)$ & Short & Very High & Satisfied \\
    LA Dijkstra proj. & $O(n m \log n + m \log m)$ & Short & Highest & Satisfied \\
    \bottomrule
  \end{widetable}
\end{table*}

\subsubsection{Breadth-first search extended in the time direction (3D BFS)}
\begin{figure}[tb]
  \begin{minipage}[b]{0.48\columnwidth}
    \centering
    \includegraphics[width=\columnwidth]{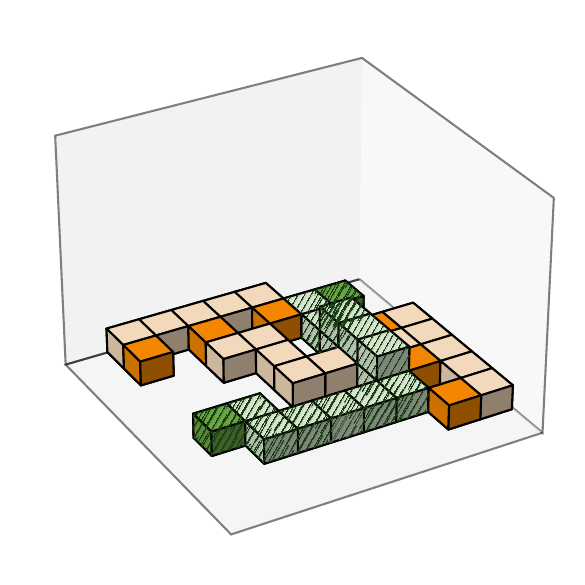}
    \vspace{-5mm}
    \caption{Example of the 3D BFS.}
    \label{fig:3d_bfs_example}
  \end{minipage}
  \begin{minipage}[b]{0.48\columnwidth}
    \centering
    \includegraphics[width=\columnwidth]{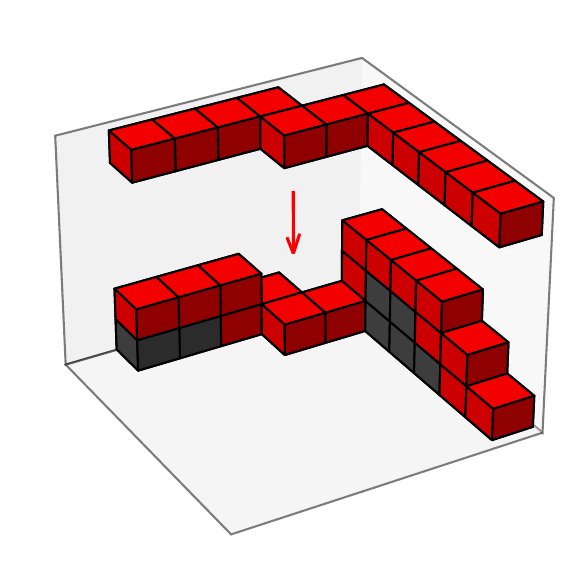}
    \caption{Example of stacking a path in the Dijkstra projection.}
    \label{fig:projection_example}
  \end{minipage}
\end{figure}

We propose a method called 3D BFS, which utilizes time direction for scheduling with a small modification to the BFS algorithm. It searches for a path in a 3D lattice, including the time axis, and connects the target data cells by one of the shortest paths. The BFS algorithm can be used to find the shortest path in the 3D lattice, as in the case of the 2D lattice.
When we choose the endpoint voxels of the lattice-surgery path in the 3D lattice, the choice of their Z coordinates is arbitrary. The 3D BFS method adopts the lowest possible voxels as endpoints and then searches for the shortest path connecting them. If there is no path connecting them, the height of the endpoint is raised, and the shortest path connecting them is searched for until it is found.
For example, by using the 3D BFS, the scheduling for the case shown in Fig.\,\ref{fig:bfs} is improved to a more efficient result shown in Fig.\,\ref{fig:3d_bfs_example}.
Note that instruction look-ahead is no longer required to mitigate scheduling delays caused by routing failures, since we can sequentially schedule all the instructions.

The pseudo-code of this procedure is shown in Alg.\,\ref{alg:3d_bfs}.
The dynamic array $A$ maintains whether voxels in the 3D lattice are occupied.
The function \textsc{Find3DShortestPath} returns one of the shortest and lowest paths connecting the target data cells, each of which corresponds to $v_1$ and $v_2$. We can find the path with polynomial time using the BFS.

\begin{algorithm}[tb]
  \caption{3D BFS scheduling}\label{alg:3d_bfs}
  \begin{algorithmic}[1]
    \Require Logical-qubit connectivity graph $G = (V, E)$.
    \Require List of $m$ instructions of two-body lattice surgery. Each instruction $o \in O$ consists of $o = (v_1, v_2, d)$ where $v_1, v_2 \in V$ and $d \in \{X,Z\}$.
    \Ensure Scheduling paths $S$ and total runtime $t$.
    \State $S \gets []$
    \State $A \gets []$
    \State Append to $A$ a map with all values set to $0$.
    \State $t \gets 0$
    \For{$i = 0, \dots, m - 1$}
    \State $o \gets O[i]$
    \State $P \gets $ \Call{Find3DShortestPath}{$o, G, A$}
    \State Append $P$ to $S$
    \For{$(v, \tau)$ in $P$}
    \State $A[\tau][v] \gets 1$
    \EndFor
    \For{$(v, \tau)$ in $P$}
    \If{$\tau = t$}
    \State Append to $A$ a map with all values set to $0$.
    \State $t \gets t + 1$
    \State \textbf{break}
    \EndIf
    \EndFor
    \EndFor
    \State return $(S, t)$
  \end{algorithmic}
\end{algorithm}

\subsubsection{Dijkstra's algorithm extended in the time direction (3D Dijkstra)}

As another method utilizing time direction, we propose the 3D Dijkstra. The basic scheme of the 3D Dijkstra is the same as that of the 3D BFS. The difference is that the 3D Dijkstra assigns the weight to the voxels and finds a path that minimizes the sum of weights.
We assign large weights as the height of voxels increases. This weighting prioritizes paths with lower voxels over those with higher voxels, and we can expect that the 3D lattice can be used more densely by packing paths from the bottom.
To find the shortest path on a weighted lattice, it is efficient to use a variant of Dijkstra's algorithm for sparse graphs~\cite{fredman1987fibonacci}: it runs in $O(V \log V + E)$ time for a general graph with $V$ vertices and $E$ edges.

We heuristically choose the weights as $2^h$, where $h$ is the height of the voxel. The reason for using exponential weights is that the ratio of the costs of adjacent layers is constant, regardless of the current height. The reasons for adopting $2$ as the base are twofold: the scheduling result was found to be efficient when measured, and the algorithm runs fast because it can be implemented with integer bit shifts.

The pseudo-code of this procedure is almost identical to that shown in Alg.\,\ref{alg:3d_bfs}. The only difference is that the function \textsc{Find3DShortestPath} should be replaced with a weighted version that uses Dijkstra's algorithm.

\subsubsection{Dijkstra projection}
As shown later in the numerical evaluation parts, the 3D BFS and 3D Dijkstra methods are high-performance but time-consuming for practical problem instances, while they are polynomial-time algorithms. They are especially time-consuming for deep circuits because their time complexity of path search depends on the height of the 3D lattice $h$, as shown in Tab.\,\ref{tab:comparison-path}. This motivates us to propose a more time-efficient algorithm, the Dijkstra projection. It searches for a 2D path on a 2D lattice and then obtains a 3D path by stacking the 2D path in the time direction, like a projection. While the 3D BFS and the 3D Dijkstra search for a path in a 3D lattice, the Dijkstra projection searches for a path only in a 2D lattice, which is likely to speed up the scheduling process.

The Dijkstra projection determines a pre-projection path in a similar way to the 3D Dijkstra. First, each cell of the qubit plane is weighted exponentially with respect to the heights of the paths already stacked. On the weighted 2D lattice, Dijkstra's algorithm searches for the path with the minimum sum of the weights. The path is then stacked in the 3D lattice along the time axis, preserving its 2D shape while ensuring that adjacent voxels of the path share a face. Fig.\,\ref{fig:projection_example} shows an example of stacking a path.

The pseudo-code of this procedure is shown in Alg.\,\ref{alg:dijkstra_projection}.
The map named $H$ corresponds to the height of 2D cells.
The function \textsc{FindShortestPathFromHeight} weights the graph $G$ with the weight obtained from the height $H$ and returns one of the shortest paths on the weighted graph. The function \textsc{LiftPath} takes a 2D path and returns a stacked 3D path as shown in Fig.\,\ref{fig:projection_example}.

\begin{algorithm}[tb]
  \caption{Dijkstra projection scheduling}\label{alg:dijkstra_projection}
  \begin{algorithmic}[1]
    \Require Logical-qubit connectivity graph $G = (V, E)$.
    \Require List of $m$ instructions of two-body lattice surgery. Each instruction $o \in O$ consists of $o = (v_1, v_2, d)$ where $v_1, v_2 \in V$ and $d \in \{X,Z\}$.
    \Ensure Scheduling paths $S$ and total runtime $t$.
    \State $S \gets []$
    \State Define $H$ as a map from $V$ to nonnegative integers and initialize all the elements with $0$.
    \State $t \gets 0$
    \For{$i = 0, \dots, m - 1$}
    \State $o \gets O[i]$ \label{line:get_instruction}
    \State $p \gets $ \Call{FindShortestPathFromHeight}{$o, G, H$}
    \State $P \gets$ \Call{LiftPath}{$p, H$}
    \State Append $P$ to $S$
    \For{$(v, \tau)$ in $P$}
    \State{$H[v] \gets \max\{H[v], \tau + 1\}$}
    \State{$t \gets \max\{t, \tau + 1\}$}
    \EndFor \label{line:outer_loop_end}
    \EndFor
    \State return $(S, t)$
  \end{algorithmic}
\end{algorithm}

\subsubsection{Look-ahead Dijkstra projection}
\label{sec:look_ahead_dijkstra_projection}

Instruction look-ahead, introduced in Sec.\,\ref{sec:look_ahead_bfs}, is a technique employed to efficiently manage the set of currently executable instructions. While this technique was initially utilized to mitigate scheduling delays caused by routing failures, we can also leverage it to provide the proposed algorithms with greater flexibility in selecting the next instruction for execution. Specifically, we integrate it with the Dijkstra projection and call it look-ahead Dijkstra projection.

Empirically, it is effective to choose an executable instruction with minimum height, where the height of an executable instruction is defined as the maximum height among its operand cells. We note that this value is static; once an instruction becomes executable, its height will not change until it is executed, because no other instructions will be executed on its operand cells.

Finally, we describe the modifications required to integrate this optimization into the Dijkstra projection. A similar approach applies to our other two algorithms. The integration involves three primary changes:
\begin{enumerate}
  \item The instruction height serves as the key for the priority queue.
  \item Since our 3D routing subroutine never fails, we only need to fetch a single executable instruction. Thus, we replace the function \textsc{GetExecutableIndexList} with a new function, \textsc{GetExecutableIndex}, which retrieves the highest-priority instruction from the priority queue in $O(\log m)$ time.
  \item Then, we modify Alg.\,\ref{alg:dijkstra_projection} by replacing Line \ref{line:get_instruction} with \textsc{GetExecutableIndex} and inserting \textsc{UpdateDependencyGraph} immediately after Line \ref{line:outer_loop_end}. It is noteworthy that these changes barely affect the total time complexity, since both functions run in $O(\log m)$ time, which is negligible compared to the $O(n)$ time required for path searching.
\end{enumerate}

\subsubsection{Modifications to meet the kink condition}

\begin{figure}[tb]
  \centering
  \begin{minipage}[b]{0.3\columnwidth}
    \centering
    \includegraphics[width=\columnwidth]{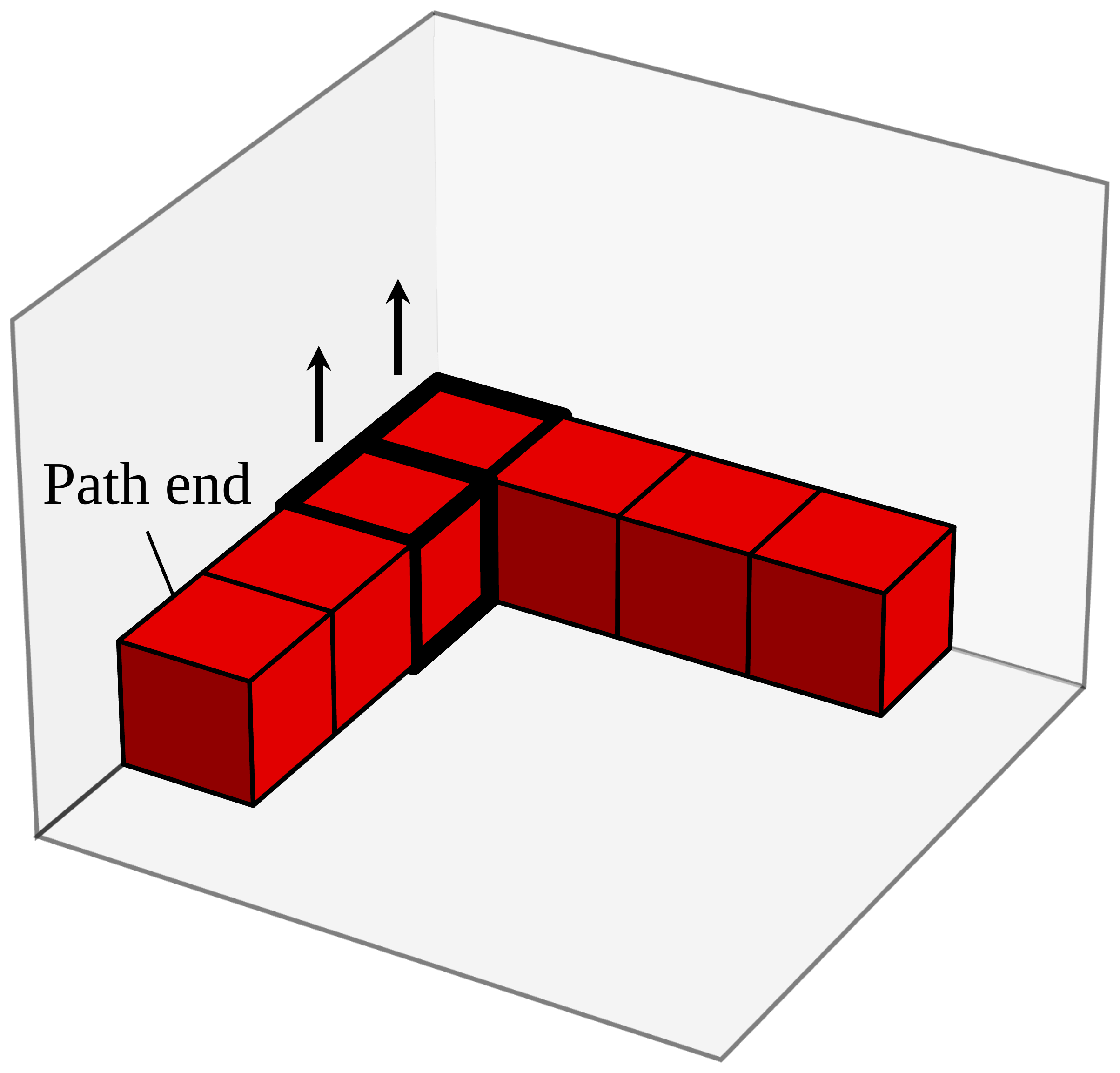}
    \subcaption{}
    \label{fig:twist_corner_before}
  \end{minipage}
  \hspace{0.1\columnwidth}
  \begin{minipage}[b]{0.3\columnwidth}
    \centering
    \includegraphics[width=\columnwidth]{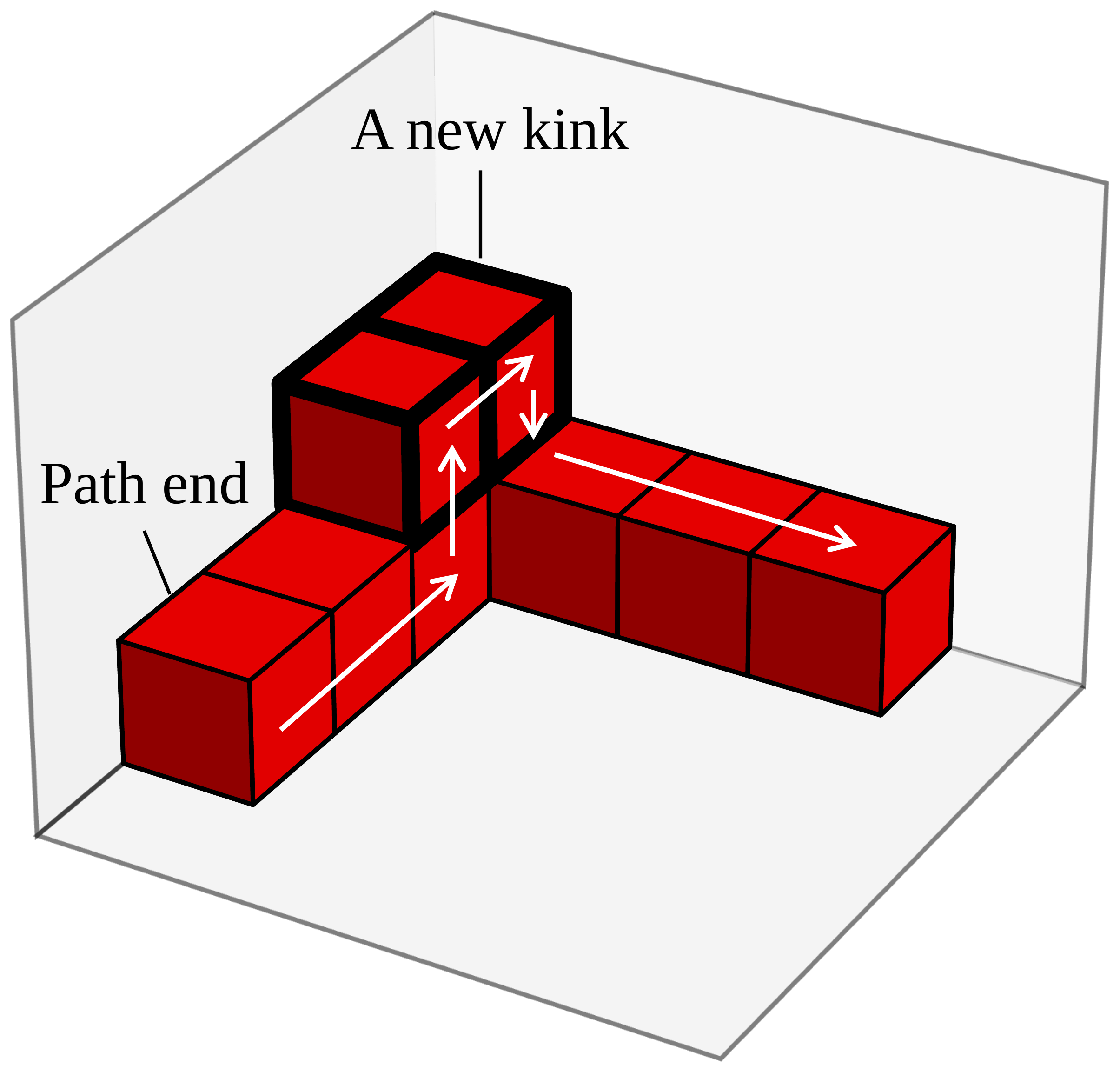}
    \subcaption{}
    \label{fig:twist_corner_after}
  \end{minipage}
  \caption{Example of creating a kink provided that the first 90-degree corner in the path is not a kink. (a) Lifting two cells of the non-kink corner creates a kink without affecting the other corners. (b) After the adjustment, a new kink is created without affecting the other corners.}
  \label{fig:twist_corner}
\end{figure}

\begin{figure}[tb]
  \centering
  \begin{minipage}[b]{0.3\columnwidth}
    \centering
    \includegraphics[width=\columnwidth]{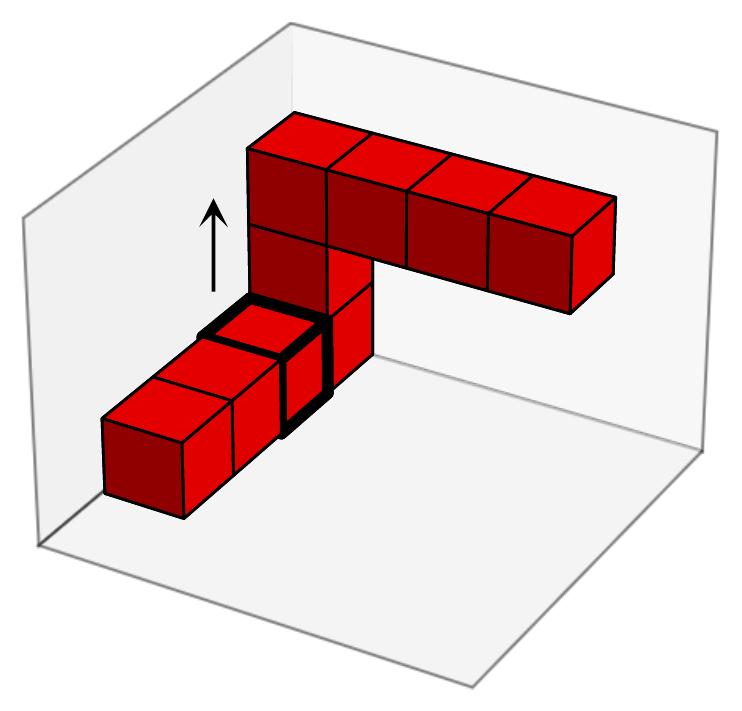}
    \subcaption{}
    \label{fig:align_kink_before}
  \end{minipage}
  \hspace{0.1\columnwidth}
  \begin{minipage}[b]{0.3\columnwidth}
    \centering
    \includegraphics[width=\columnwidth]{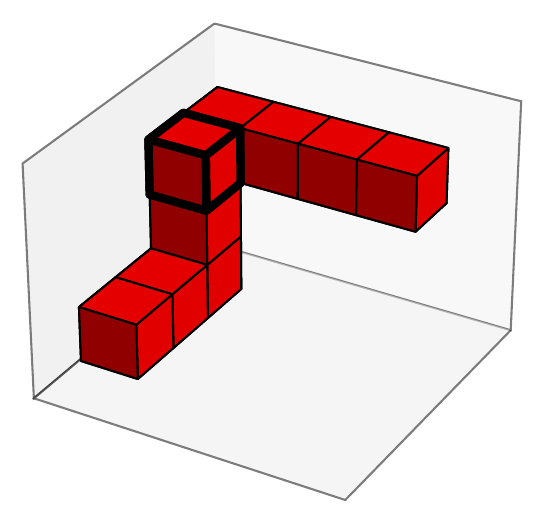}
    \subcaption{}
    \label{fig:align_kink_after}
  \end{minipage}
  \caption{Example of vertically aligning two cells adjacent to a kink. (a) To erase the kink, the lower cell should be raised and aligned with the higher cell. (b) After the alignment, the corner is no longer a kink. However, it may create or erase another adjacent kink, in which case it fails to flip the parity.}
  \label{fig:align_kink}
\end{figure}

We now discuss how we can let algorithms find paths satisfying kink conditions. Here, we focus on the Dijkstra projection algorithm and its look-ahead variant, and show how to adjust a 3D path of lattice surgery so that the parity of kinks becomes even. Our approach is to add a process to modify 3D paths after projection.
We propose a procedure that flips the kink parity of a projected path, provided that it possesses at least one 90-degree corner. Note that this condition is satisfied when modifying a projected path for an $XX$ or $ZZ$ measurement that has an odd number of kinks. This procedure is employed in the numerical evaluation because it runs in time linear in the path length, which makes it highly efficient.

The core concept for path modification involves lifting corner voxels, as illustrated in Figs.\,\ref{fig:twist_corner} and \ref{fig:align_kink}. However, the challenge lies in the fact that this modification might inadvertently create or erase another kink, thereby failing to flip the parity. To circumvent this challenge, we propose the following procedure that works on the first corner in the path.

\begin{enumerate}
  \item Scan the path, starting from either end, to identify the first 90-degree corner.
  \item If the corner is not a kink, lift the corner voxels to create a new kink, as shown in Fig.\,\ref{fig:twist_corner}. Since the first corner is chosen, this action does not affect the other corners and thus always flips the kink parity.
  \item If the corner is a kink, attempt to flip the parity by aligning it, as shown in Fig.\,\ref{fig:align_kink}. If this adjustment corrects the parity, terminate the procedure.
  \item If the parity remains odd after the attempt, apply Step 2 to the non-kink corner that results from the previous step. As Step 2 is guaranteed to flip the parity, this action ensures that the parity is flipped.
\end{enumerate}

As this procedure only assumes that the projected path has a corner, it similarly applies to routing logical CNOT gates in most cases. This assumption is fulfilled when, for instance, the data cells are assumed to have the same boundary orientations. We adopted this configuration for our numerical experiment, as described in Appendix~\ref{app:select_conversion_detail}.

\begin{figure}[tb]
  \centering
  \includegraphics[width=0.45\columnwidth]{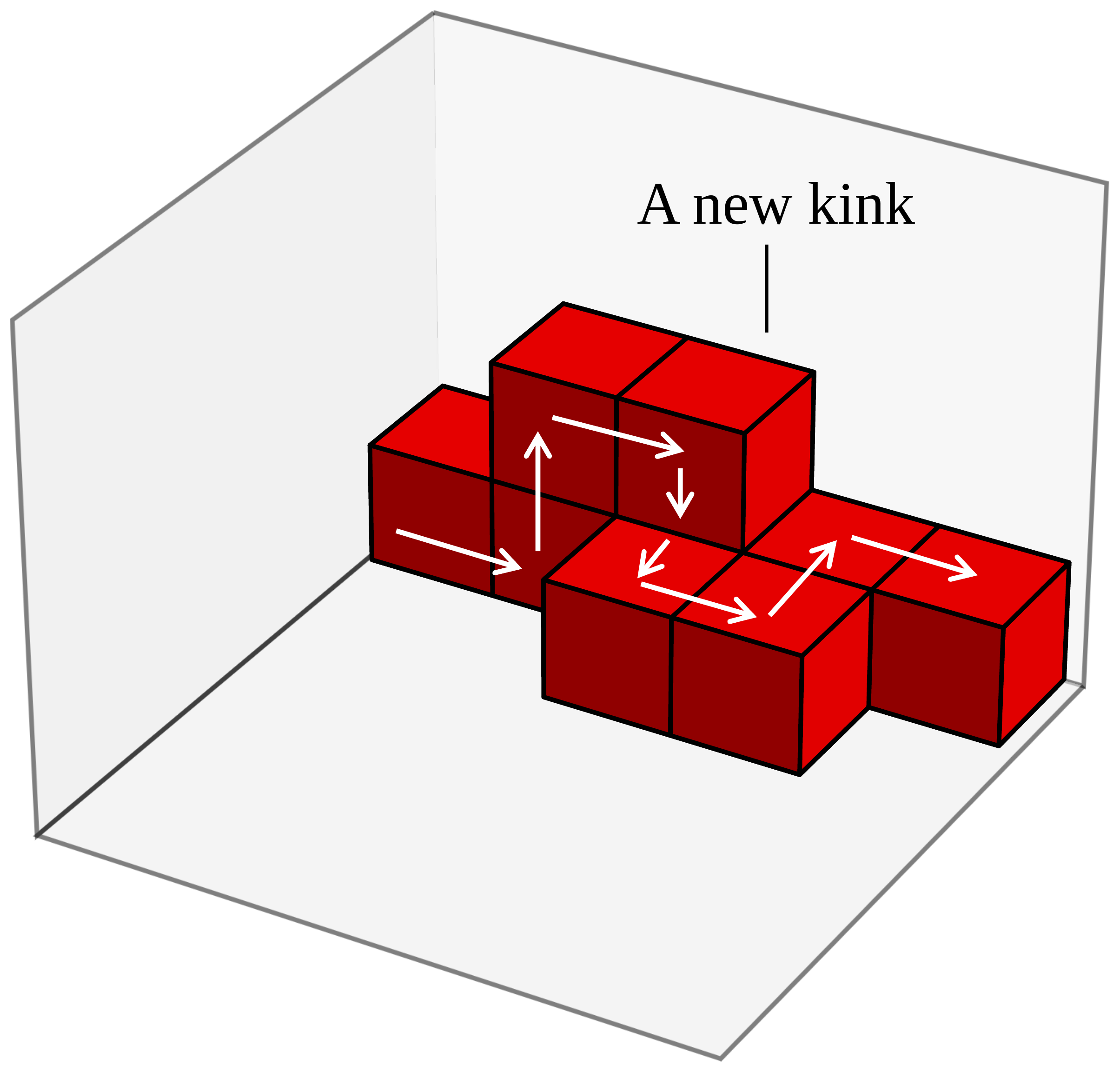}
  \caption{Twisting a straight path to create an extra kink.}
  \label{fig:straight_twist}
\end{figure}

Conversely, addressing a projected path for a CNOT gate without corners requires a totally different approach, as lifting voxels of such a path will not create any kinks. A viable alternative in this scenario is to twist a path to create a kink, as depicted in Fig.\,\ref{fig:straight_twist}. While this modification is minimal, requiring only four extra voxels, it involves nearby cells and demands finding an available space to insert a kink.

There is ample room for further proposals of kink-modification strategies, and comparing these proposals is left for future work.

\subsection{Time complexity analysis}
\label{sec:time_complexity}

This section derives the time complexities shown in Tabs.\,\ref{tab:comparison-path} and \ref{tab:comparison-method}. We first analyze the time complexity of the path finding subroutines shown in Tab.\,\ref{tab:comparison-path}. We begin by recalling that the BFS algorithm and Dijkstra's algorithm~\cite{fredman1987fibonacci} find a shortest path in an unweighted and weighted graph, respectively. Their respective running times for a graph with $V$ vertices and $E$ edges are $O(V + E)$ and $O(V \log V + E)$. Since a 2D lattice with $n$ cells has $V = O(n)$ vertices and $E = O(n)$ edges, the path search subroutines of the BFS and the Dijkstra projection run in $O(n)$ and $O(n \log n)$ time, respectively. Similarly, since a 3D lattice with height $h$ has $V = O(nh)$ vertices and $E = O(nh)$ edges, the path search subroutines of the 3D BFS and the 3D Dijkstra are $O(nh)$ and $O(nh \log nh)$, respectively.

We then analyze the overall time complexity of our scheduling methods, as shown in Tab.\,\ref{tab:comparison-method}. Before examining each method, we initially observe that the additional time complexity required for instruction look-ahead is $O(m \log m)$, since both fetching the next executable instruction and updating the dependency graph upon execution require $O(\log m)$ time.

Regarding the BFS, the bottleneck of the iterative procedure is $\textsc{FindShortestPath}$, which runs in $O(n)$ time. Since the failure in routing resets the qubit plane, the number of calls to $\textsc{FindShortestPath}$ can be bounded above by $2m$. Thus, the total runtime complexity is $O(n m)$.

We then analyze the runtime of the look-ahead BFS, which does not admit a favorable bound. Unlike in the BFS, the number of times $\textsc{FindShortestPath}$ is called can only be bounded as $O(n m)$. This bound is obtained because the outer loop iterates $O(m)$ times and the inner loop iterates $\left|L\right| = O(n)$ times. Furthermore, this bound may be tight in a pathological case where a single path partitions the qubit plane and subsequent executable instructions attempt to connect two logical qubits lying on different components. Thus, by incorporating the runtime of instruction look-ahead, the total runtime complexity can be bounded by $O(n^2 m + m \log m)$. Despite this poor theoretical bound, the algorithm runs almost as fast as the BFS in the numerical experiments.

Next, we analyze the 3D BFS and the 3D Dijkstra simultaneously as they share the algorithm except for the path search subroutine. Since the height $h$ of the 3D lattice can be evaluated as $O(m)$, the path search subroutine runs in $O(n m)$ for the 3D BFS and $O(n m \log n m)$ for the 3D Dijkstra. Since the routing subroutine never fails to find a path, the number of iterations is $m$, resulting in the total runtime of $O(n m^2)$ for the 3D BFS and $O(n m^2 \log n m)$ for the 3D Dijkstra.

Lastly, we analyze the time complexity of the Dijkstra projection and its look-ahead variant. As shown in Alg.\,\ref{alg:dijkstra_projection}, the overall time complexity of the Dijkstra projection is dominated by the function \textsc{FindShortestPathFromHeight}, which executes in $O(n \log n)$ time. We note that the function \textsc{LiftPath} will not be a bottleneck, since it runs in time linear to the path length $\left|p\right|$, even when modifying the kink parity to satisfy the condition. Given that the path length $\left|p\right|$ can be bounded above by $n$, this procedure runs in $O(n)$ time. Consequently, we have a total time complexity of $O(n m \log n)$ for the original algorithm and $O(n m \log n + m \log m)$ when combined with instruction look-ahead.

\section{Performance evaluation}
\label{sec:evaluation}
In this section, we evaluate the proposed methods (3D BFS, 3D Dijkstra, Dijkstra projection, and look-ahead Dijkstra projection) and compare them with the baseline methods (BFS and look-ahead BFS).
We used random circuits and circuits compiled from quantum phase estimation algorithms for benchmarking.
For quantum phase estimation, we first converted the circuit into an instruction sequence consisting of Pauli-$XX$ and $ZZ$ measurements and compared the performance of our methods in Sec.\,\ref{sec:select_circuit}. We also compared our methods with the existing algorithm for large-scale compilation, the Edge-Disjoint Paths Compilation (EDPC) algorithm~\cite{beverland2022surface}, in Sec.\,\ref{sec:comparison_edpc}. To satisfy their requirements and reproduce their experimental settings, we converted the circuit into another instruction sequence consisting of CNOT operations and $T$-gates and computed the scheduling results on their qubit plane pattern. See each section and Appendix~\ref{app:select_conversion_detail} for further details of these conversions and settings.

We compare these methods using two metrics. For a performance measure of the scheduling results, we used \emph{throughput}, the average number of instructions processed in parallel per code beat.
The other one is execution time, which is the required time for executing scheduling algorithms on classical computers. We evaluated the execution times on a computer equipped with an 11th Gen Intel(R) Core(TM) i5-1135G7 @ \SI{2.40}{\giga\hertz} with \SI{16}{\giga\byte} memory.
The scheduling program was implemented in C++ and optimized with the \texttt{O3} option using \texttt{g++} (Ubuntu 13.3.0-6ubuntu2$\sim$24.04) 13.3.0. For comparison with the EDPC algorithm, we also benchmarked the existing Julia code of the EDPC algorithm. We ran it on Julia Version 1.12.1 (2025-10-17) and measured the execution time with the \texttt{@btime} macro.

\subsection{Random circuits}
\label{sec:random_circuits}
For the random circuit benchmark, we set the number of total instructions to 1000, and the two logical qubits connected by each instruction were chosen uniformly at random. The data cells were placed in locations where the rows and columns are both even-numbered, as shown in Fig.\,\ref{fig:qubit_plane}. The number of logical qubits with data lined up on one side of the square qubit plane was defined as \emph{qubit plane size}. For example, Fig.\,\ref{fig:qubit_plane} corresponds to the example of a qubit plane size equal to four. The performance was evaluated by changing the qubit plane size from $2$ to $20$.
For evaluating the throughput and the execution time, we evaluated methods with ten random circuits generated with different seed values and averaged the results.

\begin{figure}[tb]
  \centering
  \includegraphics[width=\columnwidth]{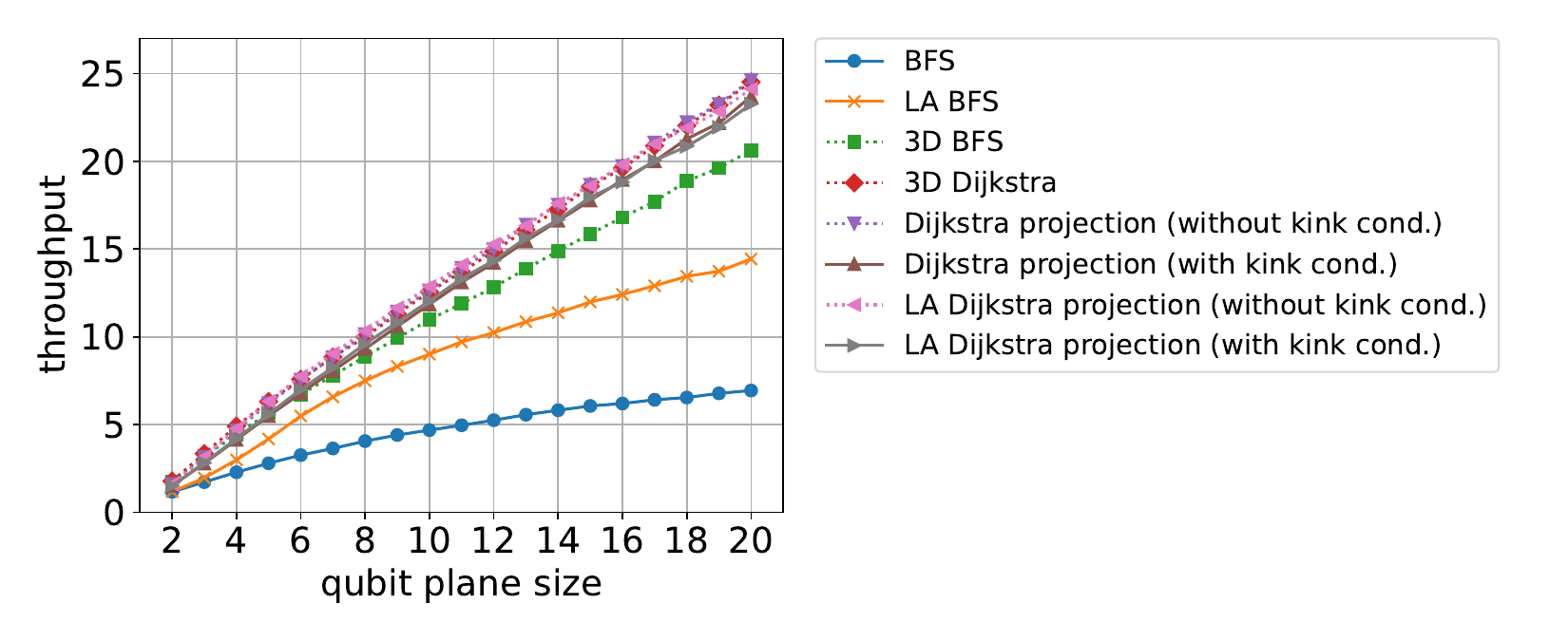}
  \caption{Throughput evaluation on random circuits.}
  \label{fig:random_circuit_throughput}
\end{figure}

We first evaluate the throughput of the baseline and proposed methods. The results are shown in Fig.\,\ref{fig:random_circuit_throughput}.
In this figure, each color corresponds to the performance of the proposed and baseline methods.
Here, we emphasize that the performance of the four methods shown as dotted lines in the figure, 3D BFS, 3D Dijkstra, Dijkstra projection (without kink condition), and look-ahead Dijkstra projection (without kink condition), do not satisfy the kink conditions. They are shown as a performance reference, and only the plots of the baseline methods, the Dijkstra projection (with kink condition), and the look-ahead Dijkstra projection (with kink condition) are valid. Since the optimization of the Dijkstra projection by instruction look-ahead hardly impacted throughput or execution time when evaluated on random circuits, we refer to both the original and optimized versions simply as Dijkstra projection throughout this setting.

The methods 3D Dijkstra and Dijkstra projection had the best throughput, followed in order by 3D BFS, look-ahead BFS, and BFS.
The small discrepancy between the Dijkstra projection and the 3D Dijkstra suggests that the limited search space of the Dijkstra projection does not significantly affect the quality of the obtained solution.
We find that the penalty of considering the kink condition is negligible since the performance difference of the Dijkstra projection with and without considering kink conditions is small.
In terms of the increase in throughput as the qubit plane size increased, the best baseline method, look-ahead BFS, showed a little plateauing, while the Dijkstra projection showed no such trend and increased at a similar rate. The throughput of the Dijkstra projection was on average $1.5$ times higher than that of the look-ahead BFS, but the improvement is likely to be even greater as the qubit plane size increases based on the above observation.

\begin{figure}[tb]
  \centering
  \includegraphics[width=\columnwidth]{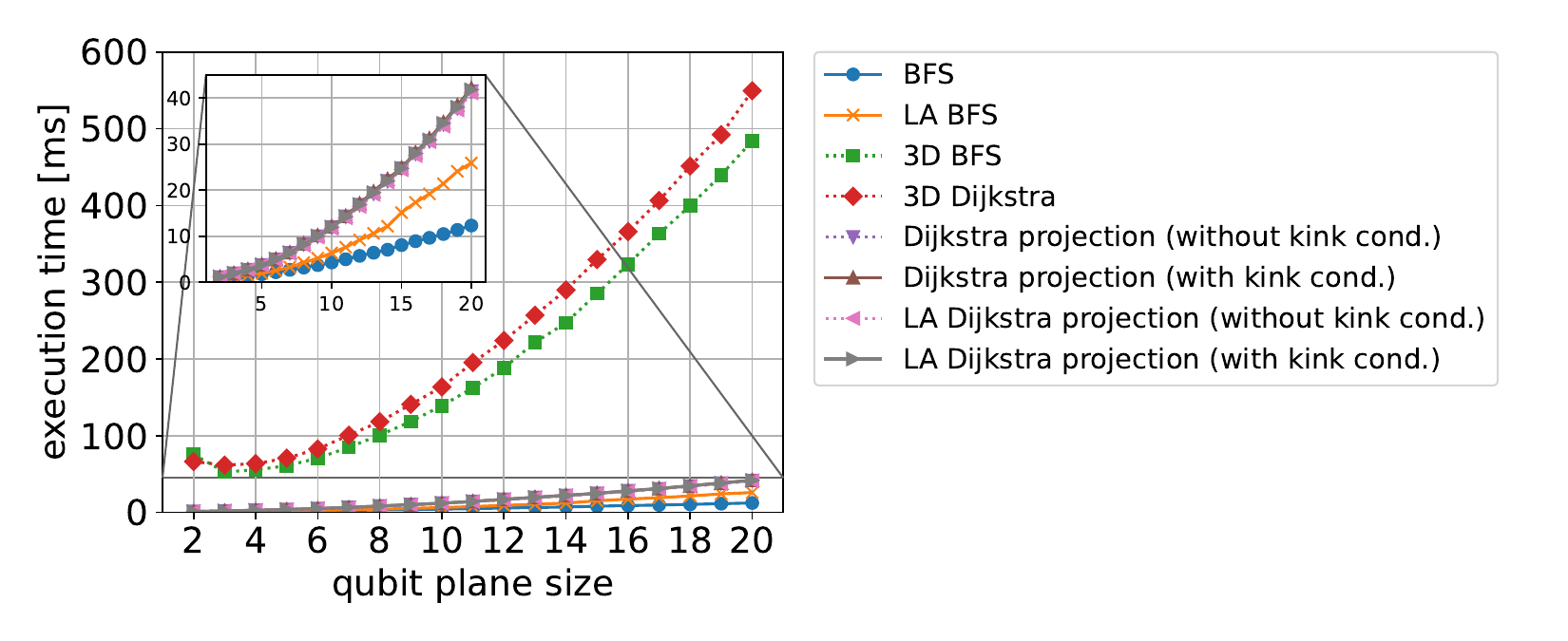}
  \caption{Evaluation of execution time on random circuits.}
  \label{fig:random_circuit_time}
\end{figure}

Next, we evaluated the execution times, which are shown in Fig.\,\ref{fig:random_circuit_time}.
We observed that the Dijkstra projection and two baseline methods show better scaling compared to the 3D BFS and the 3D Dijkstra. This is because the first three methods search on the 2D lattice, while the other two search on the 3D lattice. These results are consistent with the expected properties of the algorithm design. We also observed that there is almost no difference between the execution times of the Dijkstra projection with and without considering kink conditions. Thus, the time for adjusting kink parity is negligible compared to the path-finding part.

\subsection{SELECT circuits}
\label{sec:select_circuit}
We then evaluated the performance for practical applications. Here, we chose the SELECT circuit as a benchmark target. The SELECT operation is one of the bottleneck components in quantum phase estimation algorithms using qubitization~\cite{low2019hamiltonian,babbush2018encoding,yoshioka2022hunting}. Its action is defined from Hamiltonian $H=\sum_{i=0}^{L-1} \alpha_i P_i$ as follows:
\begin{align}
  \sum_{i=0}^{L-1} \ket{i}\bra{i} \otimes P_i,
\end{align}
where $L$ is an integer, $\ket{i}$ is a computational basis of an $L$-dimensional Hilbert space, and $P_i$ is a Pauli operator. In this benchmark, a SELECT circuit is synthesized for a Hamiltonian of a $16 \times 16$ 2D Heisenberg model (i.e., 256 spins) with the nearest-neighbor interaction. There are several ways to translate SELECT circuits into a sequence of basic operations. We used the methods proposed by Yoshioka \textit{et al.}~\cite{yoshioka2022hunting} and generated SELECT circuits with two configurations, which we named SELECT 0 and SELECT 5. These suffixes indicate the parallelism parameter, where SELECT 0 is a serial implementation of the SELECT circuit and SELECT 5 is parallelized into $2^5=32$ threads. SELECT 5 requires more logical qubits and instructions than SELECT 0, but ideally has a larger throughput due to parallelization. SELECT 0 required $25$ qubit plane size and $59{,}344$ instructions, and SELECT 5 required $55$ qubit plane size and $61{,}492$ instructions. We note that these instances are solved by quantum computers sufficiently faster than the state-of-the-art algorithm for classical computing~\cite{yoshioka2022hunting}. Thus, the performance for this instance reflects the realistic scenario of practical use of FTQCs.

In the following evaluation, we compare four methods: BFS, look-ahead BFS, Dijkstra projection, and look-ahead Dijkstra projection. We omitted the results of the 3D BFS and the 3D Dijkstra since they cannot finish the compilation within 1 hour.

The throughput of the four methods is shown in Tab.\,\ref{tab:SELECT_throughput}.
\begin{table}[tb]
  \centering
  \caption{Throughput evaluation on SELECT circuits.}
  \label{tab:SELECT_throughput}
  \renewcommand{~}{\hphantom{0}}
  \vspace{-3mm}
  \begin{tabular}{lrr} \toprule
                                                            & \multicolumn{2}{c}{throughput}            \\
    \cmidrule(rl){2-3}
                                                            & SELECT 0                       & SELECT 5 \\ \midrule

    BFS                                                     & 1.59                           & 1.76     \\
    look-ahead BFS                                          & 1.77                           & 20.74    \\
    Dijkstra projection (without kink condition)            & 2.90                           & 60.58    \\
    Dijkstra projection (with kink condition)               & 2.90                           & 57.68    \\
    look-ahead Dijkstra projection (without kink condition) & 2.90                           & 80.28    \\
    look-ahead Dijkstra projection (with kink condition)    & 2.90                           & 78.43    \\ \bottomrule
  \end{tabular}
\end{table}
The Dijkstra projection and its optimized variant show better throughput than the two baseline methods. Compared to the look-ahead BFS, the look-ahead Dijkstra projection shows improvement by a factor of $1.6$ for SELECT 0 and $3.8$ for SELECT 5. We observed higher throughput improvement for SELECT 5, which is an expected result considering that SELECT 5 has a structure like a parallel circuit and leaves much room for parallelization. We can see that the degradation by considering kink conditions is at most 5\%, which is not negligible, but does not lose the advantage of the Dijkstra projection.

The execution times of the methods are shown in Tab.\,\ref{tab:SELECT_time}.
\begin{table}[tb]
  \centering
  \caption{Evaluation of execution time on SELECT circuits.}
  \label{tab:SELECT_time}
  \renewcommand{~}{\hphantom{0}}
  \vspace{-3mm}
  \begin{tabular}{lrr} \toprule
                                                            & \multicolumn{2}{c}{execution time [s]}            \\
    \cmidrule(rl){2-3}
                                                            & SELECT 0                               & SELECT 5 \\ \midrule
    BFS                                                     & 0.58                                   & 2.78     \\
    look-ahead BFS                                          & 0.65                                   & 6.84     \\
    Dijkstra projection (without kink condition)            & 3.55                                   & 26.07    \\
    Dijkstra projection (with kink condition)               & 3.57                                   & 26.56    \\
    look-ahead Dijkstra projection (without kink condition) & 3.60                                   & 19.55    \\
    look-ahead Dijkstra projection (with kink condition)    & 3.59                                   & 19.75    \\ \bottomrule
  \end{tabular}
\end{table}
The order of execution times of the four methods for the SELECT circuits was basically the same as for the random circuit. On the other hand, the look-ahead Dijkstra projection demonstrated a significant speedup compared to the original Dijkstra projection when applied to SELECT 5. Further profiling indicated that these execution times were dominated by path searching, which strongly suggests that the instruction look-ahead technique results in substantially earlier termination of the path search subroutine. This is likely because fetching the instruction with the lowest height by instruction look-ahead flattens the current scheduling outcome with less biased height distribution, thereby shortening the subsequent lattice-surgery paths and the time required to find them.

All four methods finished in practical execution time. The additional time for considering kink conditions is at most 2\%. Thus, the time for considering kink conditions is also negligible in this case. Also, when applied to the SELECT 0 circuit, the 3D BFS and the 3D Dijkstra cannot finish within an hour while the Dijkstra projection finishes within a few seconds, again indicating that the Dijkstra projection excels in execution time among the methods that utilize the time direction.

\subsection{Comparison with the existing methods}
\label{sec:comparison_edpc}

In the following evaluation, four of our methods---BFS, look-ahead BFS, Dijkstra projection, and look-ahead Dijkstra projection---were compared with the Edge-Disjoint Paths Compilation (EDPC) algorithm~\cite{beverland2022surface}. We adopted a problem setting that satisfies the requirements of the EDPC algorithm. Specifically, the instruction sequence consists solely of CNOT gates and $T$-gates. Each $T$-gate is realized via $T$-gate teleportation between the target cell and a magic state factory cell aligned at the perimeter of the qubit plane. (For further details, see Appendix~\ref{app:select_conversion_detail}.) Under this setting, SELECT 0 required $17$ qubit plane size and $37{,}856$ instructions, and SELECT 5 required $30$ qubit plane size and $39{,}050$ instructions.

\begin{table}[tb]
  \centering
  \caption{Throughput evaluation on SELECT circuits. The grayed rows annotated with an asterisk represent reference values that disregard the kink condition, optimistically assuming that  CNOT operations complete via a 2D path in one  code beat. Although  these raw infeasible values are used to compare net routing performance, halving these grayed values will provide the corresponding feasible values, reflecting that CNOT operations realistically require two code beats. The remaining methods adhere to the kink condition and yield a feasible solution.}
  \label{tab:SELECT_throughput_EDPC}
  \renewcommand{~}{\hphantom{0}}
  \vspace{-3mm}
  \begin{tabular}{lrr}\toprule
                                                                                  & \multicolumn{2}{c}{throughput}            \\
    \cmidrule(rl){2-3}
                                                                                  & SELECT 0                       & SELECT 5 \\ \midrule

    \rowcolor{lightgray} EDPC*                                                    & 1.09                           & 9.06     \\
    EDPC                                                                          & 0.66                           & 4.74     \\
    \rowcolor{lightgray} BFS*                                                     & 1.06                           & 1.26     \\
    \rowcolor{lightgray} look-ahead BFS*                                          & 1.12                           & 17.26    \\
    \rowcolor{lightgray} Dijkstra projection (without kink condition)*            & 1.85                           & 17.34    \\
    Dijkstra projection (with kink condition)                                     & 1.85                           & 15.48    \\
    \rowcolor{lightgray} look-ahead Dijkstra projection (without kink condition)* & 1.85                           & 36.29    \\
    look-ahead Dijkstra projection (with kink condition)                          & 1.85                           & 31.39    \\ \bottomrule
  \end{tabular}
\end{table}
The throughput of these methods is shown in Tab.\,\ref{tab:SELECT_throughput_EDPC}. The grayed results represent configurations that violate the kink condition, rendering them infeasible. Since a CNOT gate can be implemented via a 2D path using two code beats~\cite{beverland2022surface}, these entries can be converted to feasible counterparts by halving the values. However, the subsequent discussion compares the raw values to evaluate routing efficiency without this overhead. Notably, since our proposed methods satisfy the kink condition with modest overheads, accounting for the feasibility would effectively double the performance advantage, making our results even more favorable compared to the baseline methods.

For the performance of the EDPC algorithm, it lies between that of the BFS and the look-ahead BFS. This is due to their different strategy of managing instruction dependency. Their algorithm initially applies topological sorting to the instruction dependency graph and repeatedly removes the instructions that  have no dependent instructions, thereby decomposing the instructions into a sequence of instruction sets consisting of mutually independent instructions. This partitioning enables a parallel compilation of the instruction sets and facilitates designing a scheduler that handles many types of instructions, whereas it also worsens the scheduling performance by effectively adding ``barriers'' in the circuit. We consider this to be one of the reasons why the look-ahead BFS outperforms EDPC, although their routing strategy is more sophisticated than our baseline methods.

Although the overall trends of our methods are consistent with the results in Sec.\,\ref{sec:select_circuit}, it is worth mentioning that the instruction look-ahead technique proved exceptionally effective in this instance. This almost closed the gap between the look-ahead BFS and the Dijkstra projection, and resulted in an outstanding performance by the look-ahead Dijkstra projection.

This gap reduction can also be attributed to the difference in CNOT gate handling. The former experiment decomposes CNOT gates into $ZZ$ and $XX$ measurements, which partially yields stair-like structures and is likely advantageous for 3D routing. However, we emphasize that the same benefits can also be observed in this experiment when we compare feasible throughputs by halving the grayed ones: the Dijkstra projection showed $1.8$-fold improvement over the look-ahead BFS, and the look-ahead Dijkstra projection demonstrated $3.6$-fold improvement.

\begin{table}[tb]
  \centering
  \caption{Evaluation of execution time on SELECT circuits. The grayed row annotated with an asterisk presents reference values, as this algorithm was implemented in Julia, whereas our methods are implemented in C\texttt{++}.}
  \label{tab:SELECT_time_EDPC}
  \renewcommand{~}{\hphantom{0}}
  \vspace{-3mm}
  \begin{tabular}{lrr} \toprule
                                                            & \multicolumn{2}{c}{execution time [s]}            \\
    \cmidrule(rl){2-3}
                                                            & SELECT 0                               & SELECT 5 \\ \midrule
    \rowcolor{lightgray} EDPC*                              & 22.19                                  & 207.52   \\    BFS                                                     & 0.31                                   & 0.90     \\
    look-ahead BFS                                          & 0.32                                   & 1.47     \\
    Dijkstra projection (without kink condition)            & 1.27                                   & 4.81     \\
    Dijkstra projection (with kink condition)               & 1.28                                   & 4.83     \\
    look-ahead Dijkstra projection (without kink condition) & 1.28                                   & 3.71     \\
    look-ahead Dijkstra projection (with kink condition)    & 1.29                                   & 3.74     \\ \bottomrule
  \end{tabular}
\end{table}
The execution times of the methods are shown in Tab.\,\ref{tab:SELECT_time_EDPC}. The substantial execution time of EDPC may stem from several reasons. We first note that their algorithm was implemented in Julia, whereas ours was implemented in C++. Although we used \texttt{@btime} to warm up the program and exclude just-in-time compilation time, this difference likely introduced overhead to their execution time. Another factor is their use of maximum flow subroutines in $T$-gate teleportation, which may have incurred high computational cost.

\subsection{Discussion}
First, we discuss characteristics seen both in the random circuits and the SELECT circuits. For throughput, the BFS, the look-ahead BFS, the 3D BFS, and the 3D Dijkstra performed well in that order, with the Dijkstra projection performing as well as the 3D Dijkstra. The look-ahead optimization in the Dijkstra projection had no impact on the random circuits, while it proved effective in SELECT circuits. It is natural that the look-ahead BFS outperforms the BFS, considering that the look-ahead BFS executes future operations and produces more spare space that would be blocked in the BFS. The primary reason the 3D BFS outperforms the look-ahead BFS is its ability to pack paths more densely into the lattice by increasing the dimension of the paths by one. Another reason is that a simple look-ahead can be reproduced by searching in 3D. The reason the 3D Dijkstra outperforms the 3D BFS is that the paths can be packed more densely into the lattice in the end by keeping the overall path low, even if it is a bit roundabout, rather than unnecessarily shortening the path by going through an excessively high layer. The reason the Dijkstra projection performs comparably to the 3D Dijkstra is likely that the paths adopted by the 3D Dijkstra tend to be along the bottom surface, as in the Dijkstra projection.

In terms of execution time, the BFS was the shortest, the look-ahead BFS was slightly longer than the BFS, the Dijkstra projection was a few times longer, and the 3D BFS and the 3D Dijkstra were much longer than the others. From these results, execution time is significantly affected by whether the search space is a 2D or 3D lattice. The fact that the look-ahead BFS is slightly slower than the BFS is a natural consequence of the additional cost of instruction look-ahead. The reason the Dijkstra projection is several times slower than the BFS and the look-ahead BFS is that, as a comparison of search algorithms, Dijkstra's algorithm is more computationally expensive than the BFS algorithm. The same argument holds for the 3D BFS and the 3D Dijkstra.

\subsection{Resource overhead analysis}
This section analyzes the overhead associated with introducing the concept of 3D routing and scheduling algorithms based on this idea, such as the Dijkstra projection. We categorize the potential penalties into four types: compilation-time overhead, error-rate overhead, qubit-count overhead, and architectural overhead.

The quantitative compilation-time overhead was shown in the previous section as execution time. The results indicate that employing the Dijkstra projection with kink constraints increases the compilation time per instruction by a factor of a few. For the SELECT instances, which are large enough to demonstrate quantum advantage~\cite{yoshioka2022hunting}, the compilation completes in approximately half a minute. This duration is negligible compared to the total execution time, as the quantum phase estimation with the corresponding SELECT circuit on a real FTQC would take over an hour~\cite{yoshioka2022hunting}. Even for larger instances, both the compilation time with the Dijkstra projection and the execution time of an actual machine scale almost linearly with the number of instructions, as summarized in Tab.~\ref{tab:comparison-method}. Therefore, the use of the Dijkstra projection introduces no practical overhead in regimes where quantum computation is meaningful.

A potential overhead in using the Dijkstra projection is an increase in the error rate. Let the number of voxels in the 3D representation be denoted as the \textit{active volume}. Because 3D routing might entail longer routing paths than 2D routing, the active volume can be larger. We denote the probability of maintaining a single logical qubit fault-tolerantly for a single code beat as $p_{\rm voxel}$ and the active volume as $A$. Then, under the assumption that $p_{\rm voxel}$ is sufficiently small and that errors are largely independent, the overall failure probability can be approximated as $p_{\rm voxel} A$.  Hence, the active volume serves as a semi-quantitative proxy of the overall failure probability, provided that the errors are approximately independent and that $p_{\rm voxel}$ is low enough to keep the overall failure risk within a practical range. We evaluated the increase in active volume using 3D-routing-based methods for the SELECT circuits, as shown in Tab.\,\ref{tab:SELECT_volume}. As a conservative analysis, our proposed method satisfying the kink condition is compared with the baseline method disregarding the kink condition. The results show that, for the SELECT 0 instance, the look-ahead Dijkstra projection with kink condition increases the active volume by about 30\% compared to the BFS without kink conditions. Conversely, the volume decreases for the SELECT 5 instance. We expect this is because improved instruction throughput due to 3D routing reduces the idle time of logical qubits. As discussed below, the total program failure rate can be kept constant by increasing the qubit count, and a 30\% increase in active volume can be compensated with a modest cost.

Next, we examine the qubit-count penalty. The total number of physical qubits in an FTQC system is the product of the number of logical-qubit cells in the qubit plane and the number of physical qubits per logical-qubit cell. In surface codes, the latter is $2d^2-1$, where $d$ is the code distance. The introduction of 3D routing does not increase the number of logical-qubit cells. Although 3D routing temporarily allocates logical qubits to separate logical instructions, they are assigned to existing cells.
In contrast, the number of physical qubits per logical-qubit cell may increase. If the active volume increases, the error-correction failure rate per voxel $p_{\rm voxel}$ must be reduced to maintain a constant overall program failure probability, which requires increasing the code distance. The error-estimation failure probability per syndrome-measurement cycle on a single cell is approximated as $0.1 (100p)^{(d+1)/2}$~\cite{litinski2019game}, where $p$ is the physical error rate per cycle. Consequently, the failure probability per voxel (per $d$ cycles) is $p_{\rm voxel} = 0.1 d (100p)^{(d+1)/2}$. Assuming $p=10^{-3}$ and $d$ is sufficiently large, increasing (decreasing) the code distance by two allows for roughly 10 times larger (smaller) active volume.
In practical regimes, the code distance $d$ is typically 23 or higher~\cite{yoshioka2022hunting}. Increasing $d$ from 23 to 25 corresponds to approximately an 18\% increase in total physical qubits. As discussed earlier, 3D routing can either increase or decrease the active volume by a few times, which corresponds to at most an 18\% variation in total physical qubits.

Finally, we consider architectural overheads when implementing 3D routing in FTQC architectures. We emphasize again that the number of logical-qubit cells remains unchanged by 3D routing. Therefore, if an FTQC machine is capable of handling all logical-qubit cells simultaneously, our method introduces no architectural overhead. In extreme cases where the total number of logical qubits $n$ is very large, it might be possible that FTQC machines cannot store $n$ logical qubits due to limitations in symbol handling, error-estimation throughput, or synchronization. However, such situations are unlikely, at least in near- and mid-term FTQCs, as achieving quantum advantage typically requires at most $10^3$ logical qubits.
Assuming a code distance of $d=23$, one code beat, i.e., the latency of a lattice-surgery instruction, corresponds to \SI{23}{\micro\second} or longer if we assume the code cycle is \SI{1}{\micro\second}~\cite{yoshioka2022hunting}. Managing $10^3$ symbols and logical qubits, as well as loading pre-compiled paths from memory every \SI{23}{\micro\second}, is a feasible requirement for classical control hardware.

\begin{table}[tb]
  \centering
  \caption{Active volume evaluation on SELECT circuits. The grayed rows annotated with an asterisk represent reference values that disregard the kink condition, optimistically assuming that  CNOT operations complete via a 2D path in one  code beat. Although  these raw infeasible values are presented to show their net routing performance, doubling these values will provide the corresponding feasible values, reflecting that CNOT operations realistically require two code beats. The remaining methods adhere to the kink condition and yield a feasible solution.}
  \label{tab:SELECT_volume}
  \renewcommand{~}{\hphantom{0}}
  \vspace{-3mm}
  \begin{tabular}{lrr}\toprule
                                                                                  & \multicolumn{2}{c}{active volume ($\times 10^6$)}            \\
    \cmidrule(rl){2-3}
                                                                                  & SELECT 0                                          & SELECT 5 \\ \midrule

    \rowcolor{lightgray} EDPC*                                                    & 10.44                                             & 5.46     \\
    EDPC                                                                          & 17.25                                             & 9.89     \\
    \rowcolor{lightgray} BFS*                                                     & 13.25                                             & 32.54    \\
    \rowcolor{lightgray} look-ahead BFS*                                          & 12.51                                             & 3.42     \\
    \rowcolor{lightgray} Dijkstra projection (without kink condition)*            & 17.47                                             & 7.35     \\
    Dijkstra projection (with kink condition)                                     & 16.43                                             & 8.21     \\
    \rowcolor{lightgray} look-ahead Dijkstra projection (without kink condition)* & 17.45                                             & 3.45     \\
    look-ahead Dijkstra projection (with kink condition)                          & 16.58                                             & 3.99     \\ \bottomrule
  \end{tabular}
\end{table}

\subsection{Robustness under Resource Pressure}

To evaluate the robustness of our proposed methods under high resource contention, we examined their performance across circuits with varying gate densities. We define the \textit{average gate density} as the number of instructions divided by the circuit depth, which quantifies the congestion level of the problem instance. This metric also serves as a theoretical upper bound for the throughput of the 2D routing methods, BFS and look-ahead BFS. As discussed in Sec.\,\ref{sec:select_circuit}, we utilized the SELECT circuits to generate different levels of congestion by varying the parallelism parameter from $0$ to $8$: SELECT 0 corresponds to a serial implementation, and SELECT $i$ is parallelized into $2^i$ threads.

\begin{figure}
  \centering
  \includegraphics[width=0.8\linewidth]{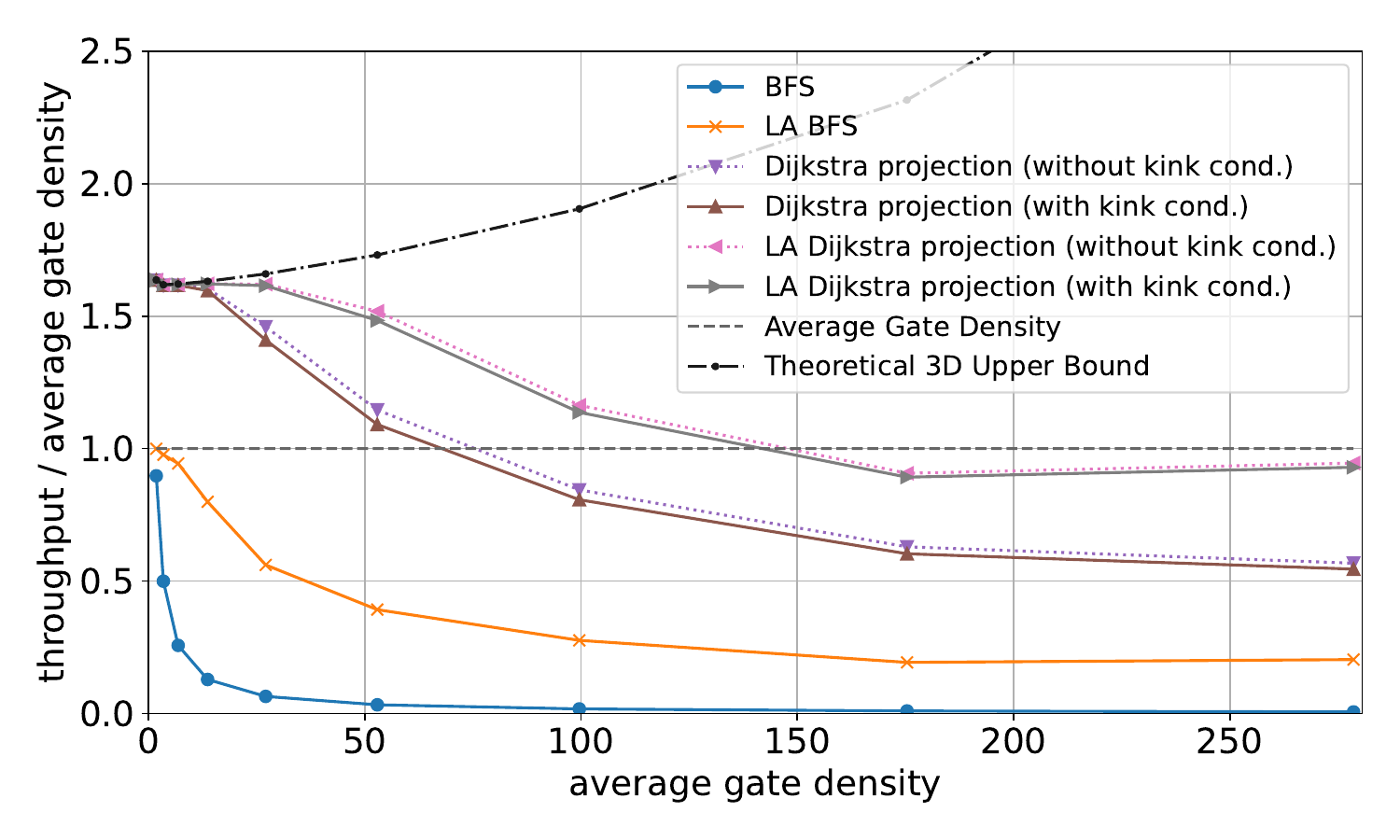}
  \caption{Degradation curves of throughput as a function of gate density. The performance is measured by the throughput normalized by the average gate density, which corresponds to an ideal throughput of conflict-free 2D routing.}
  \label{fig:degradation_throughput}
\end{figure}

\begin{figure}
  \centering
  \includegraphics[width=0.8\linewidth]{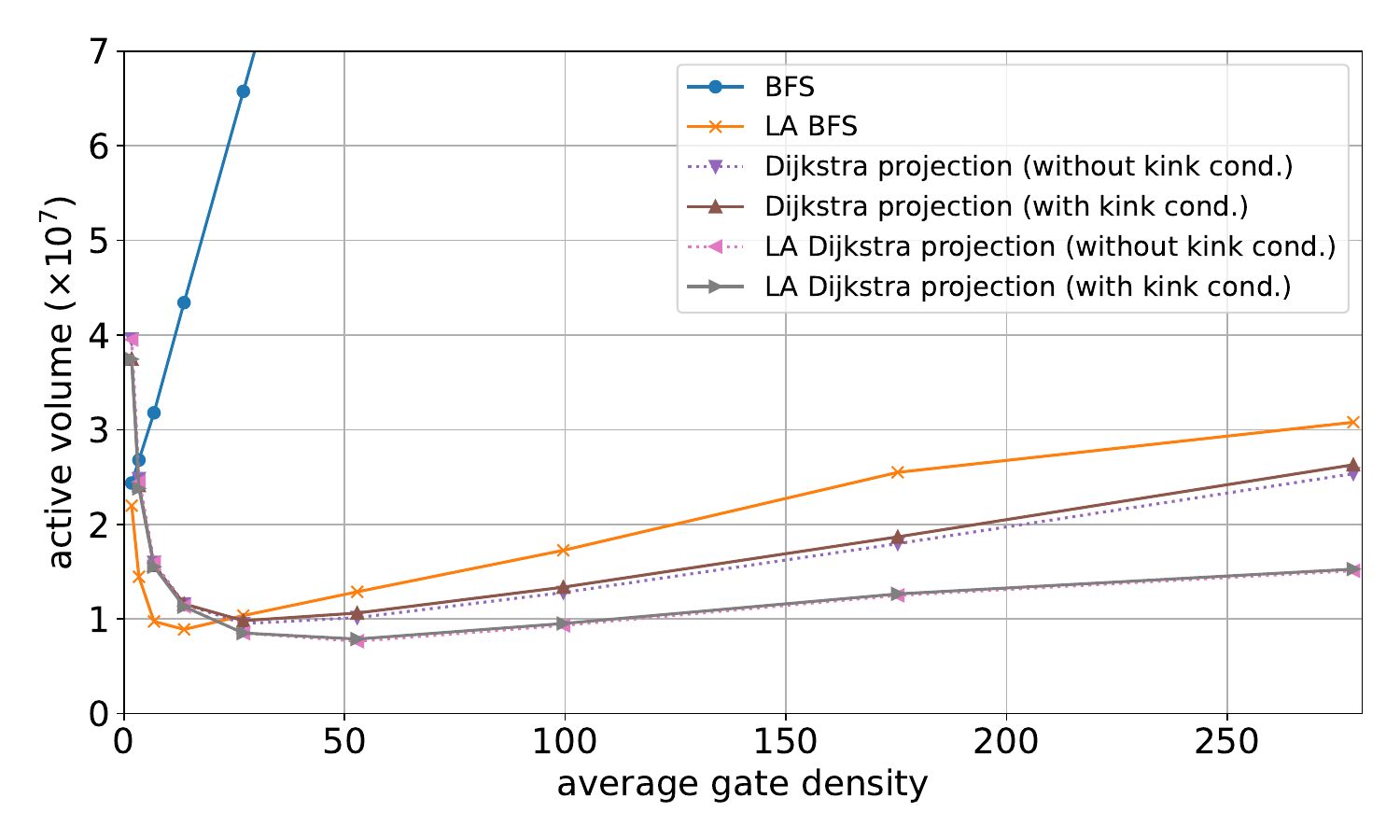}
  \caption{Degradation curves of active volume as a function of gate density.}
  \label{fig:degradation_volume}
\end{figure}

Fig.\,\ref{fig:degradation_throughput} illustrates the performance degradation of the scheduling methods as resource pressure intensifies. The performance is quantified by the throughput normalized by the average gate density, which is equivalent to the ratio of the circuit depth to the total execution time in code beats. To assess the impact of routing conflicts in the proposed methods, we also included the normalized performance of the conflict-free 3D routing, that reserves only path endpoints and ignores all intermediate segments. This benchmark serves as a theoretical upper bound on the performance of 3D routing methods.

As expected, all methods exhibit a decline in efficiency as gate density increases. Notably, at low gate densities, the proposed methods achieve near-optimal performance, outperforming the circuit depth. This can be attributed to their advantages in stair-like structures, as discussed in Sec.\,\ref{subsec:solution_improvement}. Increasing gate density leads to performance degradation across all routing methods, resulting in a growing discrepancy compared to the theoretical 3D upper bound due to intensified path conflicts. However, it is noteworthy that the proposed methods remain comparable to the circuit depth, demonstrating their resilience to high resource pressure.

Fig.\,\ref{fig:degradation_volume} shows the active volume of the routing methods in the same numerical experiment. While all methods experience degradation, the proposed methods maintain an active volume close to that of the look-ahead BFS across different gate densities. Furthermore, they outperform the look-ahead BFS in higher densities due to their shorter execution time in code beat.

\section{Conclusion and future outlook}
\label{sec:conclusion}
In this paper, we proposed efficient and high-performance algorithms for lattice-surgery instructions. We showed that the mapping of lattice-surgery instructions can be reduced to 3D path search problems. Based on this mapping, we proposed the Dijkstra projection and its look-ahead variant to perform efficient scheduling of lattice-surgery instructions. We evaluated their performance on several settings, and they showed significant improvements in throughput compared to the baseline, while the increase in execution time is modest.

In our numerical experiment, only the Pauli-$XX$ and $ZZ$ measurements were considered, and other logical operations were ignored for simplicity. However, we need to consider other types of instructions in practice, such as single-qubit Clifford gates, lattice-surgery instructions acting on many logical-qubit cells, and magic-state teleportations.
In particular, many-body lattice-surgery operations are vital since they are used to remove Clifford unitary operations from the circuits~\cite{litinski2019game} and used in several existing works~\cite{litinski2019game,silva2024multi}. In Appendix~\ref{app:manybody_lattice_surgery}, we provide the extension of our techniques to many-body lattice surgery.

One may also be concerned that logical-qubit cells are blocked during the reaction time in magic-state teleportation. Typical $T$-gate teleportation circuits require $S$-gate feedback according to the measurement results, and we cannot perform operations on cells that may be targets of the feedback during the latency. Nevertheless, we expect this is not a problem since we can relax this blocking by using auto-corrected $T$-gate teleportation~\cite{litinski2019game,fowler2012time}, which converts feedback operations from $S$-gate on the computing space to the choice of Pauli measurement basis on the ancillary space.

Combining these theoretical ideas and providing consistent and full-stack compilation frameworks, such as Refs.\,\cite{beverland2022assessing,zhu2024ecmas,zhu2025ecmas+,watkins2023high}, would be the next direction of this topic. To illustrate the contexts intended by our method, we describe an example full-stack compilation workflow in Appendix~\ref{app:compilation_framework}. As explained in this appendix, our method is compatible with lattice-surgery-based FTQC pipelines under the following four assumptions. (1) The positions and connectivity of physical qubits are fixed and cannot be changed during computation. (2) Logical qubits are encoded using surface codes, and logical operations are implemented via code deformation and lattice surgery. (3) All required surface-code logical qubits are allocated at the beginning of execution and are not deallocated during program execution. (4) Magic states are generated and managed within abstracted modules placed at fixed locations, and we ignore probabilistic behavior at runtime for simplicity. We believe that these assumptions are common when evaluating architectures with superconducting qubits~\cite{beverland2022assessing,google2023suppressing,google2025quantum}, and the third and fourth assumptions would be removed as they are assumed to simplify the compilation and evaluation process. On the other hand, the other two assumptions do not necessarily hold when integrating recent technologies such as qubit shuttling in neutral-atom platforms~\cite{zhou2025low}, non-topological or concatenated quantum error-correcting codes~\cite{gidney2025yoked,yoder2025tour,xu2024constant}, and generalized lattice-surgery operations for such codes~\cite{cohen2022low}. Extending our method to a wider range of compilation frameworks and evaluating the performance of our methods in broader situations are expected, which are left for future work.

There are several other directions for future work. For the voxel weighting in the 3D Dijkstra and the Dijkstra projection, we proposed the weighting that worked best among those we tried, but we expect that there is room for further improvement.
Also, the theoretical upper and lower bounds of the idea, separating large instructions into small fragments, may be calculated in more detail in the future. By comparing these bounds with our numerical results, we can measure the room for improvement.
While we introduced a method to satisfy the restriction of kink parity for lattice-surgery instructions, there may be a more efficient way to control the number of kinks. Integrating kink restrictions in the path-finding algorithms more naturally would be another future topic.

Integrating our method into distributed FTQC would also be essential. The integration method strongly depends on the paradigm of distributed FTQC. If the number of qubits per node is sufficiently small, logical qubits are constructed through communication among several nodes~\cite{nickerson2013topological} or realized via measurement-based FTQC~\cite{bombin2021interleaving,bourassa2021blueprint}. In such cases, the routing problem is equivalent to that in the single-node scenario, and our method can be applied without modification. The situation becomes more complex when the number of qubits per node is moderate. In this regime, each node can host several logical qubits, while two-qubit logical operations are performed via logical entanglement distribution and teleportation~\cite{pattison2025constant,maeda2025logical}. Since this inter-node teleportation can be regarded as a sparse connection between 3D voxels, we believe our method can be extended to such cases. Nevertheless, the detailed implementation of distributed FTQC architectures remains immature and will be an important subject for future work.

The results of this research extend the lattice-surgery scheduling problem, which has been addressed as a graph problem in 2D space, to a 3D path search by adding a time direction axis. Therefore, the present results not only enable computation on fault-tolerant quantum computation with higher efficiency than previously possible but also suggest that graph algorithms can be applied to optimize the compilation of programs for fault-tolerant quantum computation. Combining the optimization methods obtained in the present study with the optimization of quantum circuits and the optimization of the placement of logical qubits, and appropriately formulating these schemes into graph problems, we expect to be able to accelerate the quantum algorithms that will be practical in fault-tolerant quantum computers in the future.

\section*{Acknowledgements}
We are grateful to Hiroki Hamaguchi and Nobuyuki Yoshioka for their valuable assistance in executing the existing EDPC algorithm code.
This work is supported by PRESTO JST Grant No.\,JPMJPR1916, MEXT Q-LEAP Grant No.\,JPMXS0120319794 and JPMXS0118068682, JST Moonshot R\&D Grant No.\,JPMJMS2061, and JST CREST Grant No.\,JPMJCR23I4.

\appendix

\section{Extension to many-body lattice surgery}
\label{app:manybody_lattice_surgery}

The proof and algorithm presented in the main text focused on 3D routing for two-body lattice surgery. Two-body lattice surgery is a crucial case because quantum circuits are typically expressed in the Clifford+T form, where S and H gates are executed via code deformation~\cite{brown2017poking,fowler2018low}, and CNOT and T gates are translated into two-body lattice surgeries~\cite{chamberland2022universal}.
On the other hand, all Clifford gates can be omitted by converting two-body lattice surgeries into many-body ones, provided there is no intermediate measurement~\cite{litinski2019game}. It is therefore important to apply the key idea of this paper, decomposing a large lattice surgery into smaller fragments, to such cases. In this section, we show that the concept of this paper can be straightforwardly extended to many-body cases.

\subsection{Conditions for 3D routing for many-body lattice surgery}
In this section, we formulate a many-body variant of the 3D routing conditions, where routing is represented as a network of voxels in three dimensions connecting more than two data cells.
There are four conditions in total. Since the latter conditions depend on properties derived from the earlier ones, we explain them sequentially.\\
\textbf{Condition 1 (Tree-shaped)}: \textit{The 3D routing must not contain any loop. Each voxel at the end of 3D routing (i.e., voxels with a single connected voxel) is called a leaf. Within the routing voxels, each leaf must be a unique voxel located at a target cell, and every target cell must contain one leaf.}

\begin{figure}[ht]
  \centering
  \begin{minipage}[b]{0.3\columnwidth}
    \centering
    \includegraphics[width=\columnwidth]{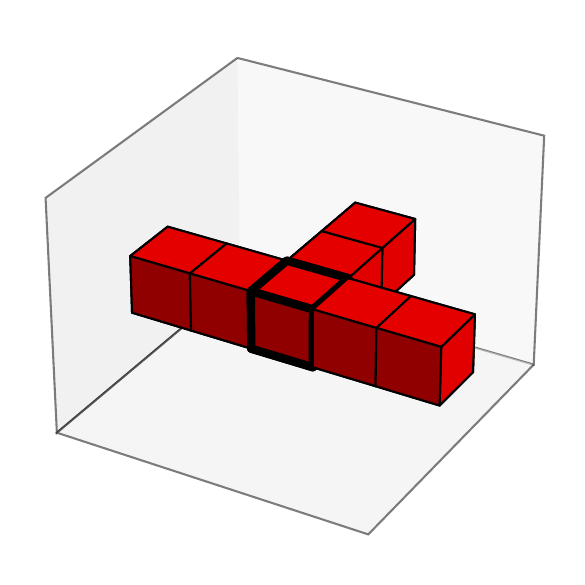}
    \subcaption{}
    \label{fig:fork_example1}
  \end{minipage}
  \begin{minipage}[b]{0.3\columnwidth}
    \centering
    \includegraphics[width=\columnwidth]{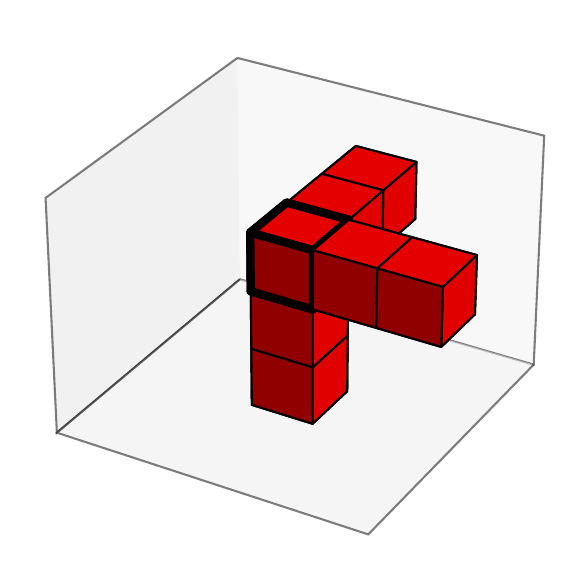}
    \subcaption{}
    \label{fig:fork_example2}
  \end{minipage}
  \begin{minipage}[b]{0.3\columnwidth}
    \centering
    \includegraphics[width=\columnwidth]{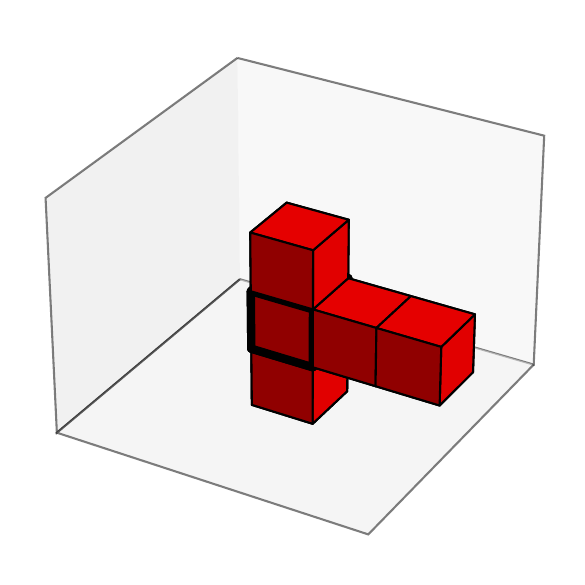}
    \subcaption{}
    \label{fig:fork_example3}
  \end{minipage}
  \caption{Examples of forks. (a) valid fork structure. (b,c) fork structures prohibited by Condition 2.}
  \label{fig:fork_example}
\end{figure}

For the second condition, we introduce the concept of \textit{fork}, a voxel connected to more than two voxels.
Figure~\ref{fig:fork_example} illustrates examples of a fork in the 3D routing, where a fork voxel is aligned with a black solid line. Then, the second condition is stated as follows.\\
\textbf{Condition 2 (No-temporal-fork)}: \textit{No voxel is connected to the top or bottom of a fork voxel.}
Thus, Fig.~\ref{fig:fork_example1} is a fork satisfying this condition while Figs.~\ref{fig:fork_example2} and \ref{fig:fork_example3} do not.

For the third condition, we need to introduce the concept of \textit{even and odd segment}. A segment is a set of voxels connected within the same XY plane in the 3D routing.
We classify all segments into even or odd as follows.
First, we select a specific leaf as a pivot leaf, and the segment connected to it is designated as an even segment. If Condition 2 is satisfied, each ancillary cell connects exactly two segments. If segments are connected to an even (odd) segment without a kink via an ancillary cell, they are labeled as an even (odd) segment. If segments are connected to an even (odd) segment with a kink via an ancillary cell, they are labeled as an odd (even) segment. Repeating this classification, all segments are labeled as either even or odd.

Then, the third and fourth conditions can be stated as follows.\\
\textbf{Condition 3 (No-leaf-in-odd-segment)}: \textit{No leaf is included in odd-segment.}\\
\textbf{Condition 4 (No-fork-in-odd-segment)}: \textit{No fork voxel is included in odd-segment.}\\
Note that one might worry that these conditions depend on the choice of the pivot leaf. However, if Condition 3 is satisfied for one choice of pivot, it remains satisfied for any other pivot choice. Therefore, the pivot leaf can be chosen arbitrarily.

\begin{figure}[ht]
  \centering
  \begin{minipage}[b]{0.45\columnwidth}
    \centering
    \includegraphics[width=\columnwidth]{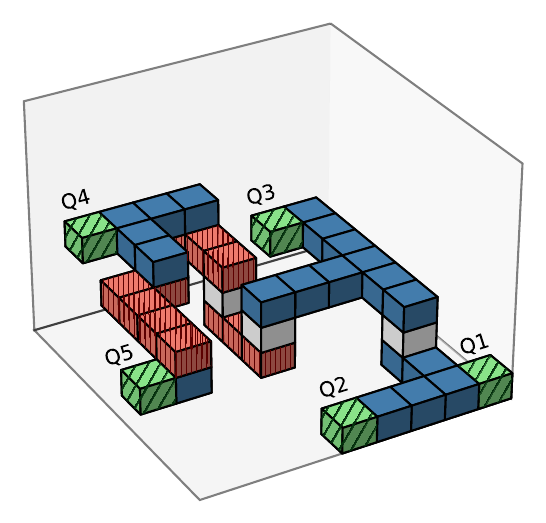}
    \subcaption{Example of segment separation.}
    \label{fig:segment_example}
  \end{minipage}
  \begin{minipage}[b]{0.4\columnwidth}
    \centering
    \includegraphics[width=\columnwidth]{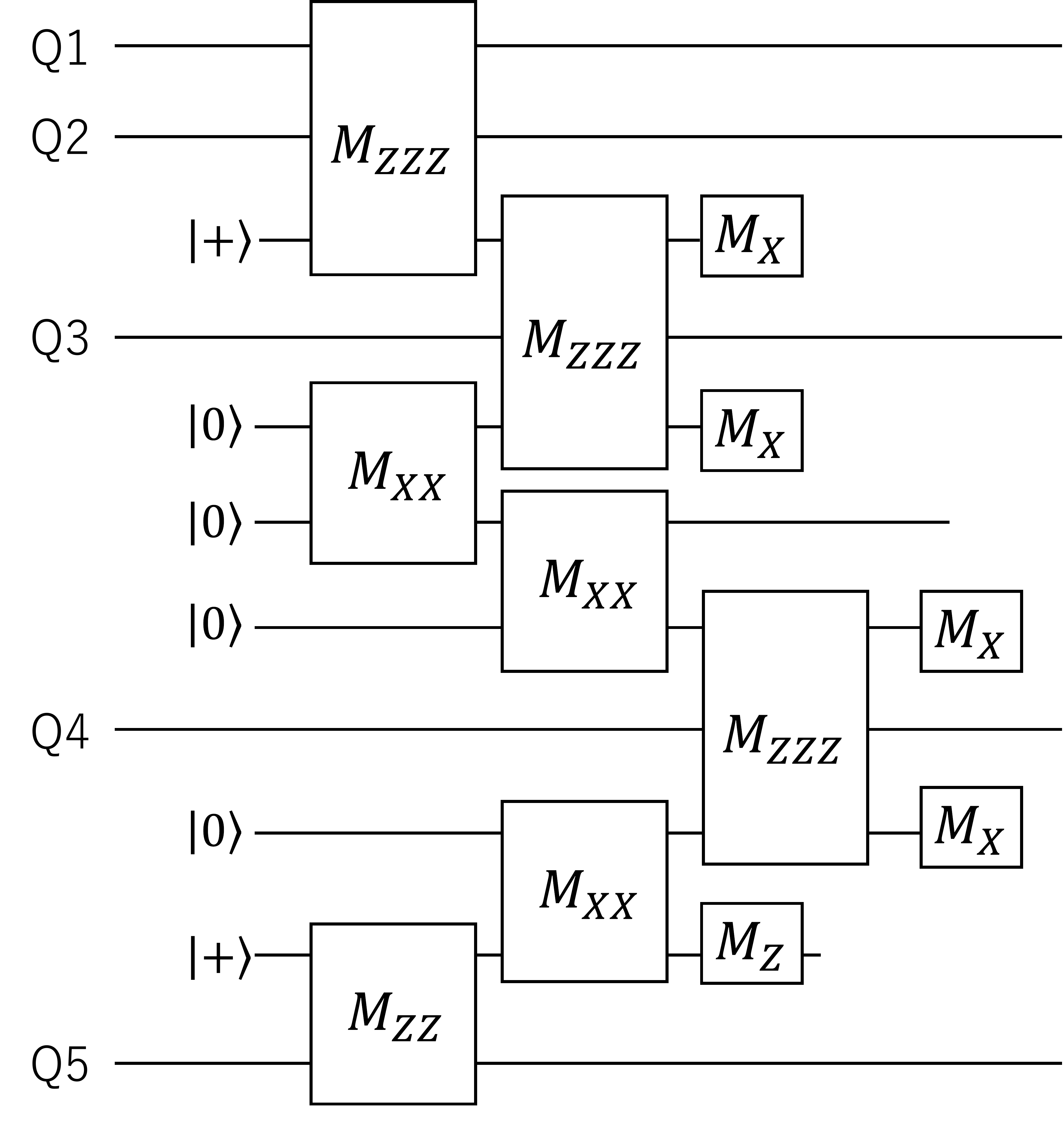}
    \subcaption{Corresponding circuits.}
    \label{fig:segment_circuit_example}
  \end{minipage}
  \caption{Example of segment separation and corresponding circuits.}
\end{figure}

Figure~\ref{fig:segment_example} shows an example of how a 3D routing that satisfies the above four conditions can be decomposed into even and odd segments. Here, diagonally hatched green voxels correspond to leaves (i.e., data cells), blue voxels are even segments, vertically hatched red voxels are odd segments, and white voxels correspond to the idling of ancillary cells. There are four even segments and three odd segments. Note that leaves are a part of even segments.

Finally, we can state the following theorem, which provides a sufficient condition for converting a 3D routing into many-body lattice surgery.
\begin{theorem}
  Suppose there are $m$ target data cells. A 3D routing satisfying the following conditions can be converted into $m$-body Pauli-$Z$($X$) measurements on $m$ target cells.
  \begin{enumerate}
    \item The 3D routing satisfies the above four conditions.
    \item The 3D routing connects $Z$($X$)-boundaries of each target data cell.
  \end{enumerate}
\end{theorem}
Thus, if all the data cells (i.e., leaves) in Fig.\,\ref{fig:segment_example} are connected via $Z$-boundaries, this 3D routing can be mapped to a five-body Pauli-$Z$ measurement.
The following subsections will show the proof of the above theorem.

Note that if we want to perform lattice surgery in which the Pauli operator contains both $X$ and $Z$, we can perform it by adjusting the position of the surface-code lattice in advance. This requires the ability to shift or rotate the shape of surface codes and to perform stabilizer Pauli measurements that involve both $X$ and $Z$~\cite{fowler2018low,litinski2019game,mcewen2023relaxing}, which would incur additional latency in scheduling.

\subsection{Proof of theorem}
This section provides the proof for many-body Pauli-$Z$ measurements. The proof for Pauli-$X$ measurement can be obtained by swapping the $X$ and $Z$ in the following explanation. First, we assign the direction of ancillary cells so that even segments are connected to a $Z$-boundary of them, and odd segments to an $X$-boundary. Given the four conditions, the 3D routing after this assignment satisfies the following properties:
\begin{itemize}
  \item Every odd segment forms a path between the $X$-boundaries of two ancillary cells.
  \item Each end of an even-segment is connected to a $Z$-boundary of an ancillary cell or target cell.
  \item Every ancillary cell bridges exactly two segments.
\end{itemize}

\begin{figure}[ht]
  \centering
  \begin{minipage}[b]{0.4\columnwidth}
    \centering
    \includegraphics[width=\columnwidth]{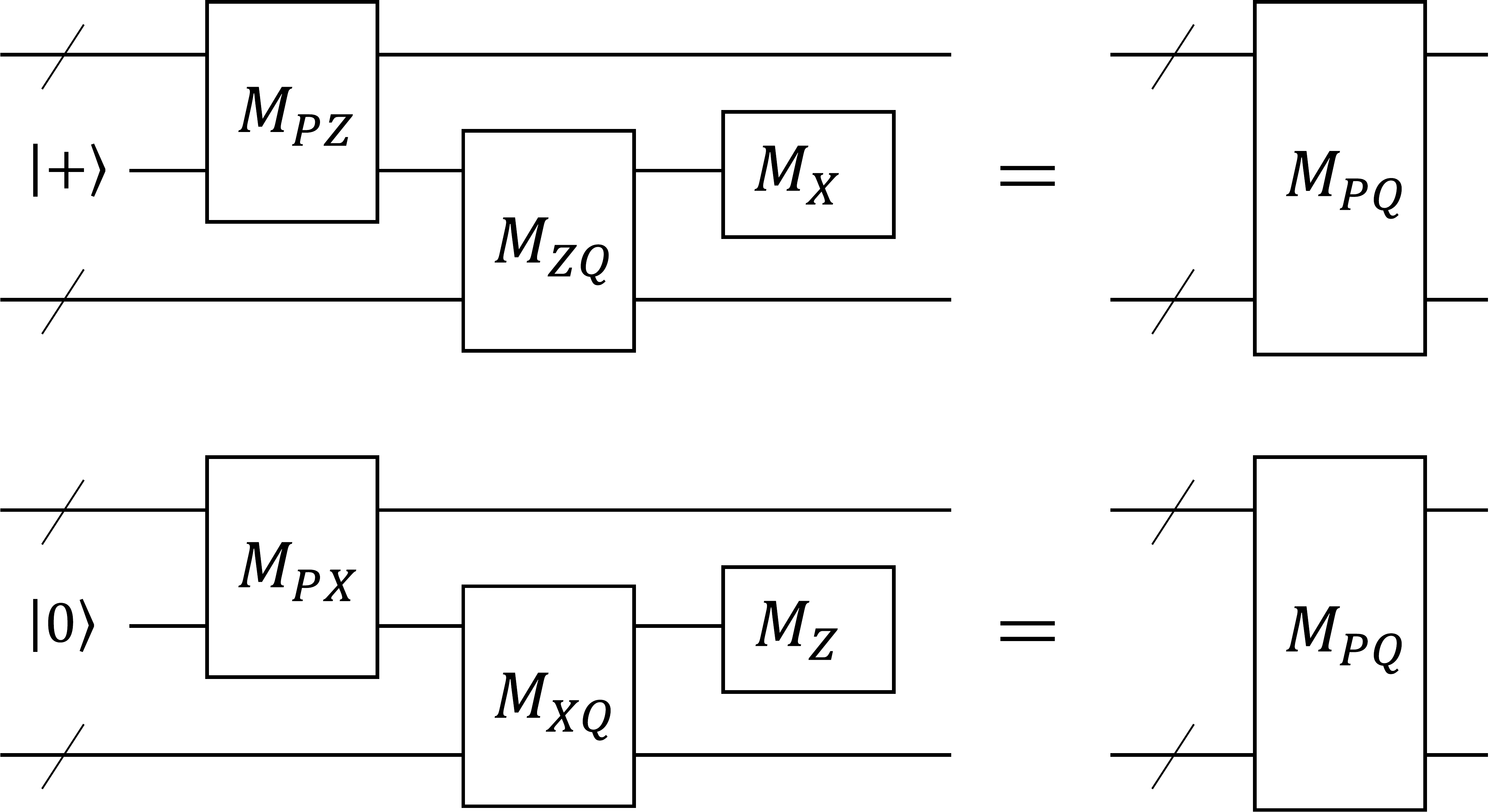}
    \subcaption{}
    \label{fig:appendix_equivalence1}
  \end{minipage}
  \hspace{0.1\columnwidth}
  \begin{minipage}[b]{0.4\columnwidth}
    \centering
    \includegraphics[width=\columnwidth]{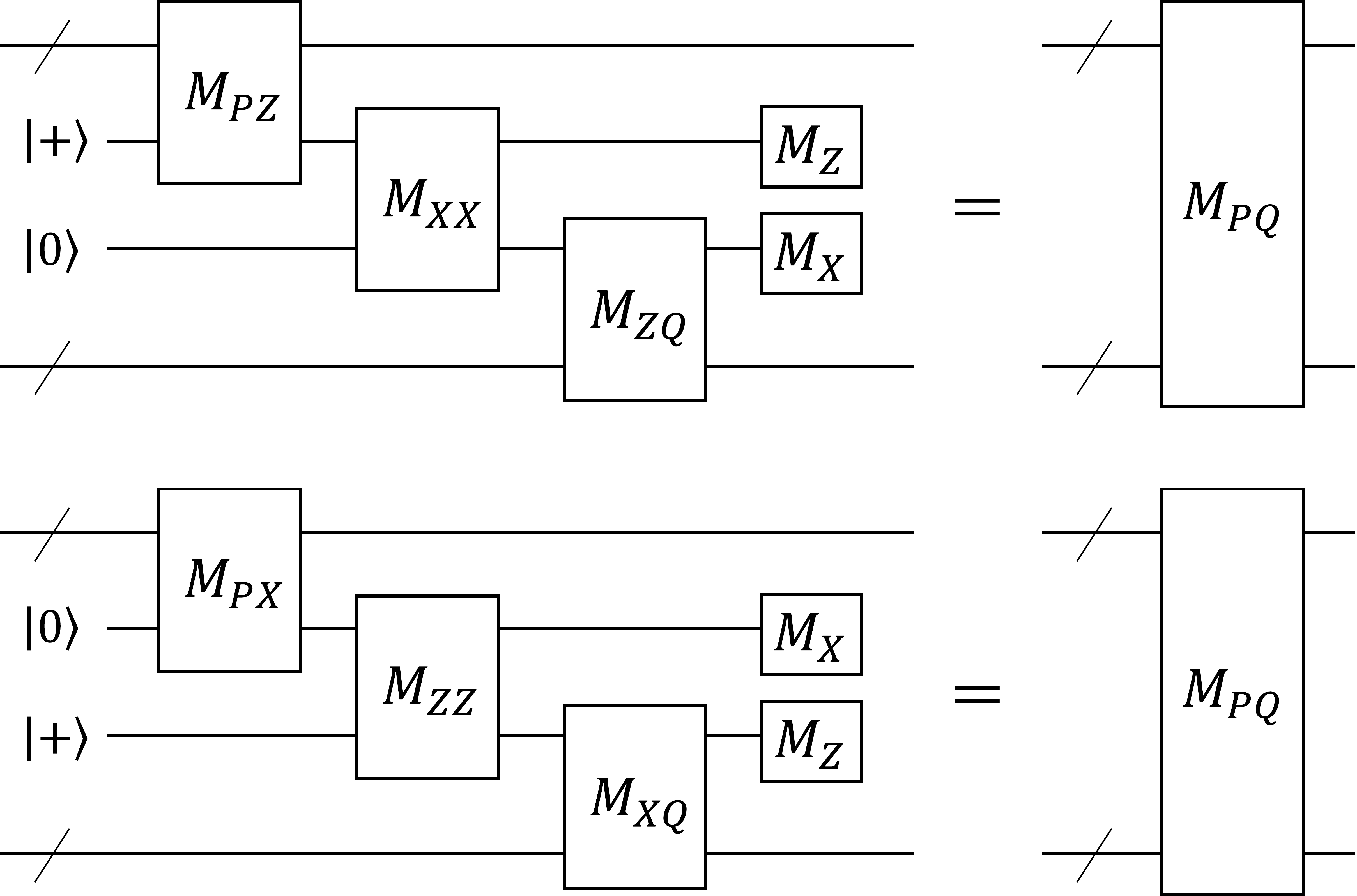}
    \subcaption{}
    \label{fig:appendix_equivalence2}
  \end{minipage}
  \\
  \begin{minipage}[b]{0.4\columnwidth}
    \centering
    \includegraphics[width=\columnwidth]{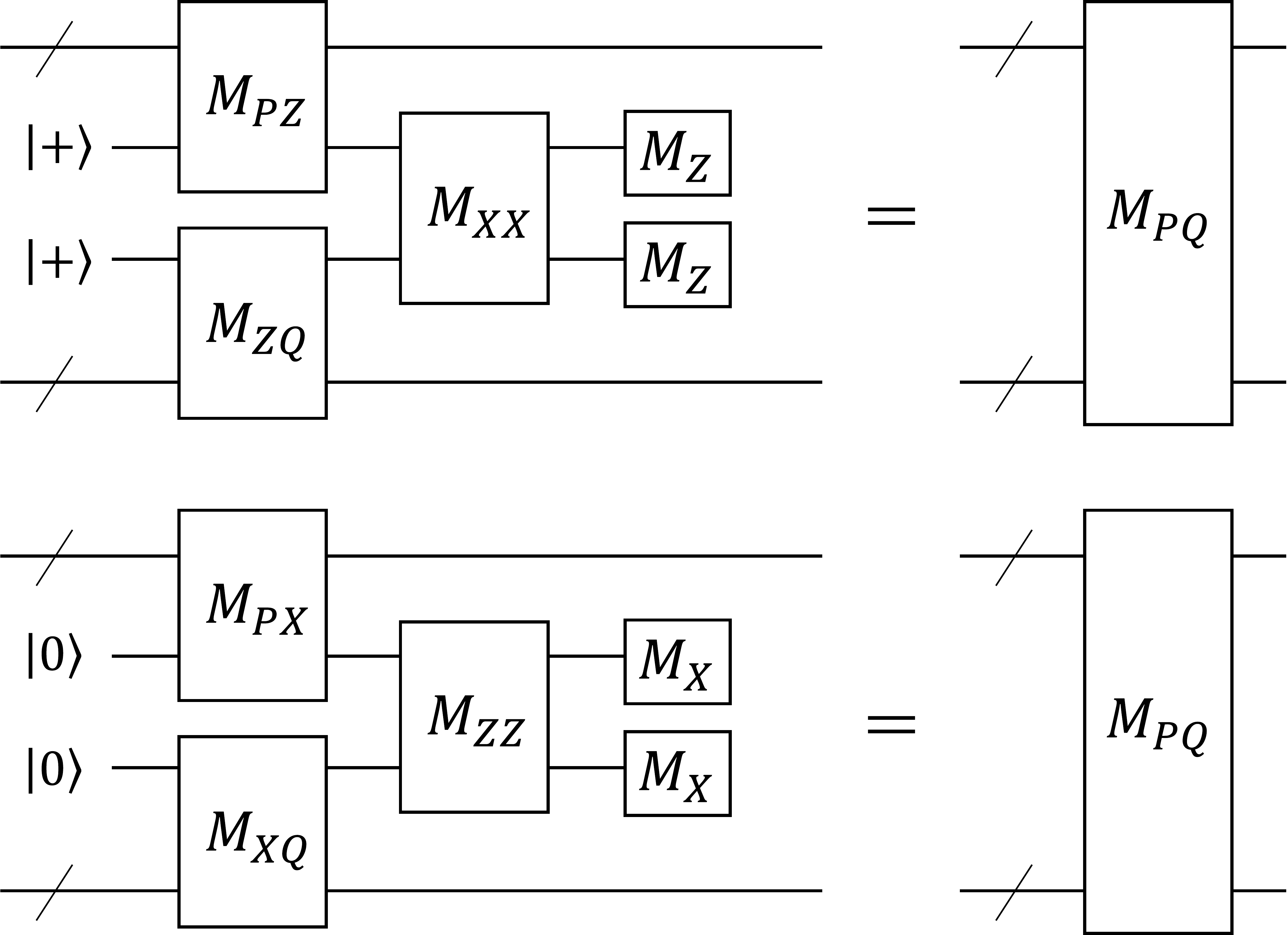}
    \subcaption{}
    \label{fig:appendix_equivalence3}
  \end{minipage}
  \hspace{0.1\columnwidth}
  \begin{minipage}[b]{0.4\columnwidth}
    \centering
    \includegraphics[width=\columnwidth]{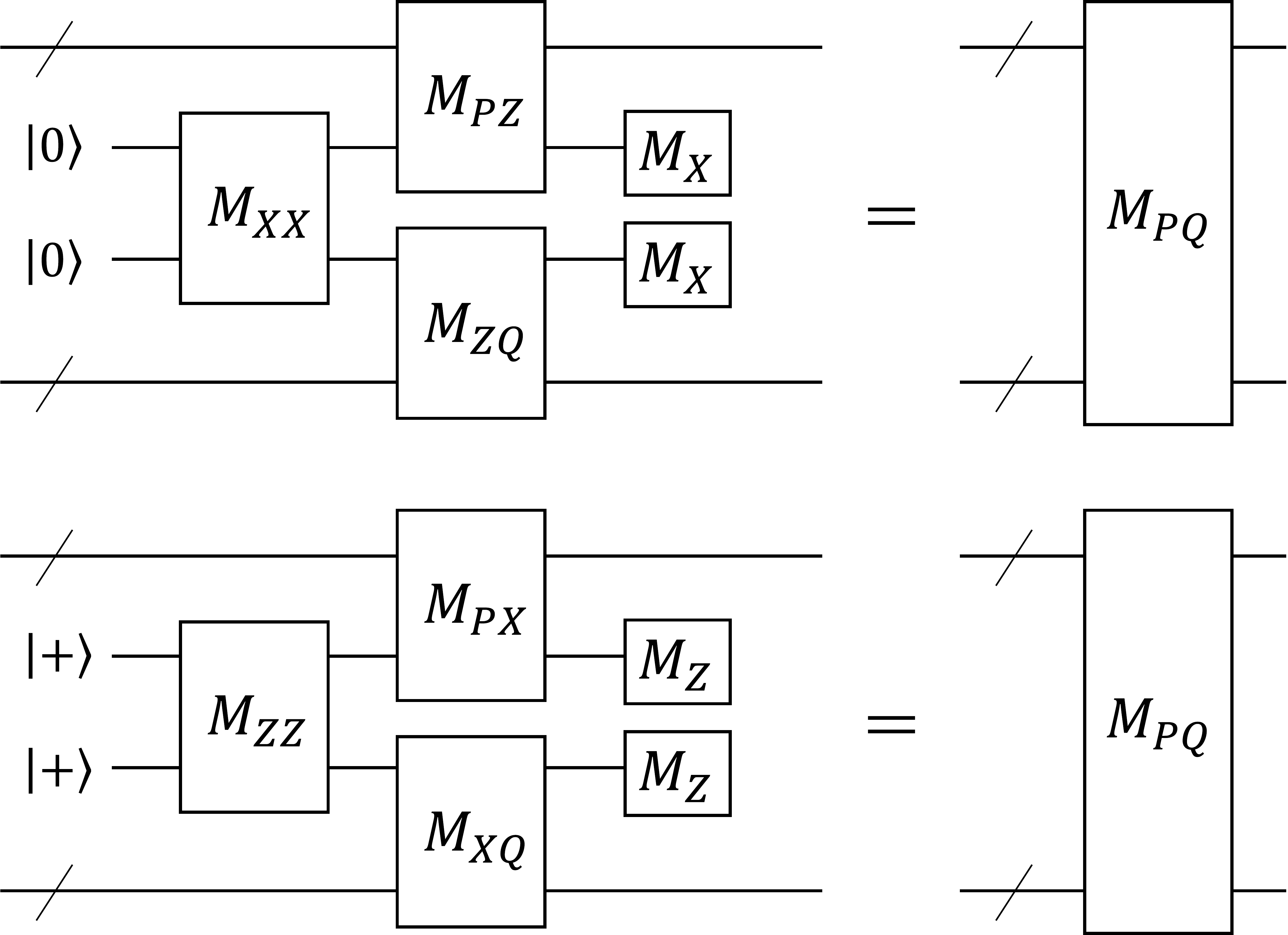}
    \subcaption{}
    \label{fig:appendix_equivalence4}
  \end{minipage}
  \caption{Equivalent processes of many-body measurements. Here, $P$ and $Q$ are an arbitrary many-body Pauli operator, and $M_{AB}$ is a Pauli measurement with projectors $(I \pm A \otimes B)/2$ on two registers. (a) Two many-body measurements that act on an ancillary logical qubit with the same basis are equal to a single many-body measurement. (b,c,d) Three alternating many-body measurements are equal to a single many-body measurement. These are many-body variants of Fig.\,\ref{fig:qc_equivalent}.}
  \label{fig:appendix_equivalent}
\end{figure}

We reduce an ancillary cell connected by two even segments into a single even segment, using the circuit equivalence shown in Fig.~\ref{fig:appendix_equivalence1}. Similarly, an ancillary cell connected by two odd segments is reduced to a single odd segment. This procedure is repeated until every ancillary cell bridges two segments of different parities.

Next, we reduce a sequence consisting of an even segment, an ancillary cell, an odd segment, another ancillary cell, and an even segment into a single even segment using the circuit equivalence shown in Figs.~\ref{fig:appendix_equivalence2}, \ref{fig:appendix_equivalence3}, ~\ref{fig:appendix_equivalence4}.
This procedure is repeated until no ancillary cells or odd segments remain. Eventually, a single even segment connects all the leaves of the original tree, which corresponds to the target many-body Pauli measurement.

Figure~\ref{fig:segment_circuit_example} shows the corresponding quantum circuits to Fig.\,\ref{fig:segment_example}. By repeating the replacement according to Fig.\,\ref{fig:appendix_equivalent}, we can convert it to five-body $Z$ measurements.

\subsection{Extension of Dijkstra projection to many-body cases}
We can also define a many-body variant of the Dijkstra projection for many-body lattice surgery on logical qubits $(Q_1, \ldots, Q_m)$ by using the two-body Dijkstra projection as a subroutine. The procedure can be summarized as follows:
\begin{enumerate}
  \item Perform the two-body Dijkstra projection for logical qubits $Q_1$ and $Q_2$.
  \item For each $i = 3, 4, \ldots, m$, perform the two-body Dijkstra projection between logical qubit $Q_i$ and the nearest $Z$-boundary of the existing even segment, and adjust the kink between them to be even.
\end{enumerate}
This procedure ensures that the final routing is tree-shaped, has no temporal forks, and that no odd segment contains any fork or leaf. Other possible implementations of the Dijkstra projection for many-body cases may exist, but their exploration is left for future work.

\section{Compilation framework and architecture of FTQC}
\label{app:compilation_framework}

This section provides a brief outline of the expected compilation procedure and the architecture of FTQC machines. Note that the explanation below serves as an example of a compilation flow; our technique can be applied to any form of compilation and any FTQC machine, as long as the underlying architecture relies on lattice-surgery techniques.

\subsection{Quantum circuit representation}
The first step is to synthesize a quantum circuit with a sequence of quantum gates defined in the form listed in Table~\ref{tab:circuit_repr}.
\begin{table}[h]
  \centering
  \caption{Quantum circuit representation}
  \label{tab:circuit_repr}
  \vspace{-3mm}
  \begin{tabular}{|l|l|l|}
    \hline
    Instruction  & Operand                             & Description              \\
    \hline \hline
    \texttt{X}   & \texttt{Q}                          & Pauli-X gate             \\
    \hline
    \texttt{Y}   & \texttt{Q}                          & Pauli-Y gate             \\
    \hline
    \texttt{Z}   & \texttt{Q}                          & Pauli-Z gate             \\
    \hline
    \texttt{H}   & \texttt{Q}                          & Hadamard gate            \\
    \hline
    \texttt{S}   & \texttt{Q}                          & Pauli-Z $\pi/4$ rotation \\
    \hline
    \texttt{T}   & \texttt{Q}                          & Pauli-Z $\pi/8$ rotation \\
    \hline
    \texttt{CX}  & \texttt{Q1} \texttt{Q1}             & CNOT gate                \\
    \hline
    \texttt{CCX} & \texttt{Q1} \texttt{Q2} \texttt{Q3} & Toffoli gate             \\
    \hline
    \texttt{RX}  & \texttt{Q} \texttt{R}               & Pauli-X rotation gate    \\
    \hline
    \texttt{RY}  & \texttt{Q} \texttt{R}               & Pauli-Y rotation gate    \\
    \hline
    \texttt{RZ}  & \texttt{Q} \texttt{R}               & Pauli-Z rotation gate    \\
    \hline
  \end{tabular}
\end{table}
In this table, the data types \texttt{Q} and \texttt{R} represent qubit symbols and rotation angles, respectively. Although control instructions such as \texttt{BRANCH}, \texttt{RETURN}, and \texttt{CALL} can be included, most quantum algorithms can be written without them. We therefore restrict ourselves to a form that omits such instructions to simplify the overall implementation.

\subsection{FTQC machine instructions}
Instructions that can be directly executed on actual FTQC machines are limited, since they must be implemented in a fault-tolerant manner. While the details of the available gate set depend on the quantum hardware, the error-correcting code, and the control strategy, a representative instruction set is shown in Table~\ref{tab:instruction_set}.
\begin{table}[h]
  \centering
  \caption{Gate representation in quantum algorithms}
  \label{tab:instruction_set}
  \vspace{-3mm}
  \begin{tabular}{|l|l|l|}
    \hline
    Instruction             & Operand                                             & Description                              \\
    \hline \hline
    \texttt{ALLOC\_QUBIT}   & \texttt{L} \texttt{Q}                               & Allocate qubits                          \\
    \hline
    \texttt{ALLOC\_FACTORY} & \texttt{L} \texttt{F}                               & Allocate magic state factory             \\
    \hline
    \texttt{INIT\_Z}        & \texttt{Q}                                          & Initialize qubits to $\ket{0}$           \\
    \hline
    \texttt{INIT\_X}        & \texttt{Q}                                          & Initialize qubits to $\ket{+}$           \\
    \hline
    \texttt{UNIT\_H}        & \texttt{Q} \texttt{L} \texttt{C}                    & Hadamard gate with rotations             \\
    \hline
    \texttt{UNIT\_S}        & \texttt{Q} \texttt{L} \texttt{C}                    & Phase gate with twist                    \\
    \hline
    \texttt{MEAS\_Z}        & \texttt{Q} \texttt{C}                               & Z-Pauli measurement                      \\
    \hline
    \texttt{MEAS\_X}        & \texttt{Q} \texttt{C}                               & X-Pauli measurement                      \\
    \hline
    \texttt{MEAS\_ZZ}       & \texttt{Q1} \texttt{Q2} \texttt{list[L]} \texttt{C} & ZZ-Pauli measurement                     \\
    \hline
    \texttt{MEAS\_XX}       & \texttt{Q1} \texttt{Q2} \texttt{list[L]} \texttt{C} & XX-Pauli measurement                     \\
    \hline
    \texttt{MEAS\_MAGIC}    & \texttt{Q} \texttt{F} \texttt{list[L]} \texttt{C}   & ZZ-Pauli measurement on target and magic \\
    \hline
    \texttt{XOR}            & \texttt{C1} \texttt{C2}                             & XOR calculation                          \\
    \hline
  \end{tabular}
\end{table}
In this format, in addition to the qubit symbol \texttt{Q}, we use the magic-state-factory symbol \texttt{F}, the location symbol \texttt{L}, and the classical-bit symbol \texttt{C}. Qubits are allocated to specific cell locations using \texttt{ALLOC\_QUBIT}. The allocated logical qubits can be initialized to $\ket{0}$ or $\ket{+}$ using the \texttt{INIT\_*} instructions. A typical FTQC requires a dedicated region for magic-state generation, which is allocated using \texttt{ALLOC\_FACTORY}.

The unitary instructions \texttt{UNIT\_*} perform logical Hadamard and phase gates acting on target cells~\cite{brown2017poking,fowler2018low}. These operations require a neighboring cell: the Hadamard gate requires any adjacent cell, while the phase gate requires one attached to the Z-boundary~\cite{beverland2022assessing}. A previously stored measurement value in \texttt{C} can be used to condition these operations.

The subsequent measurement instructions (\texttt{MEAS\_*}) are used to perform logical Pauli measurements on the target qubits and to store the outcomes in \texttt{C}. A single-qubit measurement does not require additional space, whereas multi-qubit measurements temporarily occupy the path between relevant cells. This path is specified as a sequence of cell locations \texttt{list[L]}. The \texttt{MEAS\_MAGIC} instruction is similar to \texttt{MEAS\_ZZ} but acts on a target logical qubit and a magic state generated by the factory.

The final instruction, \texttt{XOR}, performs classical-bit arithmetic on measurement outcomes, typically to compute conditions from multiple measurement results.

It is also possible to omit the path information from the operand and instead determine paths between logical qubits at runtime.
The drawback of this approach is that the path-finding process must be sufficiently fast, which can become challenging for large connectivity graphs or when the stabilizer-measurement cycle is short. On the other hand, this approach offers several advantages: (1) reduced object code size, (2) reusable functions independent of allocation status, and (3) dynamic path selection according to the runtime information, such as success timing of magic-state or entanglement generation.
We believe the optimal design choice depends on the machine architecture, and we leave this aspect as future work.

\subsection{Translation from circuit forms to instructions}
This section describes how to translate a quantum circuit representation into FTQC instructions without location symbols. Quantum gates \texttt{H} and \texttt{S} are directly replaced with \texttt{UNIT\_H} and \texttt{UNIT\_S}, respectively. We decompose \texttt{CCX} into (\texttt{CX}, \texttt{T}, \texttt{UNIT\_H}), following known constructions~\cite{nielsen2010quantum}. Next, we decompose \texttt{CX} into (\texttt{INIT\_X}, \texttt{MEAS\_ZZ}, \texttt{MEAS\_XX}, \texttt{MEAS\_Z}) and conditional (\texttt{X}, \texttt{Z}) operations using the method in Ref.~\cite{fowler2018low}. This decomposition requires an ancillary logical qubit, so we assign one ancillary qubit symbol for each data logical qubit. We then decompose the Pauli rotations (\texttt{RX}, \texttt{RY}, \texttt{RZ}) into (\texttt{UNIT\_S}, \texttt{UNIT\_H}, \texttt{T}) using the Ross–Selinger decomposition~\cite{ross2014optimal}. The \texttt{T} gate is further decomposed via magic-state consumption using (\texttt{MEAS\_MAGIC}, \texttt{MEAS\_X}) and conditional (\texttt{UNIT\_S}, \texttt{X}) gates, following Ref.~\cite{fowler2018low}.
Finally, we remove Pauli unitary operations \texttt{X}, \texttt{Y}, and \texttt{Z} and replace them with \texttt{XOR} operations using the Pauli frame technique.
After these steps, the quantum circuit is fully decomposed into \texttt{INIT\_*}, \texttt{MEAS\_*}, \texttt{UNIT\_*}, and \texttt{XOR} instructions without location operands.

Next, we perform placement and routing~\cite{yoshioka2022hunting}, which fills in the location operands (\texttt{L}) to make the executable compatible with a specific machine layout. For simplicity, we assume that all required logical qubits are allocated initially and never deallocated. In our benchmark, logical qubits are arranged on the pattern shown in Fig.~\ref{fig:qubit_plane}. The allocation rule for each evaluation is detailed in the next section.
Finally, we assign ancillary locations and paths for \texttt{UNIT\_*} and \texttt{MEAS\_*} instructions. While this process can be performed greedily, our method can be used to improve throughput.

\subsection{Instance synthesis and simplification used in this paper}
\label{app:select_conversion_detail}
For the numerical experiments in the main text, we directly synthesized the \texttt{SELECT} circuits provided in Refs.~\cite{babbush2018encoding,yoshioka2022hunting} according to the target Hamiltonian of the 2D Heisenberg model.
The \texttt{SELECT} circuits consist only of (\texttt{S}, \texttt{H}, \texttt{CX}, \texttt{CCX}), thus we do not need to mind the other gates, such as rotation gates. It is known that \texttt{CCX} in these circuits can be decomposed with four \texttt{T} gates if we know the target logical qubit is given as or returned to $\ket{0}$~\cite{babbush2018encoding,divincenzo1998quantum}.

For numerical evaluation in Sec.\,\ref{sec:select_circuit}, we performed further decompositions and removed all instructions irrelevant to lattice surgery. We also assigned the magic operand of \texttt{MEAS\_MAGIC} to one of the magic state factories in a cyclic manner, thereby reducing \texttt{MEAS\_MAGIC} to \texttt{MEAS\_ZZ}. After this conversion, we obtained an instruction sequence consisting of \texttt{MEAS\_XX} and \texttt{MEAS\_ZZ}. Using this sequence, we assigned the logical qubits and magic state factories to the data cells from the top left to the bottom right. We assume that these data cells have identical boundary orientations. To rule out the influence of magic state factories and quantify the lattice-surgery routing cost, we assumed that magic state factories produce a magic state every code beat and that the number of factories equals the number of logical qubit operands.

Regarding the comparison in Sec.\,\ref{sec:comparison_edpc}, we constructed an instruction sequence consisting of \texttt{CX} and \texttt{MEAS\_MAGIC} for compatibility with EDPC. Also, we adopted the settings of the EDPC algorithm in Ref.\,\cite{beverland2022surface}, which are outlined below. Logical qubits are assigned from top left to bottom right. It is assumed that the data cells have identical boundary orientations. Magic state factories are placed at the position $(i,j)$ such that $i$ and $j$ are even and satisfying any of the following: $i=0$, $i=W-1$, $j=0$, or $j=W-1$, where $W$ is the width of the qubit-plane cells. See Fig.~10 of Ref.\,\cite{beverland2022surface} for illustrations. It is also assumed that a magic state factory produces a single magic state every code beat.

\bibliographystyle{quantum}
\bibliography{reference_doi}

\newcommand{\MakeDoiLinksExplicit}{}
\MakeDoiLinksExplicit
\begin{thebibliography}{10}

\bibitem{shor1999polynomial}
Peter~W. Shor.
\newblock ``Polynomial-time algorithms for prime factorization and discrete
  logarithms on a quantum computer''.
\newblock SIAM Review {\bf 41}, 303--332~(1999).
\newblock
  doi:~\href{https://dx.doi.org/10.1137/S0036144598347011}{10.1137/S0036144598347011}.

\bibitem{kitaev1995quantum}
A.~Yu. Kitaev.
\newblock ``Quantum measurements and the abelian stabilizer problem''~(1995).
\newblock
  doi:~\href{https://dx.doi.org/10.48550/arXiv.quant-ph/9511026}{10.48550/arXiv.quant-ph/9511026}.
\newblock
  \href{http://arxiv.org/abs/quant-ph/9511026}{arXiv:quant-ph/9511026}.

\bibitem{lloyd1996universal}
Seth Lloyd.
\newblock ``Universal quantum simulators''.
\newblock Science {\bf 273}, 1073--1078~(1996).
\newblock
  doi:~\href{https://dx.doi.org/10.1126/science.273.5278.1073}{10.1126/science.273.5278.1073}.

\bibitem{kitaev1997quantum}
A.~Yu. Kitaev.
\newblock ``Quantum computations: algorithms and error correction''.
\newblock Russian Mathematical Surveys {\bf 52}, 1191--1249~(1997).
\newblock
  doi:~\href{https://dx.doi.org/10.1070/RM1997v052n06ABEH002155}{10.1070/RM1997v052n06ABEH002155}.

\bibitem{bravyi1998quantum}
Sergey~B. Bravyi and Alexei~Yu. Kitaev.
\newblock ``Quantum codes on a lattice with boundary''~(1998).
\newblock
  doi:~\href{https://dx.doi.org/10.48550/arXiv.quant-ph/9811052}{10.48550/arXiv.quant-ph/9811052}.
\newblock
  \href{http://arxiv.org/abs/quant-ph/9811052}{arXiv:quant-ph/9811052}.

\bibitem{fowler2012surface}
Austin~G. Fowler, Matteo Mariantoni, John~M. Martinis, and Andrew~N. Cleland.
\newblock ``Surface codes: Towards practical large-scale quantum computation''.
\newblock Physical Review A {\bf 86}, 032324~(2012).
\newblock
  doi:~\href{https://dx.doi.org/10.1103/PhysRevA.86.032324}{10.1103/PhysRevA.86.032324}.

\bibitem{horsman2012surface}
Clare Horsman, Austin~G. Fowler, Simon Devitt, and Rodney Van~Meter.
\newblock ``Surface code quantum computing by lattice surgery''.
\newblock New Journal of Physics {\bf 14}, 123011~(2012).
\newblock
  doi:~\href{https://dx.doi.org/10.1088/1367-2630/14/12/123011}{10.1088/1367-2630/14/12/123011}.

\bibitem{fowler2018low}
Austin~G. Fowler and Craig Gidney.
\newblock ``Low overhead quantum computation using lattice surgery''~(2018).
\newblock
  doi:~\href{https://dx.doi.org/doi.org/10.48550/arXiv.1808.06709}{doi.org/10.48550/arXiv.1808.06709}.
\newblock  \href{http://arxiv.org/abs/1808.06709}{arXiv:1808.06709}.

\bibitem{beverland2022assessing}
Michael~E. Beverland, Prakash Murali, Matthias Troyer, Krysta~M. Svore, Torsten
  Hoeffler, Vadym Kliuchnikov, Guang~Hao Low, Mathias Soeken, Aarthi Sundaram,
  and Alexander Vaschillo.
\newblock ``Assessing requirements to scale to practical quantum
  advantage''~(2022).
\newblock
  doi:~\href{https://dx.doi.org/10.48550/arXiv.2211.07629}{10.48550/arXiv.2211.07629}.
\newblock  \href{http://arxiv.org/abs/2211.07629}{arXiv:2211.07629}.

\bibitem{babbush2018encoding}
Ryan Babbush, Craig Gidney, Dominic~W. Berry, Nathan Wiebe, Jarrod McClean,
  Alexandru Paler, Austin Fowler, and Hartmut Neven.
\newblock ``Encoding electronic spectra in quantum circuits with linear t
  complexity''.
\newblock Physical Review X {\bf 8}, 041015~(2018).
\newblock
  doi:~\href{https://dx.doi.org/10.1103/PhysRevX.8.041015}{10.1103/PhysRevX.8.041015}.

\bibitem{yoshioka2022hunting}
Nobuyuki Yoshioka, Tsuyoshi Okubo, Yasunari Suzuki, Yuki Koizumi, and Wataru
  Mizukami.
\newblock ``Hunting for quantum-classical crossover in condensed matter
  problems''.
\newblock npj Quantum Information {\bf 10}, 45~(2024).
\newblock
  doi:~\href{https://dx.doi.org/10.1038/s41534-024-00839-4}{10.1038/s41534-024-00839-4}.

\bibitem{javadiabhari2014scaffcc}
Ali JavadiAbhari, Shruti Patil, Daniel Kudrow, Jeff Heckey, Alexey Lvov,
  Frederic~T. Chong, and Margaret Martonosi.
\newblock ``{ScaffCC: A framework for compilation and analysis of quantum
  computing programs}''.
\newblock In Proceedings of the 11th ACM Conference on Computing Frontiers.
\newblock CF '14. Association for Computing Machinery~(2014).
\newblock
  doi:~\href{https://dx.doi.org/10.1145/2597917.2597939}{10.1145/2597917.2597939}.

\bibitem{Qiskit}
{Qiskit contributors}.
\newblock ``Qiskit: An open-source framework for quantum computing''.
\newblock
  url:~\href{https://doi.org/10.5281/zenodo.2573505}{doi.org/10.5281/zenodo.2573505}.

\bibitem{sivarajah2020t}
Seyon Sivarajah, Silas Dilkes, Alexander Cowtan, Will Simmons, Alec Edgington,
  and Ross Duncan.
\newblock ``{t|ket>: a retargetable compiler for NISQ devices}''.
\newblock Quantum Science and Technology {\bf 6}, 014003~(2020).
\newblock
  doi:~\href{https://dx.doi.org/10.1088/2058-9565/ab8e92}{10.1088/2058-9565/ab8e92}.

\bibitem{watkins2023high}
George Watkins, Hoang~Minh Nguyen, Keelan Watkins, Steven Pearce, Hoi-Kwan Lau,
  and Alexandru Paler.
\newblock ``A high performance compiler for very large scale surface code
  computations''.
\newblock {Quantum} {\bf 8}, 1354~(2024).
\newblock
  doi:~\href{https://dx.doi.org/10.22331/q-2024-05-22-1354}{10.22331/q-2024-05-22-1354}.

\bibitem{herr2017optimization}
Daniel Herr, Franco Nori, and Simon~J. Devitt.
\newblock ``Optimization of lattice surgery is np-hard''.
\newblock Npj quantum information {\bf 3}, 35~(2017).
\newblock
  doi:~\href{https://dx.doi.org/10.1038/s41534-017-0035-1}{10.1038/s41534-017-0035-1}.

\bibitem{molavi2023compilation}
Abtin Molavi, Amanda Xu, Swamit Tannu, and Aws Albarghouthi.
\newblock ``Compilation for surface code quantum computers''~(2023).
\newblock
  doi:~\href{https://dx.doi.org/10.48550/arXiv.2311.18042}{10.48550/arXiv.2311.18042}.
\newblock  \href{http://arxiv.org/abs/2311.18042}{arXiv:2311.18042}.

\bibitem{litinski2019game}
Daniel Litinski.
\newblock ``A game of surface codes: Large-scale quantum computing with lattice
  surgery''.
\newblock Quantum {\bf 3}, 128~(2019).
\newblock
  doi:~\href{https://dx.doi.org/10.22331/q-2019-03-05-128}{10.22331/q-2019-03-05-128}.

\bibitem{litinski2019magic}
Daniel Litinski.
\newblock ``Magic state distillation: Not as costly as you think''.
\newblock Quantum {\bf 3}, 205~(2019).
\newblock
  doi:~\href{https://dx.doi.org/10.22331/q-2019-12-02-205}{10.22331/q-2019-12-02-205}.

\bibitem{lao2018mapping}
Lingling Lao, Bas van Wee, Imran Ashraf, J.~van Someren, Nader Khammassi, Koen
  Bertels, and Carmen~G. Almudever.
\newblock ``Mapping of lattice surgery-based quantum circuits on surface code
  architectures''.
\newblock Quantum Science and Technology {\bf 4}, 015005~(2018).
\newblock
  doi:~\href{https://dx.doi.org/10.1088/2058-9565/aadd1a}{10.1088/2058-9565/aadd1a}.

\bibitem{beverland2022surface}
Michael Beverland, Vadym Kliuchnikov, and Eddie Schoute.
\newblock ``Surface code compilation via edge-disjoint paths''.
\newblock PRX Quantum {\bf 3}, 020342~(2022).
\newblock
  doi:~\href{https://dx.doi.org/10.1103/PRXQuantum.3.020342}{10.1103/PRXQuantum.3.020342}.

\bibitem{low2019hamiltonian}
Guang~Hao Low and Isaac~L. Chuang.
\newblock ``Hamiltonian simulation by qubitization''.
\newblock Quantum {\bf 3}, 163~(2019).
\newblock
  doi:~\href{https://dx.doi.org/10.22331/q-2019-07-12-163}{10.22331/q-2019-07-12-163}.

\bibitem{fowler2012towards}
Austin~G. Fowler, Adam~C. Whiteside, and Lloyd C.~L. Hollenberg.
\newblock ``Towards practical classical processing for the surface code''.
\newblock Physical review letters {\bf 108}, 180501~(2012).
\newblock
  doi:~\href{https://dx.doi.org/10.1103/PhysRevLett.108.180501}{10.1103/PhysRevLett.108.180501}.

\bibitem{acharya2022suppressing}
Rajeev Acharya, Igor Aleiner, Richard Allen, Trond~I. Andersen, Markus Ansmann,
  Frank Arute, Kunal Arya, Abraham Asfaw, Juan Atalaya, Ryan Babbush, et~al.
\newblock ``Suppressing quantum errors by scaling a surface code logical
  qubit''.
\newblock Nature {\bf 614}, 676--681~(2023).
\newblock
  doi:~\href{https://dx.doi.org/10.1038/s41586-022-05434-1}{10.1038/s41586-022-05434-1}.

\bibitem{brown2017poking}
Benjamin~J. Brown, Katharina Laubscher, Markus~S. Kesselring, and James~R.
  Wootton.
\newblock ``Poking holes and cutting corners to achieve clifford gates with the
  surface code''.
\newblock Physical Review X {\bf 7}, 021029~(2017).
\newblock
  doi:~\href{https://dx.doi.org/10.1103/PhysRevX.7.021029}{10.1103/PhysRevX.7.021029}.

\bibitem{fredman1987fibonacci}
Michael~L. Fredman and Robert~Endre Tarjan.
\newblock ``Fibonacci heaps and their uses in improved network optimization
  algorithms''.
\newblock J. ACM {\bf 34}, 596–615~(1987).
\newblock
  doi:~\href{https://dx.doi.org/10.1145/28869.28874}{10.1145/28869.28874}.

\bibitem{silva2024multi}
Allyson Silva, Xiangyi Zhang, Zak Webb, Mia Kramer, Chan-Woo Yang, Xiao Liu,
  Jessica Lemieux, Ka-Wai Chen, Artur Scherer, and Pooya Ronagh.
\newblock ``Multi-qubit lattice surgery scheduling''.
\newblock In 19th Conference on the Theory of Quantum Computation,
  Communication and Cryptography (TQC 2024).
\newblock Pages 1--1.
\newblock Schloss Dagstuhl--Leibniz-Zentrum f{\"u}r Informatik~(2024).
\newblock
  url:~\href{https://doi.org/10.4230/lipics.tqc.2024.1}{doi.org/10.4230/lipics.tqc.2024.1}.

\bibitem{fowler2012time}
Austin~G. Fowler.
\newblock ``Time-optimal quantum computation''~(2012).
\newblock
  doi:~\href{https://dx.doi.org/10.48550/arXiv.1210.4626}{10.48550/arXiv.1210.4626}.
\newblock  \href{http://arxiv.org/abs/1210.4626}{arXiv:1210.4626}.

\bibitem{zhu2024ecmas}
Mingzheng Zhu, Hao Fu, Jun Wu, Chi Zhang, Wei Xie, and Xiang-Yang Li.
\newblock ``Ecmas: Efficient circuit mapping and scheduling for surface code''.
\newblock In 2024 IEEE/ACM International Symposium on Code Generation and
  Optimization (CGO).
\newblock Page 158–169.
\newblock IEEE~(2024).
\newblock
  doi:~\href{https://dx.doi.org/10.1109/cgo57630.2024.10444874}{10.1109/cgo57630.2024.10444874}.

\bibitem{zhu2025ecmas+}
Mingzheng Zhu, Hao Fu, Haishan Song, Jun Wu, Chi Zhang, Wei Xie, and Xiangyang
  Li.
\newblock ``Ecmas+: Efficient circuit mapping and scheduling for surface code
  encoded circuit on quantum cloud platform''.
\newblock ACM Transactions on Architecture and Code Optimization {\bf 22},
  1–25~(2025).
\newblock  doi:~\href{https://dx.doi.org/10.1145/3760783}{10.1145/3760783}.

\bibitem{google2023suppressing}
{Google Quantum AI}.
\newblock ``Suppressing quantum errors by scaling a surface code logical
  qubit''.
\newblock Nature {\bf 614}, 676–681~(2023).
\newblock
  doi:~\href{https://dx.doi.org/10.1038/s41586-022-05434-1}{10.1038/s41586-022-05434-1}.

\bibitem{google2025quantum}
{Google Quantum AI and Collaborators}.
\newblock ``Quantum error correction below the surface code threshold''.
\newblock Nature {\bf 638}, 920–926~(2024).
\newblock
  doi:~\href{https://dx.doi.org/10.1038/s41586-024-08449-y}{10.1038/s41586-024-08449-y}.

\bibitem{zhou2025low}
Hengyun Zhou, Chen Zhao, Madelyn Cain, Dolev Bluvstein, Nishad Maskara, Casey
  Duckering, Hong-Ye Hu, Sheng-Tao Wang, Aleksander Kubica, and Mikhail~D.
  Lukin.
\newblock ``Low-overhead transversal fault tolerance for universal quantum
  computation''.
\newblock Nature {\bf 646}, 303–308~(2025).
\newblock
  doi:~\href{https://dx.doi.org/10.1038/s41586-025-09543-5}{10.1038/s41586-025-09543-5}.

\bibitem{gidney2025yoked}
Craig Gidney, Michael Newman, Peter Brooks, and Cody Jones.
\newblock ``Yoked surface codes''.
\newblock Nature Communications{\bf 16}~(2025).
\newblock
  doi:~\href{https://dx.doi.org/10.1038/s41467-025-59714-1}{10.1038/s41467-025-59714-1}.

\bibitem{yoder2025tour}
Theodore~J. Yoder, Eddie Schoute, Patrick Rall, Emily Pritchett, Jay~M.
  Gambetta, Andrew~W. Cross, Malcolm Carroll, and Michael~E. Beverland.
\newblock ``Tour de gross: A modular quantum computer based on bivariate
  bicycle codes''~(2025).
\newblock
  doi:~\href{https://dx.doi.org/10.48550/ARXIV.2506.03094}{10.48550/ARXIV.2506.03094}.

\bibitem{xu2024constant}
Qian Xu, J.~Pablo Bonilla~Ataides, Christopher~A. Pattison, Nithin Raveendran,
  Dolev Bluvstein, Jonathan Wurtz, Bane Vasić, Mikhail~D. Lukin, Liang Jiang,
  and Hengyun Zhou.
\newblock ``Constant-overhead fault-tolerant quantum computation with
  reconfigurable atom arrays''.
\newblock Nature Physics {\bf 20}, 1084–1090~(2024).
\newblock
  doi:~\href{https://dx.doi.org/10.1038/s41567-024-02479-z}{10.1038/s41567-024-02479-z}.

\bibitem{cohen2022low}
Lawrence~Z. Cohen, Isaac~H. Kim, Stephen~D. Bartlett, and Benjamin~J. Brown.
\newblock ``Low-overhead fault-tolerant quantum computing using long-range
  connectivity''.
\newblock Science Advances{\bf 8}~(2022).
\newblock
  doi:~\href{https://dx.doi.org/10.1126/sciadv.abn1717}{10.1126/sciadv.abn1717}.

\bibitem{nickerson2013topological}
Naomi~H Nickerson, Ying Li, and Simon~C Benjamin.
\newblock ``Topological quantum computing with a very noisy network and local
  error rates approaching one percent''.
\newblock Nature communications {\bf 4}, 1756~(2013).
\newblock
  url:~\href{https://doi.org/10.1038/ncomms2773}{doi.org/10.1038/ncomms2773}.

\bibitem{bombin2021interleaving}
Hector Bombin, Isaac~H Kim, Daniel Litinski, Naomi Nickerson, Mihir Pant,
  Fernando Pastawski, Sam Roberts, and Terry Rudolph.
\newblock ``Interleaving: Modular architectures for fault-tolerant photonic
  quantum computing''~(2021).
\newblock
  url:~\href{https://doi.org/10.48550/arXiv.2103.08612}{doi.org/10.48550/arXiv.2103.08612}.

\bibitem{bourassa2021blueprint}
J~Eli Bourassa, Rafael~N Alexander, Michael Vasmer, Ashlesha Patil, Ilan
  Tzitrin, Takaya Matsuura, Daiqin Su, Ben~Q Baragiola, Saikat Guha, Guillaume
  Dauphinais, et~al.
\newblock ``Blueprint for a scalable photonic fault-tolerant quantum
  computer''.
\newblock Quantum {\bf 5}, 392~(2021).
\newblock
  url:~\href{https://doi.org/10.22331/q-2021-02-04-392}{doi.org/10.22331/q-2021-02-04-392}.

\bibitem{pattison2025constant}
Christopher Pattison, Gefen Baranes, Juan~Pablo Bonilla~Ataides, Mikhail~D
  Lukin, and Hengyun Zhou.
\newblock ``Constant-rate entanglement distillation for fast quantum
  interconnects''.
\newblock In Proceedings of the 52nd Annual International Symposium on Computer
  Architecture.
\newblock Pages 257--270.
\newblock ~(2025).
\newblock
  url:~\href{https://doi.org/10.1145/3695053.373106}{doi.org/10.1145/3695053.373106}.

\bibitem{maeda2025logical}
Yuya Maeda, Yasunari Suzuki, Toshiki Kobayashi, Takashi Yamamoto, Yuuki
  Tokunaga, and Keisuke Fujii.
\newblock ``Logical entanglement distribution between distant 2d array
  qubits''~(2025).
\newblock
  url:~\href{https://doi.org/10.48550/arXiv.2503.14894}{doi.org/10.48550/arXiv.2503.14894}.

\bibitem{chamberland2022universal}
Christopher Chamberland and Earl~T Campbell.
\newblock ``Universal quantum computing with twist-free and temporally encoded
  lattice surgery''.
\newblock PRX Quantum {\bf 3}, 010331~(2022).
\newblock
  url:~\href{https://doi.org/10.1103/PRXQuantum.3.010331}{doi.org/10.1103/PRXQuantum.3.010331}.

\bibitem{mcewen2023relaxing}
Matt McEwen, Dave Bacon, and Craig Gidney.
\newblock ``Relaxing hardware requirements for surface code circuits using
  time-dynamics''.
\newblock Quantum {\bf 7}, 1172~(2023).
\newblock
  url:~\href{https://doi.org/10.22331/q-2023-11-07-1172}{doi.org/10.22331/q-2023-11-07-1172}.

\bibitem{nielsen2010quantum}
Michael~A Nielsen and Isaac~L Chuang.
\newblock ``Quantum computation and quantum information''.
\newblock Cambridge university press. ~(2010).
\newblock
  url:~\href{https://doi.org/10.1017/CBO9780511976667}{doi.org/10.1017/CBO9780511976667}.

\bibitem{ross2014optimal}
Neil~J Ross and Peter Selinger.
\newblock ``Optimal ancilla-free clifford+ t approximation of
  z-rotations''~(2014).
\newblock
  url:~\href{https://doi.org/10.48550/arXiv.1403.2975}{doi.org/10.48550/arXiv.1403.2975}.

\bibitem{divincenzo1998quantum}
David~P DiVincenzo.
\newblock ``Quantum gates and circuits''.
\newblock Proceedings of the Royal Society of London. Series A: Mathematical,
  Physical and Engineering Sciences {\bf 454}, 261--276~(1998).
\newblock
  url:~\href{https://doi.org/10.1098/rspa.1998.0159}{doi.org/10.1098/rspa.1998.0159}.

\end{thebibliography}

\end{document}